\documentclass[pdflatex,sn-mathphys-num]{sn-jnl}% Math and Physical Sciences Numbered Reference Style 
\graphicspath{{./figures/}}

\usepackage{enumitem}
\usepackage{xspace}
\usepackage{fancyhdr}
\usepackage{afterpage}
\usepackage[usenames,dvipsnames]{xcolor}
\usepackage{dcolumn}
\usepackage{comment}
\usepackage{listings}
\usepackage{makeidx}
\usepackage{longtable}
\usepackage{chemformula}
\usepackage[utf8]{inputenc}
\usepackage{comment}

\usepackage[dvips]{epsfig}
\usepackage{epstopdf}
\usepackage{graphicx}% Include figure files

\usepackage{booktabs}

\usepackage{mathtools}
\usepackage{amsmath}
\usepackage{amssymb}
\usepackage{bm}
\usepackage{amsfonts}
\usepackage{commath} %http://ctan.mackichan.com/macros/latex/contrib/commath/commath.pdf
\usepackage{url}
\usepackage{natbib} % nature bib style
\newcommand{\qmax}{\ensuremath{Q_{\mathrm{max}}}\xspace}

\newcommand{\q}{\ensuremath{Q}\xspace}
\newcommand{\ir}{\ensuremath{r}\xspace}

\newcommand{\iaa}{\AA\ensuremath{^{-1}}\xspace}

\newcommand{\sjba}[1]{}

\newcommand{\mta}[1]{}

\newcommand{\fig}[1]{Fig.~\ref{fig:#1}}

\graphicspath{ {./figures/} }

\definecolor{dgreen}{HTML}{008000}

\definecolor{codegreen}{rgb}{0,0.6,0}
\definecolor{codegray}{rgb}{0.5,0.5,0.5}
\definecolor{codepurple}{rgb}{0.58,0,0.82}
\definecolor{backcolour}{rgb}{0.95,0.95,0.92}
\lstdefinestyle{mystyle}{
    backgroundcolor=\color{backcolour},
    commentstyle=\color{codegreen},
    keywordstyle=\color{magenta},
    numberstyle=\tiny\color{codegray},
    stringstyle=\color{codepurple},
    basicstyle=\ttfamily\footnotesize,
    breakatwhitespace=false,
    breaklines=true,
    captionpos=b,
    keepspaces=true,
    numbers=left,
    numbersep=5pt,
    showspaces=false,
    showstringspaces=false,
    showtabs=false,
    tabsize=2
}
% ...
% ...
% ...
\newcounter{saveenumi}

\newcommand{\be}{\begin{enumerate}}
\newcommand{\ee}{\end{enumerate}}
\newcommand{\bes}{\begin{enumerate}[wide, labelwidth=!, labelindent=0pt, label=\textbf{\textcolor{blue}{\arabic*}.}]}
\newcommand{\ees}{\end{enumerate}}

\newcommand{\cmi}{{\sc Diffpy-CMI}\xspace}

\definecolor{sobcolor}{HTML}{176d1b}
\definecolor{tillscolor}{RGB}{20, 168, 42}

\newcommand{\twoplus}{$^{2+}$\xspace}
\newcommand{\threeplus}{$^{3+}$\xspace}
\newcommand{\fourplus}{$^{4+}$\xspace}

\begin{document}                  % DO NOT DELETE THIS LINE

     %-------------------------------------------------------------------------
     % The introductory (header) part of the paper
     %-------------------------------------------------------------------------

\title[Interpretable Multimodal Machine Learning Analysis of X-ray Absorption Near-Edge Spectra and Pair Distribution Functions]{Interpretable Multimodal Machine Learning Analysis of X-ray Absorption Near-Edge Spectra and Pair Distribution Functions}

\author[1]{\fnm{Tanaporn} \sur{Na Narong}}\email{tn2539@columbia.edu}
\author[1]{\fnm{Zoe N.} \sur{Zachko}}\email{zz2828@columbia.edu}
\author*[2]{\fnm{Steven B.} \sur{Torrisi}}\email{steven.torrisi@tri.global}
\author*[1]{\fnm{Simon J. L.} \sur{Billinge}}\email{sb2896@columbia.edu}

\affil*[1]{\orgdiv{Department of Applied Physics and Applied Mathematics}, \orgname{Columbia University}, \orgaddress{\street{500 W 120th St}, \city{New York}, \postcode{10027}, \state{NY}, \country{United States}}}

\affil[2]{\orgdiv{Energy \& Materials Division}, \orgname{Toyota Research Institute}, \orgaddress{\street{4440 El Camino Real}, \city{Los Altos}, \postcode{94022}, \state{CA}, \country{United States}}}

\abstract{
We used interpretable machine learning to combine information from multiple heterogeneous spectra: X-ray absorption near-edge spectra (XANES) and atomic pair distribution functions (PDFs) to extract local structural and chemical environments of transition metal cations in oxides.
Random forest models were trained on simulated XANES, PDF, and both combined to extract oxidation state, coordination number, and mean nearest-neighbor bond length.
XANES-only models generally outperformed PDF-only models, even for structural tasks, although using the metal’s differential PDFs (dPDFs) instead of total PDFs narrowed this gap.
When combined with PDFs, information from XANES often dominates the prediction.
Our results demonstrate that XANES contains rich structural information and highlight the utility of species-specificity.
This interpretable, multimodal approach is quick to implement with suitable databases and offers valuable insights into the relative strengths of different modalities, guiding researchers in experiment design and identifying when combining complementary techniques adds meaningful information to a scientific investigation.
}

\maketitle                        % DO NOT DELETE THIS LINE
     %-------------------------------------------------------------------------
     % The main body of the paper
     %-------------------------------------------------------------------------

\section{Introduction}
%\begin{enumerate}
    %\item \done
    
    Extracting local atomic structure is crucial for characterizing materials and understanding their properties \cite{egami;b;utbp12}. 
    This is often achieved by analyzing scattering or spectroscopy data, but sometimes it is not possible to obtain the relevant details from one method alone due to limited data quality or problem complexity \cite{billi;s07,billi;p10}. 
    Researchers often fit an internal model of a system to experimental observations via optimization \cite{mcgreevy1988reverse, billingeRealspaceRietveldFull1998b}. For complex cases, it may be desirable to combining information from two or more different experiments to obtain the desired information.
    This is common in scientific literature where many characterizations are done on the same sample, but a greater challenge is to use multiple experiments to refine parameters from a single model, a process known as complex modeling \cite{juhas;aca15} or multimodal data fusion \cite{lahat2015multimodal}.
    This has proved more challenging than expected due to the heterogeneous nature of information in different experiments and possible incompatibility of systematic errors \cite{farrowunpublished}.
    Few attempts have been made to combine data from two different structural probes in a rational co-refinement. 
    One approach combined pair distribution functions (PDFs) \cite{egami;b;utbp12, billinge2023atomic} and small-angle x-ray scattering (SAXS) \cite{farrow2014robust}, while another merged extended X-ray absorption fine structure (EXAFS) and
    total scattering/PDF data in reverse Monte Carlo (RMC) refinements \cite{tucke;jpcm07, krayzman2008simultaneous}.
    However in general, traditional methods still struggle to integrate heterogeneous information from different experiments, in large part because we lack \emph{a priori} knowledge on how to weight the contributions from each measurement in the cost function. 
    
    Data-driven methods are an attractive alternative, in which a precise representation of the system is bypassed and the model may simply predict properties of interest from data.
    Here we explore whether machine learning (ML) can address the challenges of combining different spectral measurement modalities, leveraging its ability to extract insights from complex datasets and handle heterogeneous inputs.
    Instead of direct structural inversion, partial characterization that reveals the coordination number or average bond length can help practitioners rule out certain candidate structures.
    Here we present an interpretable ML approach to extract and combine local structure information from two distinct measurement methods: X-ray near-edge absorption spectra (XANES) and atomic pair distribution functions (PDFs).

    %\item \done
    
    Both XANES and PDF are powerful and popular techniques to probe local atomic environments \cite{yano2009x,egami;b;utbp12}.
    They are also both well-suited to a database approach. Theoretical X-ray absorption spectra are relatively inexpensive to obtain using FEFF \cite{rehr2010parameter} and are a part of the Materials Project (MP) database \cite{mathew2018high,jain2013commentary}, and PDFs are straightforward to compute with \cmi \cite{juhas;aca15,billinge2023atomic} from the atomic coordinates.
    These two methods present an interesting pair of modalities because their information content may overlap despite their differing physical origins.
    X-ray absorption at an element's absorption edge produces XANES features that reflect the electronic structure of the resonant atom, while X-ray diffraction and scattering generate PDF patterns that reveal local geometrical arrangements of atoms. 
    Understanding the information content in XANES and PDF, and their complementarity, can inform experimental design: when to use which technique and when it is useful to combine them. 
    Experiments, particularly conducted at beamlines at national user facilities, are valuable resources. Gaining early insights how to design experimental campaigns for the optimal information yield can help the scientific community make best use of expensive resources.  
    We show how interpretable ML can do this straightforwardly by revealing information that off-the-shelf nonlinear models can extract from different spectrum types.

    %\item \done

    One major advantage of using ML is the ability to extract obscure information from complex spectral data.
    Many ML models have been developed and implemented for XANES analysis.
    Review articles by Timoshenko \& Frenkel \cite{timoshenko2019inverting} and Penfold et al. \cite{penfold2024machinelearning} and references therein cover recent applications of ML for prediction and for understanding structure-spectra relationship from XANES.
    ML techniques that rely on labeled example data (supervised learning) have been applied to inverse problems with XANES, such as mapping from spectral data to material properties. Such models include but are not limited to heavily parameterized function approximators like convolutional neural networks (CNNs) and decision tree ensembles (i.e. random forest models) \cite{lecun2015deeplearning, breiman2001random}.
    Supervised ML techniques such as convolutional neural networks (CNNs) and random forest models were proven effective at determining the absorbing atom's environment including coordination number \cite{timoshenko2017supervised, carbone_classification_2019,zheng_random_2020}, oxidation state \cite{torrisi_random_2020}, and nearest-neighbor bond lengths \cite{torrisi_random_2020}.
    More recently, with generative AI on the rise, diffusion models were employed to generate 3D structure candidates of amorphous carbon conditioned on XANES \cite{kwon2023spectroscopyguided}.
    
    With emphasis on interpretability of the results, some of these studies identified regions of the XANES spectra that contributed the most insights for the model prediction task, either by including only the pre-edge region as input \cite{carbone_classification_2019} or by evaluating the feature importance of random forest models \cite{torrisi_random_2020, zheng_random_2020}.
    Tetef \textit{et al.} did similar analyses to compare information content in XANES and valence-to-core x-ray emission spectra (VtC-XES) \cite{tetef2021unsupervised, tetef2022informed}.
    Alternatively, Guda \textit{et al.} featurized XANES based on spectral descriptors, such as edge position and energy, and analyzed their correlations with properties \cite{guda_understanding_2021}.
    Additionally, unsupervised ML techniques, such as principal component analysis (PCA), have been employed to examine the structure-spectrum relationship within a low-dimensional latent space \cite{guda_understanding_2021, tetef2021unsupervised, liang2023decoding}.

    %\item \done
    
    Compared to XANES, fewer ML models have been developed to extract structural information from PDF. 
    Although determining structure directly from PDF (the nanostructure inverse problem \cite{juhas;aca15}) remains a challenging task, ML has enabled progress on crystals and nanoparticles.
    For crystal structures, Liu et al. developed a CNN for space group identification \cite{liu2019using}.
    Trained on calculated PDFs, their model was robust \cite{lan2022robustness} against different parameter choices in PDF calculations \cite{billinge2023atomic}.
    For nanoparticles, Kjær and colleagues developed DeepStruc, a graph-based conditional variational autoencoder (CVAE) algorithm \cite{kjaerDeepStrucStructureSolution2023} that can solve monometallic nanoparticle structures from calculated and experimental PDFs.
    Other ML prediction tasks from PDF include that of Ref. \cite{zhang2023pair}, where unsupervised techniques including PCA and non-negative matrix factorization (NMF) were applied to feature engineer PDF for prediction of defect concentrations in Ti oxides.
    On the other hand, some studies have integrated ML methods to improve the model refinements, the conventional approach for PDF analysis, by using gradient boosted decision trees \cite{anker2022extracting} to select the best starting model for fitting, or combining ML and DFT \cite{klove2023machinelearningbased} to automate the model selection and improve the optimization algorithm.
    NMF has been shown effective for decomposition of PDFs into components that resemble partial (differential) PDFs of different chemical components of the system \cite{liu2021validation} and can be easily implemented \cite{thatcher2022nmfmapping}.
    More recently, with generative AI, Guo \emph{et al.} developed models to solve a structure solution from x-ray diffraction (XRD) data from nanoparticles, outputting a continuous electron density map \cite{guo2023endtoend} or predicting a unit cell with discrete atomic coordinates \cite{guo2024diffusion}. 

    %\item \done 
    
    Unlike traditional methods, ML can readily combine heterogeneous inputs \cite{baltruvsaitis2018multimodal, hastie2009elements}, made even easier when using techniques like random forests \cite{breiman2001random} that can flexibly handle diverse input types off-the-shelf with minimal numerical issues.
    Very recently, applications of ML in materials science started incorporating multimodal inputs to improve predictive performance.
    On XANES, Carbone \textit{et al.} showed that combining XANES spectra from multiple species (C, H, and N) improved the model's performance in predicting chemical motifs of small organic molecules \cite{carbone2024accurate}.
    Szymanski \textit{et al.} combined XRD patterns and PDFs as input into their neural networks and improved the accuracy for space group identification \cite{szymanski_integrated_2024-2}. They found that models trained on PDF or XRD separately often failed on different samples, but combining them enhanced performance.
    We note in passing that combining inputs has been done in the context of conventional modeling, for example the {\sc RMCprofile}  method \cite{tucke;jpcm07} where combined PDF and XRD was needed to obtain models that were sufficiently well ordered.  
    In this case the inputs are two representations of the same information related through a Fourier transform, albeit where the information is differently weighted in the two representations.
    However, combining inputs with different physical origins and properties is more challenging, limiting the number of such studies.
    Unlike prior multimodal ML models \cite{carbone2024accurate, szymanski_integrated_2024-2} that used complementary inputs with similar origins and did not explore interpretability, our work aims to overcome heterogeneity between any arbitrary modalities and investigate how information is balanced between them, hence we focus on interpretability and relative importance of the input features.

    %\item \done
    
    In this work, we investigate with interpretable ML the information content and complementarity of XANES and PDF. We also explore whether a multimodal approach results in improvement, i.e., whether using XANES and PDF together as inputs to the model improves predictions of targets that describe local materials structure.
    The system we chose for the demonstration was transition metal oxides. 
    We curated four separate datasets, corresponding to the four metals (Ti, Mn, Fe, Cu), consisting of structures, theoretically calculated XANES spectra \cite{mathew2018high} (K-edge of the transition metal atom) available on the Materials Project (MP) database \cite{jain2013commentary}, and PDFs computed with~\cmi \cite{juhas;aca15,billinge2023atomic}.
    Separately, on each dataset we trained random forest models \cite{breiman2001random} to extract the metal's oxidation state, coordination number, and average nearest-neighbor bond lengths.
    These targets encode properties of the local atomic environment of the transition metal atoms, ranging from electronic to purely structural characteristics.
    Similar sets of prediction tasks have been performed on XANES and used to evaluate contributions of different features and regions of the spectra, particularly by Torrisi~\emph{et al.} \cite{torrisi_random_2020} and Guda~\emph{et al.} \cite{guda_understanding_2021}, but these tasks have not been the focus of previous PDF studies with ML.
    This work offers not only the database-level analysis of PDFs but also a systematic comparison of XANES and PDFs across various prediction tasks, providing valuable insights for practitioners of both experimental techniques.

    %\item \done

    The information content, or fitness to a particular problem, depends on both the input and the output of interest.
    We chose to use random forest models because off-the-shelf implementations allow us to study the feature importance \cite{breiman2001random} during training.
    Feature importance is a score assigned to each input feature based on how much it contributes to the trained model's prediction, giving us an unbiased assessment of which modality and where in the XANES and PDF spectra contribute the most to reducing training loss for each prediction task.
    When combining XANES and PDF, feature importance plots also reveal how the information is balanced between the two data sources and when one input dominates.

    %\item \done

    Here our main result is to compare the random forest model performances when trained on the different inputs: XANES, PDF, and both of them combined, on the three prediction tasks: oxidation state, coordination number, and mean nearest-neighbor bond length.
    We used F1 scores (harmonic mean of precision and recall) as a metric for classification tasks (oxidation state and coordination number) and root mean square errors (RMSEs) for the bond length regression.
    Surprisingly, models trained on XANES outperformed those trained on PDFs in most cases, even for structural targets like bond length.
    We showed that this advantage went away when site-specific differential-PDFs (dPDFs) were used as the PDF inputs, indicating that the advantage of XANES comes from its species sensitivity (assuming that the edge energies are well-separated from other species that are present, which is commonly the case for K-edges of transition metals).
    In 3/4 cases, combining XANES and dPDFs improved the model performances over using XANES alone, and in one case outperformed dPDF. This suggests that dPDFs contribute complementary local structural information to the model.
    
    This study illustrates the power of ML models to easily combine heterogeneous measured spectra that contain distinct and complementary information to accomplish a classification or regression task. 
    In multimodal analysis, one of the greatest challenges is to know how to weight the contributions of the different inputs.  
    Our approach, of investigating the feature importance of the models in their tasks, may suggest how practitioners may approach the weighting of two modalities.
    These insights enhance our understanding of the data and enable us to assess the relative importance of different measurements for success in a particular task.
    Our investigation into the differing performance of XANES, PDF, and dPDF provides a potential template for future studies which seek to understand the difference in utility of different spectral modes.

\section{Results}
\subsection{Oxidation state classification}
%\item \done 

First we report results for the classification of oxidation state from XANES and PDF separately. A random forest model was trained to classify three oxidation states, $CS=+2$, +3, and +4 for Ti, Mn, and Fe, and $CS=+1$, +2, and +3 for Cu. 
Due to heavy class imbalances in the oxidation state, we use the F1 score as a performance metric as the standard accuracy score (true positive rate) alone can be misleading. 
The F1 score of each class is the harmonic mean of the precision $P$ and the recall $R$, ranging from 0 (worst) to 1 (best), where
\begin{align}
    P &= \frac{TP}{TP + FP} \\
    R &= \frac{TP}{TP + FN} \\
    F1 &= \frac{2PR}{P + R} = \frac{2TP}{2TP + FP + FN},
\end{align}
and $TP$, $FP$, and $FN$ are true positive, false positive, and false negative rates, respectively.
We compute F1 scores for each class and a weighted mean across all classes that takes into account the proportion of the class in the dataset.
The test scores shown in all figures are the weighted mean F1 scores.
As a basis for comparison and evaluation of our models, we also compute baseline scores (weighted mean F1) achieved by a trivial classifier that, naively, always predicts the modal class.
A well-performing model should achieve a significantly higher F1 score than this baseline.
The procedure for data procurement, labeling, and model training is described in detail in the Methods section, Sec.~\ref{sec:methods}.

\begin{figure}
    \centering
    \includegraphics[width=\textwidth]{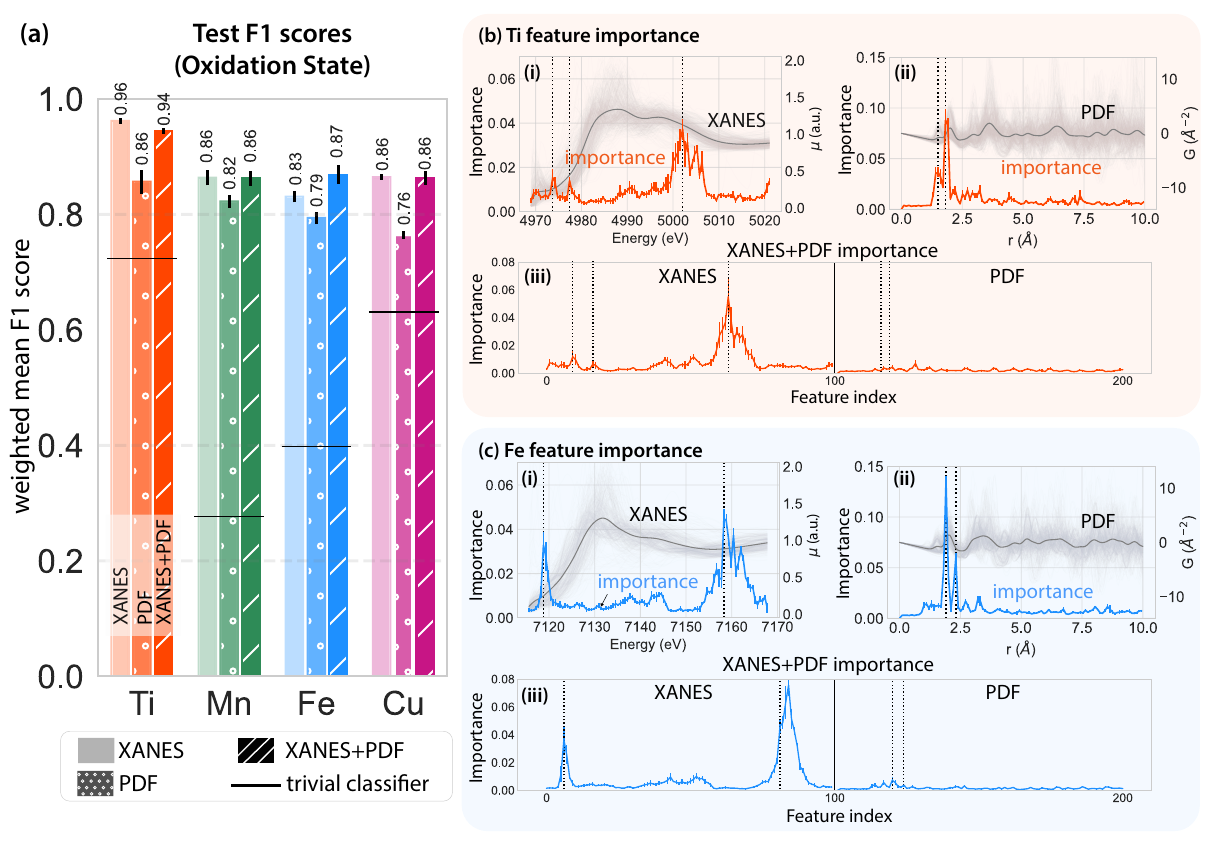}
    \caption{Oxidation state classification results: (a) test F1 scores for XANES (left bar of each triplet, no pattern), PDF (middle bar with circular pattern), and XANES+PDF models (right bar with striped pattern).
    Results are shown for the four datasets, from left to right: Ti (orange), Mn (green), Fe (blue), and Cu (magenta). 
    Black horizontal lines indicate F1 scores from a trivial classifier that labels all samples as the modal class. 
    Error bars on the test scores represent one standard deviation.
    Feature importance plots are shown for (b) Ti and (c) Fe datasets, for (i) XANES features, (ii) PDF features, and (iii) XANES+PDF features.
    In (i) and (ii), the XANES and PDF spectra are plotted in light gray, with scales defined by the secondary $y-$axis on the right. 
    The dark gray line represents the average spectrum.
    Vertical dotted lines mark the locations of prominent features from the XANES- and PDF-only importance plots. 
    To highlight changes in feature importance for the combined XANES+PDF models, the same lines are shown in (iii) for comparison.}
    \label{fig:cs}
\end{figure}
As \fig{cs}(a) shows, XANES and PDF models performed well giving F1 scores between 0.83 and 0.96, and 0.76 and 0.86, respectively, depending on the transition metal.
These scores are well above the baseline F1 scores computed for a trivial classifier that labels all samples as the modal class, shown in the figure as black horizontal lines.
These baselines scores are what we might expect if the input data did not contain any predictive information for the task. 
The baseline scores are high for Ti (0.73) and Cu (0.63) because these datasets are heavily imbalanced, hence labeling all samples as the dominant class yields high scores, but these baseline scores are still significantly lower than the F1 scores achieved by our XANES and PDF models.
This indicates that both PDF and XANES data contain rich information that is predictive of the oxidation state of the ions, as expected.

XANES models outperformed PDF models for all metals. 
This is aligned with our intuition given XANES spectra's direct dependence on the metal atom's electronic structure.
Results vary slightly for different metals, with both XANES and PDF models performing best on the Ti dataset. 
XANES performs slightly less well, but comparably across the other metals (F1 in the range 0.83 to 0.86), Mn, Fe and Cu.
In contrast, PDF performs progressively less well going from Ti to Mn to Fe and then to Cu, with F1 dropping from 0.86 to 0.76 across this series.
It is not immediately clear what is behind the different performance on the different metals.

%\item \done

Next, we look at the feature importance to investigate where in the XANES and PDF spectra information that predicts for the oxidation state resides. 
The feature importance score is a measure of each feature's contribution to the decrease in Gini impurity \cite{louppe2013understanding}, weighted by the fraction of samples in the training data that used the feature in decision-making for classification.
The importance is then averaged across all the decision trees in the forest.
Decision trees partition the input parameter space into sub-regions through a series of decisions. Each decision compares whether a feature value is above or below a threshold, which is selected to minimize the Gini impurity (or another loss function, such as RMSE for regression tasks). 
To estimate feature importance, input features can be ordered based on how much they contribute to the final prediction across all training samples. 
Note that feature importance is determined at training time and is considered a property of the model and the training data.
In our feature importance plots, the score on the $y$-axis represents the individual importance of each feature, and the scores across the input vector sum up to 1.

We chose this impurity-based approach over the permutation method, which measures the mean decrease in accuracy when each feature column is shuffled, because our spectral features, especially XANES, are correlated with each other.   
When multiple features are correlated, permuting one barely affects the model accuracy because the model still gets information from other correlated features.
Hence, the permutation method tends to underestimate the contribution of the correlated features.
Indeed, we found that this method `turned off' important regions of the XANES spectra (see SI for results on Ti).
This affected XANES more than PDF because XANES features have higher multicollinearity.
All further results use the impurity method.

\fig{cs}(b) and (c) show impurity-based feature importance for Ti and Fe, respectively.
The plots for Mn and Cu are shown in the SI (Figs.~S3 and S4).
The top left (i) and right (ii) plots in each pane show the feature importance when using XANES and PDF, respectively, separately as input. 
To help understand which regions of the spectra are highly predictive we show the XANES and PDF spectra from the respective datasets plotted as gray lines, with the darker gray line showing the average spectrum.
For all metals, the post-edge region of XANES is a strong predictor for oxidation state.
For Fe, additionally, the pre-edge region shows comparable feature importance to the post-edge region in the oxidation state prediction.
This Fe pre-edge importance is consistent with the findings in Ref. \cite{wilke2001oxidation}, where the pre-edge characteristics were used to differentiate Fe\twoplus from Fe\threeplus.
Analysis done on Fe silicates by Guda \emph{et al.} highlights the importance of both the white-line peak position and post-edge "pit" energy for Fe oxidation state prediction \cite{guda_understanding_2021}, but only the latter is observed here for our oxide dataset.

%\item \done

While the XANES feature-importance plots show that the predictive information is quite spread out, the most important PDF features concentrate around the nearest-neighbor peak (\fig{cs}(b)(ii) and (c)(ii)).  
For example, the most important PDF features for Fe coincide with the leading and trailing edges of the nearest-neighbor peak. 
We also see the same thing for Mn and Cu (see SI).
This is in agreement with the known correlation between the metal cation oxidation state and the nearest-neighbor bond length \cite{brown1973empirical, brown1985bond}.
A change in oxidation state results in a shift in the nearest neighbor peak in the PDF, which will produce the largest change in the spectrum at the positions of the steepest slopes on the leading and trailing edge of this sharp peak, and this might explain why these features have the most discriminating power.
Specifically for Fe, according to Shannon's table of ionic radii \cite{shannon1976revised}, the Fe-O bond lengths for Fe\twoplus are between 2.01 and 2.18~\AA, whereas Fe\threeplus and Fe\fourplus form shorter bonds with lengths in the range 1.89-2.045~\AA, and 1.985~\AA, respectively.
Therefore, we might expect the PDF values in the vicinity of $r=1.9$ and 2.3~\AA\ to be indicative of the Fe oxidation state, as we see in practice.

All peaks in the PDF will be affected by this change in nearest neighbor bond-length, but the higher-\ir peaks each individually contribute less information to the prediction, as indicated by their lower feature importance.
This is presumably because the effects such as shifts on the higher-\ir peaks is smaller, the peaks themselves are smaller and there is more peak overlap and more averaging across different structures, all of which reduce the information content of any particular point in the curve.
However, a strong takeaway is that the feature importance computed from the Random Forest models provides rich information about where in the PDF spectrum information lies for differentiating oxidation state.

%\item \done

Despite the XANES managing quite well to predict oxidation state on its own, it is interesting to explore if adding the PDF data leads to overall better predictions.
We find that combining XANES and PDF did not improve the oxidation state prediction over the XANES-only models for 3/4 metals except Fe, as evidence in the F1 histograms in \fig{cs}(a) (hatched plots). 
Only for Fe is there any increase in performance over XANES alone and the improvement is barely larger than the uncertainties. 

Multimodal feature importance plots show that information from XANES dominated the oxidation state prediction whereas the PDF features contributed very little, as shown in \fig{cs}(b)(iii) and (c)(iii) for Ti and Fe, respectively (Mn and Cu plots are given in SI).
The first 100 features (left half) correspond to points in the XANES spectrum and the second hundred (right half) correspond to points in the PDF.
On average, XANES features account for 80\% of the total feature importance with some variability across the four datasets.
This is somewhat expected since we know the XANES data is rich in information about oxidation state of the metal ions \cite{henderson2014x, waychunas_synchrotron_1987-1}.
Note that the feature importance scores are normalized such that they sum up to 1 across the input vector, so the scaling changed slightly as the combined XANES+PDF input has 200 features instead of 100.

The vertical dotted lines mark the locations of important XANES and PDF features when they were used separately.
These locations correspond to local maxima of feature importance, extracted with the find\_peaks function in scipy \cite{virtanen2020scipy}.
As discussed above, we know that the PDF nearest-neighbor peak (e.g., see the dotted lines in the \fig{cs}(b)(ii) and (c)(ii)) contains rich information about the oxidation state of the metal ion, but this PDF information is largely `turned off' when XANES data are available.
With both XANES and PDF present, the XANES feature importance barely changed for Ti and Cu (see SI), and the new peaks still coincide with the dotted lines. 

%\item \done

For Fe and Mn (see SI), on the other hand, the post-edge XANES importance peak is taller and shifts slightly to the right when PDF is also included, but it is inconclusive whether this improved the prediction.
For Fe, quite impressively, the combined model correctly predicted 22\% of the samples that were previously misclassified when using XANES and PDF separately, indicating that the model gained some complementary information from the multimodal input. 
We did not observe this effect on Mn despite the similar trend in feature importance.
This is also true for Ti, and for Cu, where only 13\% of the wrong predictions were corrected.

\subsection{Coordination number classification}

%\item \done

We now report the results of applying the same approach to predict coordination number. 
Our datasets (Ti, Mn, Fe, and Cu) contain structures with metal coordination numbers $CN=4$, 5, and 6, where the steps for computing CN is described in Sec.~\ref{sec:methods}.
Since coordination number is a structural quantity, and PDF is a structural technique, one might expect PDF to perform well on this task, with XANES to also contain important information.
As with the oxidation state classification, we used F1 scores as our performance metric.

\begin{figure}
    \centering
    \includegraphics[width=\textwidth]{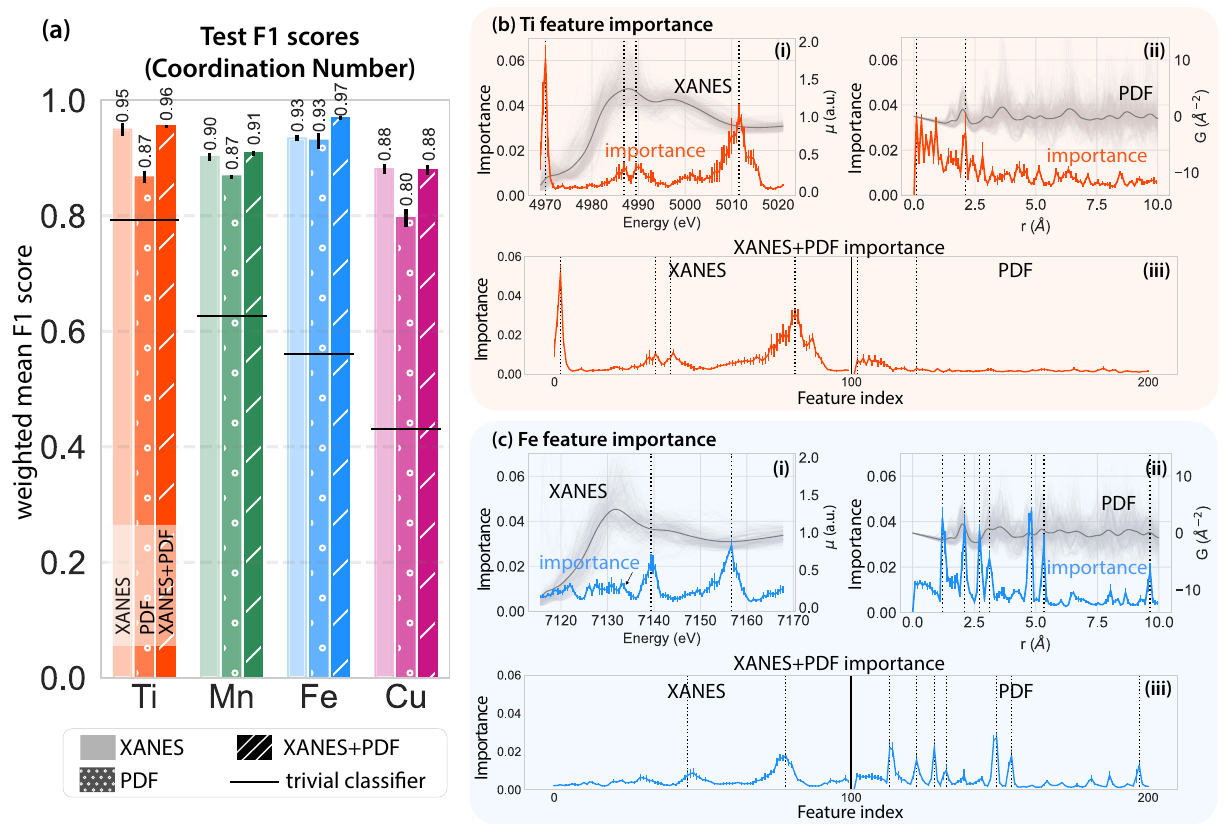}
    \caption{Coordination number classification results: (a) test F1 scores for XANES (left bar of each triplet, no pattern), PDF (middle bar with circular pattern), and XANES+PDF models (right bar with striped pattern).
    Results are shown for the four datasets, from left to right: Ti (orange), Mn (green), Fe (blue), and Cu (magenta). 
    Black horizontal lines indicate F1 scores from a trivial classifier that labels all samples as the modal class. 
    Error bars on the test scores represent one standard deviation.
    Feature importance plots are shown for (b) Ti and (c) Fe datasets, for (i) XANES features, (ii) PDF features, and (iii) XANES+PDF features.
    In (i) and (ii), the XANES and PDF spectra are plotted in light gray, with scales defined by the secondary $y-$axis on the right. 
    The dark gray line represents the average spectrum.
    Vertical dotted lines mark the locations of prominent features from the XANES- and PDF-only importance plots. 
    To highlight changes in feature importance for the combined XANES+PDF models, the same lines are shown in (iii) for comparison.}
    \label{fig:cn}
\end{figure}

\fig{cn}(a) shows weighted mean F1 scores of the XANES (left bar) and PDF (middle bar) models for each of the four metals. 
Surprisingly, although coordination number is closely related to the height of the PDF nearest-neighbor peak, XANES outperformed PDF for most metals (Ti, Mn, and Cu).
Fe is the only exception here where XANES and PDF models achieved the same F1 score of 0.93.
Both XANES and PDF models performed very well with F1 scores above the trivial classifier baseline, with XANES scores in the range 0.88 (Cu) to 0.95 (Ti) and PDF scores ranging from 0.80 (Cu) to 0.93 (Fe).
Most scores here are higher than the F1 scores for oxidation state classification (see \fig{cs}(a)).
This indicates that both XANES and PDF contain rich information about the transition metal's coordination number.
%

%\item \done
We also investigated whether some aspect of our PDF featurization led to information loss.
Initially, we had normalized all the PDFs in the training and testing sets to the range $[-1, 1]$.
This rescaling might have caused the PDF models' underperformance, as the amplitude of the first PDF peak is known to contain information about coordination number and we are applying a different scaling to this peak for each PDF pattern in the dataset.
To test this hypothesis, we retrained the models using unscaled PDFs and PDFs normalized to $[0, 1]$. 
Using unscaled PDFs slightly improved the coordination number prediction for all metals but still fell short of XANES performance.
We also experimented with different normalization methods on XANES, finding that unscaled XANES performed best.
Therefore, we chose to use raw XANES and raw PDFs in all of our models including the oxidation state prediction discussed previously (\fig{cs}) and the results shown in \fig{cn}.

%\item \done 

%\item \done

The pre-edge region of XANES is a strong predictor for coordination number for 3/4 metals (Ti, Mn, and Cu).
However, for Cu (see SI Fig.~S6), unlike any other metals we investigated, the post-edge region of XANES did not contain important information on the coordination number. 
Previous studies that predicted coordination numbers from XANES \cite{carbone_classification_2019, torrisi_random_2020, zheng_random_2020, guda_understanding_2021} have reported similar interpretability trends.
These include the already known pre-edge importance, particularly high for Ti \cite{farges_ti_1997} but low for Fe \cite{jackson2005multi}, along with the notable edge importance observed for Mn and Cu \cite{torrisi_random_2020, zheng_random_2020}. 
For Mn (see SI Fig.~S5), Torrisi~\emph{et al.} reported strong pre-edge significance \cite{torrisi_random_2020}, while our results showed a more evenly distributed importance across the full energy range.

On the other hand, the PDF importance plots look very different for Ti and Fe. 
While the most distinguishing information lies within the nearest-neighbor peak for Ti, the Fe PDF importance extends all the way to $r\sim 10$~\AA.
We also observed similar PDF `fingerprints' for Mn, but not in Cu (see SI).
This indicates that, at least for Fe and Mn, the nearest-neighbor coordination geometry has a ripple effect on the arrangement of atoms further out, making the PDF a good measure of the metal coordination number.
One of the most important PDF features for Fe is at $r = 2.1$~\AA, aligning closely with the nearest-neighbor peak on the average-PDF plot (dark gray line in \fig{cn}(b)(ii)) and consistent with the Fe-O bond length range 1.95-2.18~\AA\ for Fe(VI).
The Fe-O bond lengths are 2.03~\AA\ or shorter for Fe(IV) and Fe(V) \cite{shannon1976revised}, so we think the PDF value at $r = 2.1$~\AA\ helped distinguish Fe(VI) from the rest.
The Mn bond lengths are slightly distinctive \cite{shannon1976revised} for different coordination numbers, but the PDF value near $r=1.7$~\AA\ might indicate the presence of shorter Mn-O bonds that are feasible for Mn(IV) but not for Mn(V) and Mn(VI), both of which form Mn-O bonds that are 1.86~\AA\ or longer.

%\item \done

Unlike the oxidation state, where the most important PDF features lie on the leading and trailing edges of the nearest-neighbor peak (\fig{cs}(b)(ii) and (c)(ii)), the key PDF feature for predicting coordination number aligns with the peak itself. 
This suggests that the ML model was extracting the coordination number from the height of the nearest-neighbor peak, consistent with what we know from the definition of the PDF.
However, it is less clear why the PDF in the higher-\ir region contains much more information for Fe and Mn, but not Ti and Cu.

%\item \done

The combined XANES and PDF gave results similar to the the oxidation state prediction, with the F1 scores showing significant improvement only for Fe but not for the other three metals (hatched bar plots in \fig{cn}(a)).
The feature importance plots for the combined XANES and PDF input for Ti and Fe are shown in \fig{cn}(b)(iii) and (c)(iii), respectively (see SI for Mn and Cu plots). 
For Ti and Cu, XANES features dominated the prediction, accounting for 78\% and 62\% of the total importance, respectively. 
For Ti, the importance of the PDF feature near 2~\AA\ is suppressed when the input includes XANES, where its pre-edge region already contains very strongly distinguishing feature for coordination number \cite{farges_ti_1997}.
On the other hand, for Fe, all of the strongest PDF peaks are retained when XANES is present without apparent shifts.
This is also true for Mn (see SI), where the PDF importance also contains a series of prominent peaks extending into the higher-\ir region.
For Mn, the PDF showed even higher importance than XANES, contributing 63\% of the total importance, yet this yielded a smaller improvement in F1 scores compared to Fe, where XANES and PDF had comparable importance when combined.

%\item \done

Unlike the other metals, the XANES-only and PDF-only models performed equally well on Fe (F1 score of 0.93), yet combining XANES and PDF noticeably improved the performance.
Analysis of the misclassified samples revealed that, for Fe, the majority (79\%) of samples misclassified by XANES were correctly classified by PDF. 
In contrast, for the other metals, XANES and PDF were more likely to misclassify the same samples, where only 30\% (or less) of the wrong predictions by XANES are recovered by PDF, hence combining the spectra did not significantly improve the performance.
This suggests that the information content in XANES and PDF that is relevant for the coordination number prediction is the most complementary for Fe, and less so in Mn, Ti, and Cu.

% 
% BL results
\subsection{Mean nearest-neighbor bond-length regression}
%\item \done

\begin{figure}
   \centering
   \includegraphics[width=\textwidth]{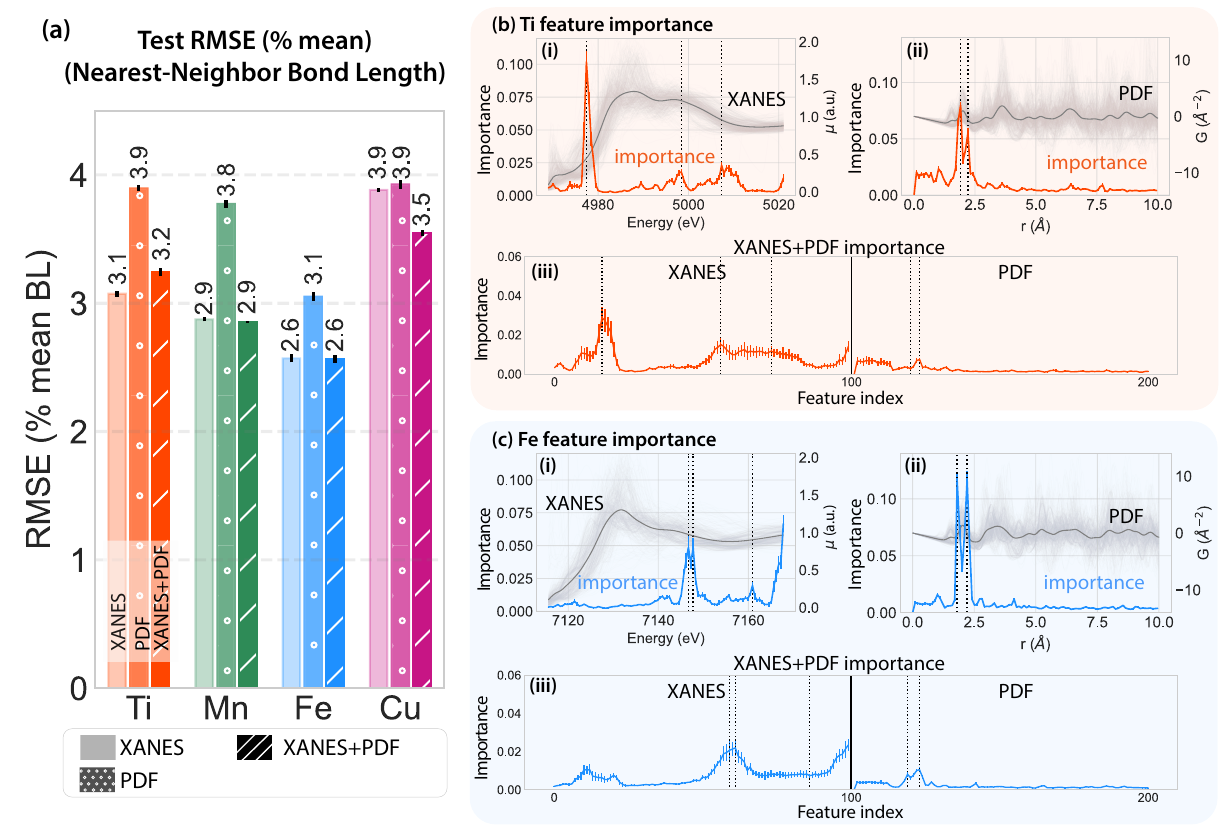}
   \caption{Bond Length Regression Results: (a) test RMSE scores expressed as a percentage of the mean bond length for XANES (left bar of each triplet, no pattern), PDF (middle bar with circular pattern), and XANES+PDF models (right bar with striped pattern).
   Results are shown for the four datasets, from left to right: Ti (orange), Mn (green), Fe (blue), and Cu (magenta).
   Note that for these RMSEs smaller numbers indicate better model performance.
   Error bars on the test scores represent one standard deviation.
   Feature importance plots are shown for (b) Ti and (c) Fe datasets, for (i) XANES features, (ii) PDF features, and (iii) XANES+PDF features.
   In (i) and (ii), the XANES and PDF spectra are plotted in light gray, with scales defined by the secondary $y-$axis on the right. 
   The dark gray line represents the average spectrum.
   Vertical dotted lines mark the locations of prominent features from the XANES- and PDF-only importance plots. 
   To highlight changes in feature importance for the combined XANES+PDF models, the same lines are shown in (iii) for comparison.}
   \label{fig:bl}
\end{figure}

In this section, we present the results for the prediction of mean nearest-neighbor bond length. 
As with coordination number, bond length is a descriptor for the structure of a material where one might expect the model trained solely on the PDF to perform well.
\fig{bl}(a) shows the RMSE of the test set expressed as a percentage of the mean nearest-neighbor bond length for each metal.
Note that shorter bars in this figure indicate better model performance.
The test RMSEs ranged from 0.053 to 0.080~\AA\ (see Table~S8 in SI), i.e., percentage errors in the range 2.6-3.9\% of the mean bond lengths.
The RMSEs indicate overall model performance rather than the precision of individual predictions of bond length. 
For comparison, models that naively extract the position of the nearest-neighbor PDF peak to estimate the mean bond length produced RMSEs between 14 to 20\%, much worse than our random forest models.

Contrary to our expectations, for the average bond-length estimates, models trained on XANES data outperformed models trained on PDF data for all of the four metals. 
XANES outperformed PDF by significant margins for Ti, Mn, and Fe and a small margin for Cu.
Both the XANES and PDF models showed similar trends across the metals. 
Both models performed best on Fe with RMSEs of 2.6\% and 3.1\%, respectively, did almost equally well on Ti and Mn, and struggled the most with Cu. 
Parity plots (Fig.~S7 in the SI) provide some insights into this trend.
The plots show that the models produced smaller errors on samples with bond lengths closer to the dataset mean, and while the Fe dataset has the fewest outliers, the Cu dataset has the most outliers, making it a more difficult dataset for the regression models.
%

%\item \done

\fig{bl}(b) and (c) show the feature importance plots for Ti and Fe, respectively. The plots for Mn and Cu are given in the SI.
For XANES, the edge features are very important for Ti, Mn, and Cu. 
Additionally, post-edge features closer to the EXAFS region contribute a lot for Mn and Fe, but less so for Ti and Cu. 
Torrisi~\emph{et al.} reported very similar interpretability trends for all metals \cite{torrisi_random_2020}, but our Mn plot had an additional post-edge peak at 6650~eV (see SI). 
Interestingly for Fe, while Guda~\emph{et al.} highlighted the significance of both the edge energy and the post-edge features for bond length prediction \cite{guda_understanding_2021}, we did not observe a clear edge importance in \fig{bl}(c)(i). 

Turning to the PDF, the most important features for all metals coincide with the leading and trailing edges of the nearest-neighbor peak, similar to what we saw in the oxidation state classification (\fig{cs}).  
This behavior can be rationalized on the basis that changes in bond-length result in shifts of the nearest neighbor peak, with the steep leading and trailing edges of the peak being most sensitive to these shifts.
This is directly seen in the feature importance when regressing on bond length directly, as well as in oxidation state, which is strongly correlated with bond length \cite{brown1973empirical, brown1985bond}.

%\item \done

Combining XANES and PDF as inputs did not consistently improve performance over the XANES models (right bars in \fig{bl}(a)). 
In some cases, errors were the same or higher, as with Ti and Fe, while for Mn, the improvement was minimal.  
The only significant improvement was for Cu, where the RMSE dropped from 3.9\% (0.079~\AA) with the XANES model to 3.5\% (0.072~\AA) with the combined model. 

As shown in \fig{bl}(b)(iii) and (c)(iii), combining XANES and PDF suppressed the PDF feature importance, with XANES features dominating the prediction for all metals.
This mirrors the trends observed in the oxidation state prediction (\fig{cs}), though for bond length, some of the PDF importance around the nearest-neighbor peak remained.
This effect is most evident for Cu (see SI), where the combined model produced the greatest reduction in the test error, even though its PDF-only model performed relatively worse compared to other metals.
Additionally, the XANES importance remains largely unchanged on adding PDF data for Fe and Cu, though we do observe some broadening in the feature importance plots for Ti and Mn.

\subsection{differential-PDF results}

%\item \done

Contrary to our initial expectations, random forest models trained on XANES outperformed those trained on PDF for all three prediction targets: oxidation state, coordination number, and mean nearest-neighbor bond length for all four transition metals studied.
Though this was expected for electronic properties (e.g., oxidation state), we were surprised by the superior performance of XANES for structural properties such as coordination number and bond length.
Because of the intimate relationship between the atomic positions (structural signal) and the resulting electronic spectrum (spectroscopic signal) it is clear that structural techniques can be sensitive to electronic properties, and vice versa. Hence, XANES contains rich local structural information.  
However, it is worth investigating why the PDF didn't perform so well in these structural tasks.

One possible advantage that XANES has is that it yields information only on the environment of the metal ion whereas the total-PDF contains possibly confounding information about the environment of other ions in the system; it is not site specific. 
We can assess the importance of this factor on simulated data, as all the model training and testing can be repeated by replacing the total-PDF input (which contains distances from every atom to every other atom) with the differential-PDFs (dPDFs) \cite{egami;b;utbp12}, which include only distances from the metal ions to all other ions in the system. 
Although it is not straightforward to measure dPDFs experimentally, they can be measured in principle \cite{aur1983local, kofalt1986differential, waseda1984novel, waseda2002anomalous, petkovLocalStructureIn02000, petkovApplicationDifferentialResonant2018a} and for our current purposes, they are easy to calculate using \cmi (see Methods section).

%\item \done

\begin{figure}
   \centering
   \includegraphics[width=\textwidth]{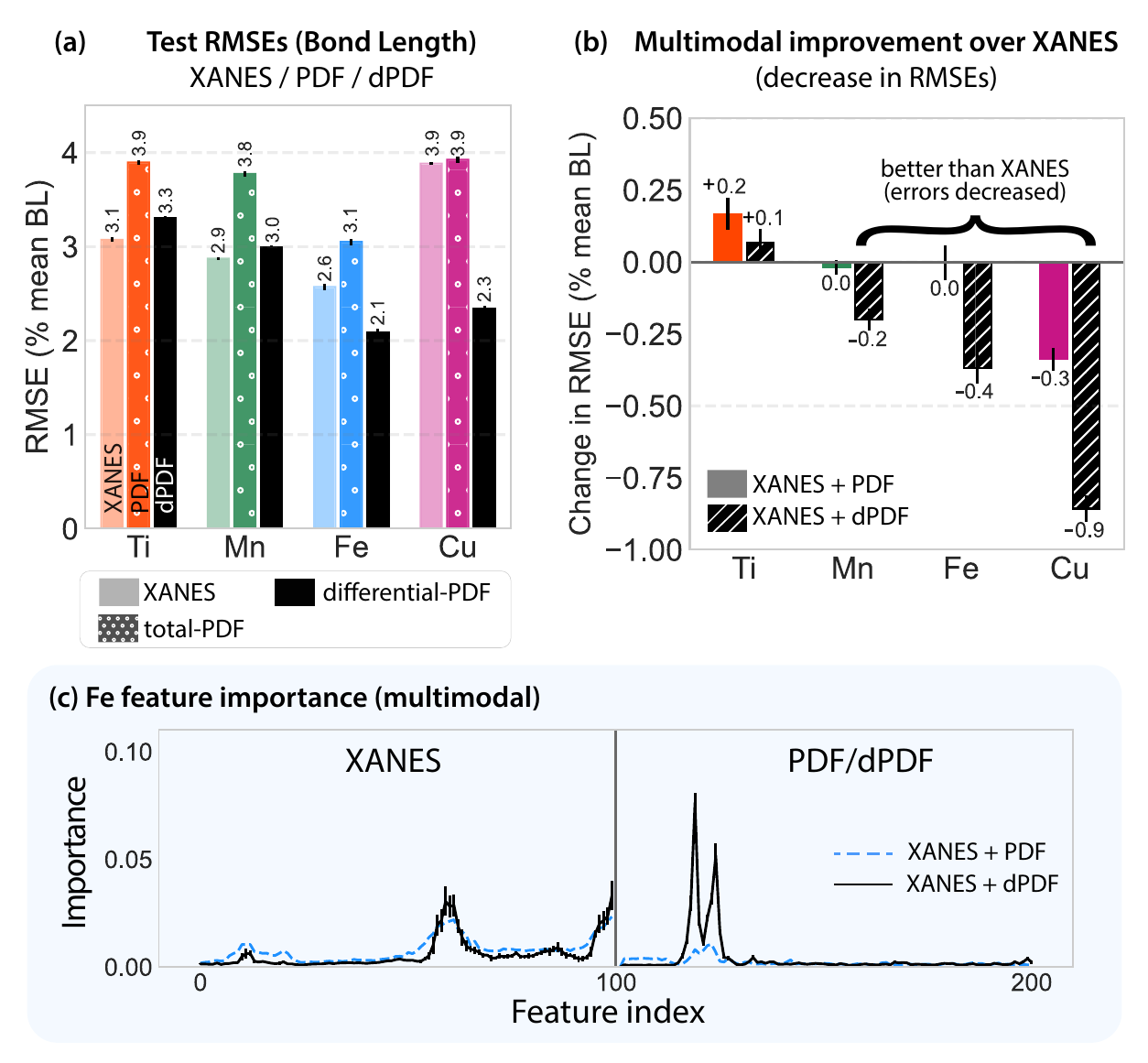}
   \caption{differential-PDF (dPDF) results on Bond Length Regression: (a) test RMSE scores expressed as a percentage of the mean bond length for XANES (left bar of each triplet, light color with no pattern), PDF (middle bar with circular pattern), and dPDF models (right bar, black).
   Results are shown for the four datasets, from left to right: Ti (orange), Mn (green), Fe (blue), and Cu (magenta). 
   Note that for these RMSEs smaller numbers indicate better model performance.
   The XANES and PDF results from \fig{bl} are reproduced here for easy comparison.
   Error bars on the test scores represent one standard deviation.
   (b) Change in RMSEs when training the models on multimodal inputs, combining XANES with PDF (solid) or dPDF (black, striped).
   Negative changes in RMSEs (for Mn, Fe, and Cu) indicate an improvement over the XANES-only models.
   (c) Feature importance plots for the multimodal inputs, comparing XANES+PDF (dashed line) and XANES+dPDF (solid line) models for Fe.} 
   \label{fig:dpdf_bl}
\end{figure}

We found that the relative importance of the information coming from the PDF was greatly enhanced when we used dPDF rather than total-PDF as the input for all three prediction targets across all metals. 
The biggest improvement was in the mean bond length regression (see bar plots in \fig{dpdf_bl}(a)), where the RMSEs decreased from 3.1-3.9\% when using total-PDFs to 2.1-3.3\% when using differential-PDFs as input.
dPDF even outperformed XANES for Fe and Cu, which contrasts with the case for the total-PDF.
For Ti and Mn, dPDF performed almost as well as XANES.
See SI Table~S8 for the test RMSEs across all models and metals.

We also found that combining XANES with dPDFs, rather than total-PDFs, produced greater multimodal improvement over the XANES-only models. 
As shown in \fig{dpdf_bl}(b), the biggest improvement was on Cu, where the RMSE dropped from 3.9\% to 3.0\% when combining XANES and dPDF, followed by Fe (though dPDF-only models for these metals still produced lower errors).
For Fe, combining total-PDF with XANES did not improve the model, but replacing the total-PDF with dPDF did.
Multimodal feature importance plots in \fig{dpdf_bl}(c) show greatly enhanced importance of the dPDF features around the nearest-neighbor peak (same location as the most important PDF features), a stark contrast to the suppressed total-PDF importance.
We also saw similar effects on other metals (see Fig.~S12 in SI), although multimodal Ti models still performed worse than XANES models.
This shows that a significant contributing factor to the poorer performance of the total-PDF was due to confounding effects coming from parts of the signal not originating from the transition metal site, rather than anything inherent in the physics of the signal, e.g., diffraction vs. spectroscopy.

%\item \done  

We saw similar results in the oxidation state prediction (see Fig.~S10 and Table~S6 in SI), where dPDF models outperformed the total-PDF models across all elements.
Significant improvement over total-PDF was seen in Ti, Mn, and Cu, with F1 score increases in the 7-11\% range, but this improvement was only modest for Fe (3\%).
For all metals, combining dPDF with XANES slightly improved the performance over using XANES alone.
For most metals here (Ti, Mn, and Cu), as with bond length multimodal results, the dPDF features around the nearest-neighbor peak retained much more notable importance when combined with XANES.
The only exception here is Fe, where replacing total-PDFs with dPDFs did not change the feature importance nor the multimodal model score significantly.

The single-modal and multimodal dPDF models were able to extract the oxidation state based on the correlation between oxidation state and nearest-neighbor bond lengths \cite{brown1973empirical, brown1985bond}.
Although the total-PDFs also contain this information, the models were more capable of extracting this information from the dPDFs for most metals here, further suggesting that dPDFs can be seen as a clearer signal for describing the coordination environment of a specific species. 

%\item \done 

Lastly, for the case of coordination number (see Fig.~S11 and Table~S6 in SI), we saw noticeable performance improvements when the models were given dPDFs in Ti and Cu, where the F1 scores increased by 6\%.
For these metals, the dPDF feature importance plots showed more specific features around the nearest-neighbor peak compared to the total-PDF counterparts.
On the other hand, the improvement was minimal for Mn and Fe, where the total-PDF models already performed very well and were able to extract information from PDF features at the higher-$r$ range. 

When combining PDF and XANES, Cu is the only metal that shows a clear advantage of using dPDFs instead of total-PDFs.
The Cu multimodal feature importance plot revealed prominent nearest-neighbor dPDF features that had completely diminished when combining the total-PDFs with XANES.
On the other hand, even though the dPDF-only model performed well on Ti, XANES still heavily dominated the coordination number prediction by the multimodal models whether we used total- or differential-PDFs.
For Mn and Fe, similar to their total-PDF importance plots, the dPDF plots highlight a series of important features over a longer $r$ range. 
Although the relative importance changed and some peaks were missing from the dPDF plots, these changes did not affect the Mn and Fe model performances significantly.

\section{Discussion}

%\item \done

This work demonstrates how to use interpretable machine learning techniques in tandem with computationally generated multimodal spectra to study the relative contributions of XANES and PDF to predictions of coordination number, oxidation state, and bond length. 
Feature importance plots reveal regions in the spectra from which the random forest models draw the most important insights and allow us to compare the relative utility of different spectra for the same prediction tasks in metal oxides. 
We anticipate that experimentalists who have access to comparable databases of reference spectra may be able to use this approach to compare the expressiveness of different modalities, both individually and in combination, without the need for a detailed forward model that can accommodate both spectral types. 
Assuming that simple ML models like random forests can capture the same trends that a more system-specific model would, this approach could also help plan experiments by assessing the value of incorporating an additional modality.

For XANES, as previously observed we find that the pre-edge region is a strong predictor of coordination number for most metals investigated (Ti, Mn, and Cu), particularly for Ti where the pre-edge peak height and position are strongly dependent on Ti coordination number \cite{farges_ti_1997}.
This lends confidence to our approach.
Pre-edge XANES also contains distinguishing information on Fe oxidation state \cite{wilke2001oxidation}.
Additionally, XANES edge energy \cite{glatzel2009electronic, guda_understanding_2021} and peak positions \cite{natoli1984distance} have been correlated with nearest-neighbor bond length, and this is reflected in the feature importance plots for most metals (Ti, Mn, and Cu), except for Fe.
Although the post-edge region of XANES are less studied by traditional methods due to fitting difficulty \cite{waychunas_synchrotron_1987-1}, our ML models and others (such as \cite{torrisi_random_2020, guda_understanding_2021}) found the post-edge XANES region to be highly predictive for most targets.
On the other hand, PDF importance plots showed that the models were able to extract most information from nearest-neighbor PDF peaks. 
While coordination number and bond length can be extracted more directly from PDF peak heights and positions, the PDF importance plots confirmed that the oxidation state prediction is enabled by its correlation with the nearest-neighbor bond lengths \cite{brown1973empirical, brown1985bond}.

To compare the relative importance of the pre-edge and post-edge regions of XANES, we also tried training the models on each region separately (see results in Sec.~6 in SI) for the coordination number prediction.
For all four metals, including Ti where the pre-edge importance is the most prominent, the models trained on post-edge XANES performed better than those trained on the pre-edge region alone, but still slightly worse than when using the whole spectrum as input (see Fig.~S13 in SI).
The results are consistent with the findings in \cite{carbone_classification_2019}, where their CNN models for coordination number prediction performed worse when trained on the pre-edge only compared to the whole spectrum, indicating that the post–edge region contains important information.
Additionally, we combined each region of XANES with PDFs and dPDFs and re-trained the models.
While the test scores did not change significantly whether we used the pre-edge, post-edge, or the whole range of XANES in the combined models (see Figs.~S14 and S16 in SI), the feature importance plots (Figs.~S15 and S17 in SI) highlight that Ti is the only metal here where both pre-edge and post-edge XANES heavily dominated the prediction in the presence of PDFs/dPDFs.
For other metals (Mn, Fe, and Cu), post-edge XANES and PDF/dPDF features are closer in importance and contain more coordination number information than pre-edge XANES.

XANES contains rich information about the absorbing atom's electronic spectrum and has traditionally been used to obtain properties like oxidation state of the ions.  
It has been used much less to obtain detailed local structural information due to the nonlinearity of the absorption response, in contrast with the absorption signal in the EXAFS region which is easily fit to the EXAFS equation \cite{rehr2000theoretical}.
However, ML techniques have increasingly enabled researchers to extract specific local structural information from XANES such as coordination environments and nearest-neighbor bond lengths for nanoparticles \cite{timoshenko2017supervised, timoshenko2020silver} and transition metal oxides \cite{carbone_classification_2019, torrisi_random_2020, zheng_random_2020}, and aromaticity \cite{tetef2021unsupervised} and chemical motifs \cite{carbone2024accurate} of small organic molecules.
And although generative AI was used to predict 3D structures of amorphous carbon \cite{kwon2023spectroscopyguided} conditioned on XANES, overall, the scope of what can be extracted from XANES is still somewhat limited to specific prediction targets and simpler chemical systems, so we might expect structural probes such as PDF to outperform XANES for purely structural information.

However, in our work, XANES models outperformed total-PDF models for all metals and prediction tasks, including geometrical properties like coordination number and bond length.
This adds to the growing body of work \cite{timoshenko2019inverting, penfold2024machinelearning} on extracting precise local structural information directly from XANES. 
The intrinsic correlation between electronic and atomic structure in XANES evidently allows random forest models to effectively leverage these relationships.
Our findings also raise the question of why XANES outperformed PDF in most cases.
It could be that the structural information is encoded more efficiently in XANES; for example, the non-linear relationship between the electronic spectrum and the structure could increase sensitivity.
In our work, the ML models are able to learn a direct mapping from the structural signal to the electronic properties, and from the electronic signal to the structural properties.

While testing the efficiency of the encoding is difficult, we have evidence for a secondary hypothesis: XANES outperforms PDF because it is a resonant method that specifically captures the local environment of the target transition metal atoms, whereas PDF includes structural information about all atoms, many of which are less relevant for our prediction tasks. 
The total-PDF averages contributions from all atoms in the system, and the performance disparity between XANES and PDF suggests that the contributions of irrelevant pairs of atoms with similar nearest-neighbor distances dilute the signal from the species of interest.
At least part of the PDF's underperformance can be attributed to this.
This is supported by the marked improvement in performance when using dPDF, which includes only the local environments of the species of interest.

This underscores the potential utility of developing methods that can extract differential PDFs experimentally, such as using isotopic substitution of neutrons \cite{soper1977neutron, enderby1995ion, zhao1998neutron} or anomalous differential diffraction methods \cite{aur1983local, kofalt1986differential, waseda1984novel, waseda2002anomalous, petkovApplicationAtomicPair2000, petkovApplicationDifferentialResonant2018a}.
These methods exist but are currently difficult and not widely used.
One reason is that the experiments and the subsequent data analysis are difficult. 
A second limitation of the anomalous differential PDF is that the \q-range available from these measurements, and therefore the real-space resolution in the PDF, is limited by the energy of the incident x-rays. 
For example, first-row transition metals such as studied here the absorption edges are below 10~keV and the accessible \q-range in an adPDF measurement would be well below $\qmax = 10$~\iaa.
An interesting future study would explore in more quantitative detail how low real-space resolution but chemically resolved adPDFs complement XANES.

Another useful extension of this work would be to do the same analysis comparing XANES and EXAFS input signals, or PDF and EXAFS.

As a caveat we note that, like any study drawing conclusions from ML, our results are tied to the dataset that was used. 
We are more confident that the observed trends would apply for similar systems (bulk ordered transition metal oxide crystals), and caution that extrapolation to other classes of system could and should be verified with comparable libraries of structures and spectra (surface structures, nanoparticles, disordered materials, non-oxide systems, and so on).
It is also important to note discrepancies between the theoretical spectra used in our study and the experimental data.
For example, experimental PDFs might have lower real-space resolution due to smaller \qmax, which depends on the X-ray energy and scattering angle (we assumed \qmax$=30$~\iaa in the calculation).
For the case of XANES, FEFF calculations \cite{rehr2010parameter} are known to perform better in the post-edge region, where multiple scattering effects dominate, than in the pre-edge region, where electronic structure dominates.
Another limitation is that we work with small defect-free ordered unit cells from the Materials Project, which represent a strongly idealized version of real materials.
In real materials, the structures that are observed may differ for a variety of reasons, including but not limited to: spontaneous symmetry breaking, distortion into larger unit cells due to the Jahn-Teller effect, or displacive disorder due to finite temperature.
Future applications of this method on individual systems of interest should attempt to include these effects as necessary in the reference structures that are used to generate input datasets.
Additionally, intermediate featurization techniques could potentially improve the model performance described in this study, but we leave this exploration for future work.

\section{Methods}
\label{sec:methods}
\subsection{Datasets}
Our datasets consist of transition metal oxide structures available on the Materials Project (MP) database \cite{jain2013commentary}. 
We curated four separate datasets, one each for the four metal elements: Ti, Mn, Fe, and Cu.
These metals had the most MP structures available that met the requirements of our prediction tasks.
In most of the oxide structures used here, the transition metal atoms' nearest neighbors are all oxygen, but other species besides the metal and oxygen are present in some structures (see Table~S4 in SI).
Depending on the targets (oxidation state, coordination number, and bond length), the sample size varied between $\approx600$ and 1300 compounds (see Tables~S1, S2, and S3 in SI).
Each dataset included the Pymatgen-compatible structures, calculated K-edge XANES spectra with the transition metal as the absorbing atom, and total- and differential-PDFs computed by us. 

The XANES spectra available on MP were computed via FEFF calculations \cite{mathew2018high}.
We used \cmi \cite{billinge2023atomic,juhas;aca15} to compute the total- and differential-PDFs of all structures in our datasets.
In detail, we used \cmi’s PDFCalculator module to compute both the total- and differential-PDFs of the structures from their crystallographic coordinates. 
We first calculated the total-PDFs in the range $0 \leq r \leq 10$~\AA\ and interpolated onto a 100-pt Nyquist-Shannon grid \cite{farrow2011nyquist} assuming measured PDFs with $Q_{max}=30$~\iaa. 
We did the same for the differential-PDFs, with the additional step of filtering for the species of interest (the transition metal in our case) when using PDFCalculator. 
We used unscaled PDFs and XANES because they gave slightly better model performances than when using scaled inputs.
Previously, Torrisi \emph{et al.} found that scaling did not significantly change pointwise model interpretability trends \cite{torrisi_random_2020}.

%\item \done

For XANES, the MP database contains multiple XANES spectra with different absorbing species for each structure; in our datasets, we only used K-edge XANES for which the transition metal element was the absorbing species.
The XANES spectra were then interpolated onto a 100-pt grid to match the length of our PDF input vectors.
The energy domain for XANES was chosen for each transition metal dataset based on where most spectra within the dataset overlapped (see Table~S4 in SI).

%\item \done

Next, we extracted target quantities (oxidation state, coordination number, and mean nearest-neighbor bond length) from the structures as the ground truths for supervised learning.
Oxidation state was extracted using Pymatgen's BVAnalyzer module.
To avoid ambiguities from averaging variable oxidation states across different sites, we further subsampled our dataset for this task and kept only structures in which all transition metal cations had the same oxidation state.
This makes it easier to understand the ability of random forest models to extract oxidation state trends as this simplifies the problem space.
Further, to make sure there is sufficient representation for each class and equal number of classes in all datasets (three classes), we included only Ti, Mn, and Fe oxides with metal cation oxidation states of 2+, 3+, and 4+, and Cu oxides with oxidation states 1+, 2+, and 3+.

%\item \done

Metal coordination numbers were obtained using Pymatgen's CrystalNN module, which uses Voronoi decomposition and solid angle weights to calculate the most probable coordination environment around a specified atom \cite{pan2021benchmark}.
We first extracted the coordination number (number of nearest neighbors) of each metal site in the structure.
Similar to our oxidation state dataset curation, we filtered for structures in which all of the metal sites had the same coordination number and only included those with coordination numbers of 4, 5, or 6 in the prediction task.
This resulted in datasets of size $N\approx$900 for each transition metal element used here. 

%\item \done

We also used the CrystalNN module to extract mean nearest-neighbor bond lengths. 
For each structure, we first extracted all the distances from each transition metal site to its nearest neighbors and computed the mean bond length of each site.
We then averaged this mean distance across all transition metal sites in the structure and used this value as our prediction target.
This ``mean of the means" approach weighs the mean bond lengths of all sites equally, independent of their coordination numbers, whereas simply taking the average of all bond lengths in a structure would over-represent those of the metal sites with higher coordination numbers.

\subsection{Model training}
%\item \done

We trained random forest classifiers for oxidation state and coordination number prediction and random forest regressors for mean nearest-neighbor bond length prediction. Both model types made use of scikit-learn \cite{pedregosa2011scikit}. 
For all targets, we held out 20\% of samples from each dataset for testing and trained on the remaining 80\%.
Initially, three models were trained for each transition metal element, with varying inputs: one model with total-PDF as input, another with XANES as input, and the third with both XANES and total-PDF as input.
We used F1 scores (class-wise and weighted mean) as the performance metric for the classification tasks and Root Mean Square Error (RMSE) for the regression task.
We report the RMSE as the percentage of mean bond length to contextualize the magnitude of the model error relative to the true values.

We used 10-fold cross validation to determine the optimal hyperparameters for each model, including the number of trees and the maximum number of features considered at each split.
The models with the best hyperparameters were trained 10 times on the same training data (80\%) using different random seeds.
We then evaluated their performance on a holdout test set (20\%) to calculate the mean and standard deviation of the test scores. 
The error bars on all bar plots represent one standard deviation.

%\item \done 

For oxidation state and coordination number, we also computed baseline F1 scores for comparison with our model scores.
Our datasets for these classification tasks consist of three imbalanced classes, often with a clear majority class, so we computed the baseline F1 scores based on a trivial classifier that labels all samples as the modal class.
The baseline score depends on the class proportions, so it is unique for each metal and each target.

%\item \done  

%\item \done 

%\item \done

We also estimated baseline RMSEs for the mean nearest-neighbor bond length prediction.
Our ``naive" model in this case estimated the mean bond length by extracting the the location (in~\AA) of the nearest-neighbor peak from the total-PDF using the peak-finding algorithm in scipy \cite{virtanen2020scipy}.
The RMSEs of these predictions were much higher than the random forest model test errors so they were not shown in \fig{bl}.
The baseline errors ranged from 14-20.7\% of the mean bond length, much larger than 2-4\% from the random forest models. 

%\item \done 

\section*{Data availability}
%\item \doing 
The code and datasets for this study are available in Zenodo and can be accessed via this link: \url{https://zenodo.org/records/13760791}, DOI: 10.5281/zenodo.13760791.

%-------------------------------------------------------------------------
     % The back matter of the paper - acknowledgements and references
     %-------------------------------------------------------------------------
\backmatter
\bmhead{Acknowledgements} 
We thank Weike Ye and Linda Hung for helpful discussions.
This work was funded by Toyota Research Institute, grant number PO-002332.  

%\end{enumerate}

\bibliography{billinge-group-bib, bg-pdf-standards, tn_ml_x-ray, tn_xanes_soln, tn_dpdf, tn_pdf}


\begin{thebibliography}{72}
\ifx \bisbn   \undefined \def \bisbn  #1{ISBN #1}\fi
\ifx \binits  \undefined \def \binits#1{#1}\fi
\ifx \bauthor  \undefined \def \bauthor#1{#1}\fi
\ifx \batitle  \undefined \def \batitle#1{#1}\fi
\ifx \bjtitle  \undefined \def \bjtitle#1{#1}\fi
\ifx \bvolume  \undefined \def \bvolume#1{\textbf{#1}}\fi
\ifx \byear  \undefined \def \byear#1{#1}\fi
\ifx \bissue  \undefined \def \bissue#1{#1}\fi
\ifx \bfpage  \undefined \def \bfpage#1{#1}\fi
\ifx \blpage  \undefined \def \blpage #1{#1}\fi
\ifx \burl  \undefined \def \burl#1{\textsf{#1}}\fi
\ifx \doiurl  \undefined \def \doiurl#1{\url{https://doi.org/#1}}\fi
\ifx \betal  \undefined \def \betal{\textit{et al.}}\fi
\ifx \binstitute  \undefined \def \binstitute#1{#1}\fi
\ifx \binstitutionaled  \undefined \def \binstitutionaled#1{#1}\fi
\ifx \bctitle  \undefined \def \bctitle#1{#1}\fi
\ifx \beditor  \undefined \def \beditor#1{#1}\fi
\ifx \bpublisher  \undefined \def \bpublisher#1{#1}\fi
\ifx \bbtitle  \undefined \def \bbtitle#1{#1}\fi
\ifx \bedition  \undefined \def \bedition#1{#1}\fi
\ifx \bseriesno  \undefined \def \bseriesno#1{#1}\fi
\ifx \blocation  \undefined \def \blocation#1{#1}\fi
\ifx \bsertitle  \undefined \def \bsertitle#1{#1}\fi
\ifx \bsnm \undefined \def \bsnm#1{#1}\fi
\ifx \bsuffix \undefined \def \bsuffix#1{#1}\fi
\ifx \bparticle \undefined \def \bparticle#1{#1}\fi
\ifx \barticle \undefined \def \barticle#1{#1}\fi
\bibcommenthead
\ifx \bconfdate \undefined \def \bconfdate #1{#1}\fi
\ifx \botherref \undefined \def \botherref #1{#1}\fi
\ifx \url \undefined \def \url#1{\textsf{#1}}\fi
\ifx \bchapter \undefined \def \bchapter#1{#1}\fi
\ifx \bbook \undefined \def \bbook#1{#1}\fi
\ifx \bcomment \undefined \def \bcomment#1{#1}\fi
\ifx \oauthor \undefined \def \oauthor#1{#1}\fi
\ifx \citeauthoryear \undefined \def \citeauthoryear#1{#1}\fi
\ifx \endbibitem  \undefined \def \endbibitem {}\fi
\ifx \bconflocation  \undefined \def \bconflocation#1{#1}\fi
\ifx \arxivurl  \undefined \def \arxivurl#1{\textsf{#1}}\fi
\csname PreBibitemsHook\endcsname

%%% 1
\bibitem[\protect\citeauthoryear{Egami and Billinge}{2012}]{egami;b;utbp12}
\begin{bbook}
\bauthor{\bsnm{Egami}, \binits{T.}},
\bauthor{\bsnm{Billinge}, \binits{S.J.L.}}:
\bbtitle{Underneath the {{Bragg}} Peaks: Structural Analysis of Complex Materials},
\bedition{2nd} edn.
\bsertitle{Pergamon Materials Series},
vol. \bseriesno{16}.
\bpublisher{{Elsevier}},
\blocation{{Amsterdam}}
(\byear{2012})
\end{bbook}
\endbibitem

%%% 2
\bibitem[\protect\citeauthoryear{Billinge and Levin}{2007}]{billi;s07}
\begin{barticle}
\bauthor{\bsnm{Billinge}, \binits{S.J.L.}},
\bauthor{\bsnm{Levin}, \binits{I.}}:
\batitle{The {{Problem}} with {{Determining Atomic Structure}} at the {{Nanoscale}}}.
\bjtitle{Science}
\bvolume{316}(\bissue{5824}),
\bfpage{561}--\blpage{565}
(\byear{2007})
\doiurl{10.1126/science.1135080}
\end{barticle}
\endbibitem

%%% 3
\bibitem[\protect\citeauthoryear{Billinge}{2010}]{billi;p10}
\begin{barticle}
\bauthor{\bsnm{Billinge}, \binits{S.J.L.}}:
\batitle{The nanostructure problem}.
\bjtitle{Physics}
\bvolume{3},
\bfpage{25}
(\byear{2010})
\doiurl{10.1103/Physics.3.25}
\end{barticle}
\endbibitem

%%% 4
\bibitem[\protect\citeauthoryear{McGreevy and Pusztai}{1988}]{mcgreevy1988reverse}
\begin{barticle}
\bauthor{\bsnm{McGreevy}, \binits{R.}},
\bauthor{\bsnm{Pusztai}, \binits{L.}}:
\batitle{Reverse monte carlo simulation: a new technique for the determination of disordered structures}.
\bjtitle{Molecular simulation}
\bvolume{1}(\bissue{6}),
\bfpage{359}--\blpage{367}
(\byear{1988})
\end{barticle}
\endbibitem

%%% 5
\bibitem[\protect\citeauthoryear{Billinge}{1998}]{billingeRealspaceRietveldFull1998b}
\begin{bchapter}
\bauthor{\bsnm{Billinge}, \binits{S.J.L.}}:
\bctitle{Real-space {{Rietveld}}: Full profile structure refinement of the atomic pair distribution function}.
In: \beditor{\bsnm{Billinge}, \binits{S.J.L.}},
\beditor{\bsnm{Thorpe}, \binits{M.F.}} (eds.)
\bbtitle{Local {{Structure}} from {{Diffraction}}},
p. \bfpage{137}.
\bpublisher{{Plenum}},
\blocation{{New York}}
(\byear{1998})
\end{bchapter}
\endbibitem

%%% 6
\bibitem[\protect\citeauthoryear{Juh{\'a}s et~al.}{2015}]{juhas;aca15}
\begin{barticle}
\bauthor{\bsnm{Juh{\'a}s}, \binits{P.}},
\bauthor{\bsnm{Farrow}, \binits{C.}},
\bauthor{\bsnm{Yang}, \binits{X.}},
\bauthor{\bsnm{Knox}, \binits{K.}},
\bauthor{\bsnm{Billinge}, \binits{S.}}:
\batitle{Complex modeling: A strategy and software program for combining multiple information sources to solve ill posed structure and nanostructure inverse problems}.
\bjtitle{Acta Crystallographica Section A: Foundations and Advances}
\bvolume{71}(\bissue{6}),
\bfpage{562}--\blpage{568}
(\byear{2015})
\doiurl{10.1107/S2053273315014473}
\end{barticle}
\endbibitem

%%% 7
\bibitem[\protect\citeauthoryear{Lahat et~al.}{2015}]{lahat2015multimodal}
\begin{barticle}
\bauthor{\bsnm{Lahat}, \binits{D.}},
\bauthor{\bsnm{Adali}, \binits{T.}},
\bauthor{\bsnm{Jutten}, \binits{C.}}:
\batitle{Multimodal data fusion: an overview of methods, challenges, and prospects}.
\bjtitle{Proceedings of the IEEE}
\bvolume{103}(\bissue{9}),
\bfpage{1449}--\blpage{1477}
(\byear{2015})
\end{barticle}
\endbibitem

%%% 8
\bibitem[\protect\citeauthoryear{Farrow}{unpublished}]{farrowunpublished}
\begin{botherref}
\oauthor{\bsnm{Farrow}, \binits{C.}}
Private communication
(unpublished)
\end{botherref}
\endbibitem

%%% 9
\bibitem[\protect\citeauthoryear{Billinge and Jensen}{2023}]{billinge2023atomic}
\begin{bbook}
\bauthor{\bsnm{Billinge}, \binits{S.}},
\bauthor{\bsnm{Jensen}, \binits{K.}}:
\bbtitle{Atomic Pair Distribution Function Analysis: A Primer}
vol. \bseriesno{22}.
\bpublisher{Oxford University Press},
\blocation{Oxford, UK}
(\byear{2023})
\end{bbook}
\endbibitem

%%% 10
\bibitem[\protect\citeauthoryear{Farrow et~al.}{2014}]{farrow2014robust}
\begin{barticle}
\bauthor{\bsnm{Farrow}, \binits{C.}},
\bauthor{\bsnm{Shi}, \binits{C.}},
\bauthor{\bsnm{Juh{\'a}s}, \binits{P.}},
\bauthor{\bsnm{Peng}, \binits{X.}},
\bauthor{\bsnm{Billinge}, \binits{S.L.J.}}:
\batitle{Robust structure and morphology parameters for {{CdS}} nanoparticles by combining small-angle {{X-ray}} scattering and atomic pair distribution function data in a complex modeling framework}.
\bjtitle{Journal of Applied Crystallography}
\bvolume{47}(\bissue{2}),
\bfpage{561}--\blpage{565}
(\byear{2014})
\doiurl{10.1107/S1600576713034055}
\end{barticle}
\endbibitem

%%% 11
\bibitem[\protect\citeauthoryear{Tucker et~al.}{2007}]{tucke;jpcm07}
\begin{barticle}
\bauthor{\bsnm{Tucker}, \binits{M.G.}},
\bauthor{\bsnm{Keen}, \binits{D.A.}},
\bauthor{\bsnm{Dove}, \binits{M.T.}},
\bauthor{\bsnm{Goodwin}, \binits{A.L.}},
\bauthor{\bsnm{Hui}, \binits{Q.}}:
\batitle{{{RMCProfile}}: Reverse {{Monte Carlo}} for polycrystalline materials}.
\bjtitle{Journal of Physics: Condensed Matter}
\bvolume{19}(\bissue{33}),
\bfpage{335218}
(\byear{2007})
\doiurl{10.1088/0953-8984/19/33/335218}
\end{barticle}
\endbibitem

%%% 12
\bibitem[\protect\citeauthoryear{Krayzman et~al.}{2008}]{krayzman2008simultaneous}
\begin{barticle}
\bauthor{\bsnm{Krayzman}, \binits{V.}},
\bauthor{\bsnm{Levin}, \binits{I.}},
\bauthor{\bsnm{Tucker}, \binits{M.G.}}:
\batitle{Simultaneous reverse {{Monte Carlo}} refinements of local structures in perovskite solid solutions using {{EXAFS}} and the total scattering pair-distribution function}.
\bjtitle{Journal of Applied Crystallography}
\bvolume{41}(\bissue{4}),
\bfpage{705}--\blpage{714}
(\byear{2008})
\doiurl{10.1107/S0021889808013277}
\end{barticle}
\endbibitem

%%% 13
\bibitem[\protect\citeauthoryear{Yano and Yachandra}{2009}]{yano2009x}
\begin{barticle}
\bauthor{\bsnm{Yano}, \binits{J.}},
\bauthor{\bsnm{Yachandra}, \binits{V.K.}}:
\batitle{X-ray absorption spectroscopy}.
\bjtitle{Photosynthesis research}
\bvolume{102},
\bfpage{241}--\blpage{254}
(\byear{2009})
\end{barticle}
\endbibitem

%%% 14
\bibitem[\protect\citeauthoryear{Rehr et~al.}{2010}]{rehr2010parameter}
\begin{barticle}
\bauthor{\bsnm{Rehr}, \binits{J.J.}},
\bauthor{\bsnm{Kas}, \binits{J.J.}},
\bauthor{\bsnm{Vila}, \binits{F.D.}},
\bauthor{\bsnm{Prange}, \binits{M.P.}},
\bauthor{\bsnm{Jorissen}, \binits{K.}}:
\batitle{Parameter-free calculations of x-ray spectra with feff9}.
\bjtitle{Physical Chemistry Chemical Physics}
\bvolume{12}(\bissue{21}),
\bfpage{5503}--\blpage{5513}
(\byear{2010})
\end{barticle}
\endbibitem

%%% 15
\bibitem[\protect\citeauthoryear{Mathew et~al.}{2018}]{mathew2018high}
\begin{barticle}
\bauthor{\bsnm{Mathew}, \binits{K.}},
\bauthor{\bsnm{Zheng}, \binits{C.}},
\bauthor{\bsnm{Winston}, \binits{D.}},
\bauthor{\bsnm{Chen}, \binits{C.}},
\bauthor{\bsnm{Dozier}, \binits{A.}},
\bauthor{\bsnm{Rehr}, \binits{J.J.}},
\bauthor{\bsnm{Ong}, \binits{S.P.}},
\bauthor{\bsnm{Persson}, \binits{K.A.}}:
\batitle{High-throuput computational x-ray absorption spectroscopy}.
\bjtitle{Scientific Data}
\bvolume{5},
\bfpage{180151}
(\byear{2018})
\doiurl{10.1038/sdata.2018.151}
\end{barticle}
\endbibitem

%%% 16
\bibitem[\protect\citeauthoryear{Jain et~al.}{2013}]{jain2013commentary}
\begin{botherref}
\oauthor{\bsnm{Jain}, \binits{A.}},
\oauthor{\bsnm{Ong}, \binits{S.P.}},
\oauthor{\bsnm{Hautier}, \binits{G.}},
\oauthor{\bsnm{Chen}, \binits{W.}},
\oauthor{\bsnm{Richards}, \binits{W.D.}},
\oauthor{\bsnm{Dacek}, \binits{S.}},
\oauthor{\bsnm{Cholia}, \binits{S.}},
\oauthor{\bsnm{Gunter}, \binits{D.}},
\oauthor{\bsnm{Skinner}, \binits{D.}},
\oauthor{\bsnm{Ceder}, \binits{G.}}, et al.:
Commentary: The materials project: A materials genome approach to accelerating materials innovation.
APL materials
\textbf{1}(1)
(2013)
\end{botherref}
\endbibitem

%%% 17
\bibitem[\protect\citeauthoryear{Timoshenko and Frenkel}{2019}]{timoshenko2019inverting}
\begin{barticle}
\bauthor{\bsnm{Timoshenko}, \binits{J.}},
\bauthor{\bsnm{Frenkel}, \binits{A.I.}}:
\batitle{``{{Inverting}}'' {{X-ray Absorption Spectra}} of {{Catalysts}} by {{Machine Learning}} in {{Search}} for {{Activity Descriptors}}}.
\bjtitle{ACS Catalysis}
\bvolume{9}(\bissue{11}),
\bfpage{10192}--\blpage{10211}
(\byear{2019})
\doiurl{10.1021/acscatal.9b03599}
\end{barticle}
\endbibitem

%%% 18
\bibitem[\protect\citeauthoryear{Penfold et~al.}{2024}]{penfold2024machinelearning}
\begin{barticle}
\bauthor{\bsnm{Penfold}, \binits{T.}},
\bauthor{\bsnm{Watson}, \binits{L.}},
\bauthor{\bsnm{Middleton}, \binits{C.}},
\bauthor{\bsnm{David}, \binits{T.}},
\bauthor{\bsnm{Verma}, \binits{S.}},
\bauthor{\bsnm{Pope}, \binits{T.}},
\bauthor{\bsnm{Kaczmarek}, \binits{J.}},
\bauthor{\bsnm{Rankine}, \binits{C.}}:
\batitle{Machine-learning strategies for the accurate and efficient analysis of x-ray spectroscopy}.
\bjtitle{Machine Learning: Science and Technology}
\bvolume{5}(\bissue{2}),
\bfpage{021001}
(\byear{2024})
\doiurl{10.1088/2632-2153/ad5074}
\end{barticle}
\endbibitem

%%% 19
\bibitem[\protect\citeauthoryear{Lecun et~al.}{2015}]{lecun2015deeplearning}
\begin{barticle}
\bauthor{\bsnm{Lecun}, \binits{Y.}},
\bauthor{\bsnm{Bengio}, \binits{Y.}},
\bauthor{\bsnm{Hinton}, \binits{G.}}:
\batitle{Deep learning.}
\bjtitle{Nature}
\bvolume{521},
\bfpage{436}--\blpage{444}
(\byear{2015})
\end{barticle}
\endbibitem

%%% 20
\bibitem[\protect\citeauthoryear{Breiman}{2001}]{breiman2001random}
\begin{barticle}
\bauthor{\bsnm{Breiman}, \binits{L.}}:
\batitle{Random forests}.
\bjtitle{Machine learning}
\bvolume{45},
\bfpage{5}--\blpage{32}
(\byear{2001})
\end{barticle}
\endbibitem

%%% 21
\bibitem[\protect\citeauthoryear{Timoshenko et~al.}{2017}]{timoshenko2017supervised}
\begin{barticle}
\bauthor{\bsnm{Timoshenko}, \binits{J.}},
\bauthor{\bsnm{Lu}, \binits{D.}},
\bauthor{\bsnm{Lin}, \binits{Y.}},
\bauthor{\bsnm{Frenkel}, \binits{A.I.}}:
\batitle{Supervised {{Machine-Learning-Based Determination}} of {{Three-Dimensional Structure}} of {{Metallic Nanoparticles}}}.
\bjtitle{The Journal of Physical Chemistry Letters}
\bvolume{8}(\bissue{20}),
\bfpage{5091}--\blpage{5098}
(\byear{2017})
\doiurl{10.1021/acs.jpclett.7b02364}
\end{barticle}
\endbibitem

%%% 22
\bibitem[\protect\citeauthoryear{Carbone et~al.}{2019}]{carbone_classification_2019}
\begin{barticle}
\bauthor{\bsnm{Carbone}, \binits{M.R.}},
\bauthor{\bsnm{Yoo}, \binits{S.}},
\bauthor{\bsnm{Topsakal}, \binits{M.}},
\bauthor{\bsnm{Lu}, \binits{D.}}:
\batitle{Classification of local chemical environments from x-ray absorption spectra using supervised machine learning}.
\bjtitle{Physical Review Materials}
\bvolume{3}(\bissue{3}),
\bfpage{033604}
(\byear{2019})
\doiurl{10.1103/PhysRevMaterials.3.033604}
\end{barticle}
\endbibitem

%%% 23
\bibitem[\protect\citeauthoryear{Zheng et~al.}{2020}]{zheng_random_2020}
\begin{botherref}
\oauthor{\bsnm{Zheng}, \binits{C.}},
\oauthor{\bsnm{Chen}, \binits{C.}},
\oauthor{\bsnm{Chen}, \binits{Y.}},
\oauthor{\bsnm{Ong}, \binits{S.P.}}:
Random {{Forest Models}} for {{Accurate Identification}} of {{Coordination Environments}} from {{X-Ray Absorption Near-Edge Structure}}.
Patterns
\textbf{1}(2)
(2020)
\doiurl{10.1016/j.patter.2020.100013}
\end{botherref}
\endbibitem

%%% 24
\bibitem[\protect\citeauthoryear{Torrisi et~al.}{2020}]{torrisi_random_2020}
\begin{barticle}
\bauthor{\bsnm{Torrisi}, \binits{S.B.}},
\bauthor{\bsnm{Carbone}, \binits{M.R.}},
\bauthor{\bsnm{Rohr}, \binits{B.A.}},
\bauthor{\bsnm{Montoya}, \binits{J.H.}},
\bauthor{\bsnm{Ha}, \binits{Y.}},
\bauthor{\bsnm{Yano}, \binits{J.}},
\bauthor{\bsnm{Suram}, \binits{S.K.}},
\bauthor{\bsnm{Hung}, \binits{L.}}:
\batitle{Random forest machine learning models for interpretable {{X-ray}} absorption near-edge structure spectrum-property relationships}.
\bjtitle{npj Computational Materials}
\bvolume{6}(\bissue{1}),
\bfpage{1}--\blpage{11}
(\byear{2020})
\doiurl{10.1038/s41524-020-00376-6}
\end{barticle}
\endbibitem

%%% 25
\bibitem[\protect\citeauthoryear{Kwon et~al.}{2023}]{kwon2023spectroscopyguided}
\begin{botherref}
\oauthor{\bsnm{Kwon}, \binits{H.}},
\oauthor{\bsnm{Hsu}, \binits{T.}},
\oauthor{\bsnm{Sun}, \binits{W.}},
\oauthor{\bsnm{Jeong}, \binits{W.}},
\oauthor{\bsnm{Aydin}, \binits{F.}},
\oauthor{\bsnm{Chapman}, \binits{J.}},
\oauthor{\bsnm{Chen}, \binits{X.}},
\oauthor{\bsnm{Carbone}, \binits{M.R.}},
\oauthor{\bsnm{Lu}, \binits{D.}},
\oauthor{\bsnm{Zhou}, \binits{F.}},
\oauthor{\bsnm{Pham}, \binits{T.A.}}:
Spectroscopy-{{Guided Discovery}} of {{Three-Dimensional Structures}} of {{Disordered Materials}} with {{Diffusion Models}}.
arXiv
(2023).
\doiurl{10.48550/arXiv.2312.05472}
\end{botherref}
\endbibitem

%%% 26
\bibitem[\protect\citeauthoryear{Tetef et~al.}{2021}]{tetef2021unsupervised}
\begin{barticle}
\bauthor{\bsnm{Tetef}, \binits{S.}},
\bauthor{\bsnm{Govind}, \binits{N.}},
\bauthor{\bsnm{Seidler}, \binits{G.T.}}:
\batitle{Unsupervised machine learning for unbiased chemical classification in {{X-ray}} absorption spectroscopy and {{X-ray}} emission spectroscopy}.
\bjtitle{Physical Chemistry Chemical Physics}
\bvolume{23}(\bissue{41}),
\bfpage{23586}--\blpage{23601}
(\byear{2021})
\doiurl{10.1039/D1CP02903G}
\end{barticle}
\endbibitem

%%% 27
\bibitem[\protect\citeauthoryear{Tetef et~al.}{2022}]{tetef2022informed}
\begin{barticle}
\bauthor{\bsnm{Tetef}, \binits{S.}},
\bauthor{\bsnm{Kashyap}, \binits{V.}},
\bauthor{\bsnm{Holden}, \binits{W.M.}},
\bauthor{\bsnm{Velian}, \binits{A.}},
\bauthor{\bsnm{Govind}, \binits{N.}},
\bauthor{\bsnm{Seidler}, \binits{G.T.}}:
\batitle{Informed {{Chemical Classification}} of {{Organophosphorus Compounds}} via {{Unsupervised Machine Learning}} of {{X-ray Absorption Spectroscopy}} and {{X-ray Emission Spectroscopy}}}.
\bjtitle{The Journal of Physical Chemistry A}
\bvolume{126}(\bissue{29}),
\bfpage{4862}--\blpage{4872}
(\byear{2022})
\doiurl{10.1021/acs.jpca.2c03635}
\end{barticle}
\endbibitem

%%% 28
\bibitem[\protect\citeauthoryear{Guda et~al.}{2021}]{guda_understanding_2021}
\begin{barticle}
\bauthor{\bsnm{Guda}, \binits{A.A.}},
\bauthor{\bsnm{Guda}, \binits{S.A.}},
\bauthor{\bsnm{Martini}, \binits{A.}},
\bauthor{\bsnm{Kravtsova}, \binits{A.N.}},
\bauthor{\bsnm{Algasov}, \binits{A.}},
\bauthor{\bsnm{Bugaev}, \binits{A.}},
\bauthor{\bsnm{Kubrin}, \binits{S.P.}},
\bauthor{\bsnm{Guda}, \binits{L.V.}},
\bauthor{\bsnm{{\v S}ot}, \binits{P.}},
\bauthor{\bsnm{{van Bokhoven}}, \binits{J.A.}},
\bauthor{\bsnm{Cop{\'e}ret}, \binits{C.}},
\bauthor{\bsnm{Soldatov}, \binits{A.V.}}:
\batitle{Understanding {{X-ray}} absorption spectra by means of descriptors and machine learning algorithms}.
\bjtitle{npj Computational Materials}
\bvolume{7}(\bissue{1}),
\bfpage{1}--\blpage{13}
(\byear{2021})
\doiurl{10.1038/s41524-021-00664-9}
\end{barticle}
\endbibitem

%%% 29
\bibitem[\protect\citeauthoryear{Liang et~al.}{2023}]{liang2023decoding}
\begin{barticle}
\bauthor{\bsnm{Liang}, \binits{Z.}},
\bauthor{\bsnm{Carbone}, \binits{M.R.}},
\bauthor{\bsnm{Chen}, \binits{W.}},
\bauthor{\bsnm{Meng}, \binits{F.}},
\bauthor{\bsnm{Stavitski}, \binits{E.}},
\bauthor{\bsnm{Lu}, \binits{D.}},
\bauthor{\bsnm{Hybertsen}, \binits{M.S.}},
\bauthor{\bsnm{Qu}, \binits{X.}}:
\batitle{Decoding structure-spectrum relationships with physically organized latent spaces}.
\bjtitle{Physical Review Materials}
\bvolume{7}(\bissue{5}),
\bfpage{053802}
(\byear{2023})
\doiurl{10.1103/PhysRevMaterials.7.053802}
\end{barticle}
\endbibitem

%%% 30
\bibitem[\protect\citeauthoryear{Liu et~al.}{2019}]{liu2019using}
\begin{barticle}
\bauthor{\bsnm{Liu}, \binits{C.-H.}},
\bauthor{\bsnm{Tao}, \binits{Y.}},
\bauthor{\bsnm{Hsu}, \binits{D.}},
\bauthor{\bsnm{Du}, \binits{Q.}},
\bauthor{\bsnm{Billinge}, \binits{S.J.L.}}:
\batitle{Using a machine learning approach to determine the space group of a structure from the atomic pair distribution function}.
\bjtitle{Acta Crystallographica Section A: Foundations and Advances}
\bvolume{75}(\bissue{4}),
\bfpage{633}--\blpage{643}
(\byear{2019})
\doiurl{10.1107/S2053273319005606}
\end{barticle}
\endbibitem

%%% 31
\bibitem[\protect\citeauthoryear{Lan et~al.}{2022}]{lan2022robustness}
\begin{barticle}
\bauthor{\bsnm{Lan}, \binits{L.}},
\bauthor{\bsnm{Liu}, \binits{C.-H.}},
\bauthor{\bsnm{Du}, \binits{Q.}},
\bauthor{\bsnm{Billinge}, \binits{S.J.L.}}:
\batitle{Robustness test of the {{spacegroupMining}} model for determining space groups from atomic pair distribution function data}.
\bjtitle{Journal of Applied Crystallography}
\bvolume{55}(\bissue{3}),
\bfpage{626}--\blpage{630}
(\byear{2022})
\doiurl{10.1107/S1600576722002990}
\end{barticle}
\endbibitem

%%% 32
\bibitem[\protect\citeauthoryear{Kj{\ae}r et~al.}{2023}]{kjaerDeepStrucStructureSolution2023}
\begin{barticle}
\bauthor{\bsnm{Kj{\ae}r}, \binits{E.T.S.}},
\bauthor{\bsnm{Anker}, \binits{A.S.}},
\bauthor{\bsnm{Weng}, \binits{M.N.}},
\bauthor{\bsnm{Billinge}, \binits{S.J.L.}},
\bauthor{\bsnm{Selvan}, \binits{R.}},
\bauthor{\bsnm{Jensen}, \binits{K.M.{\O}.}}:
\batitle{{{DeepStruc}}: Towards structure solution from pair distribution function data using deep generative models}.
\bjtitle{Digital Discovery}
\bvolume{2}(\bissue{1}),
\bfpage{69}--\blpage{80}
(\byear{2023})
\doiurl{10.1039/D2DD00086E}
\end{barticle}
\endbibitem

%%% 33
\bibitem[\protect\citeauthoryear{Zhang et~al.}{2023}]{zhang2023pair}
\begin{barticle}
\bauthor{\bsnm{Zhang}, \binits{S.}},
\bauthor{\bsnm{Gong}, \binits{J.}},
\bauthor{\bsnm{Chu}, \binits{S.}},
\bauthor{\bsnm{Xiao}, \binits{D.Z.}},
\bauthor{\bsnm{{Reeja-Jayan}}, \binits{B.}},
\bauthor{\bsnm{McGaughey}, \binits{A.J.H.}}:
\batitle{Pair distribution function analysis for oxide defect identification through feature extraction and supervised learning}.
\bjtitle{APL Machine Learning}
\bvolume{1}(\bissue{2}),
\bfpage{026115}
(\byear{2023})
\doiurl{10.1063/5.0130681}
\end{barticle}
\endbibitem

%%% 34
\bibitem[\protect\citeauthoryear{Anker et~al.}{2022}]{anker2022extracting}
\begin{barticle}
\bauthor{\bsnm{Anker}, \binits{A.S.}},
\bauthor{\bsnm{Kj{\ae}r}, \binits{E.T.S.}},
\bauthor{\bsnm{Juelsholt}, \binits{M.}},
\bauthor{\bsnm{Christiansen}, \binits{T.L.}},
\bauthor{\bsnm{Skj{\ae}rv{\o}}, \binits{S.L.}},
\bauthor{\bsnm{J{\o}rgensen}, \binits{M.R.V.}},
\bauthor{\bsnm{Kantor}, \binits{I.}},
\bauthor{\bsnm{S{\o}rensen}, \binits{D.R.}},
\bauthor{\bsnm{Billinge}, \binits{S.J.L.}},
\bauthor{\bsnm{Selvan}, \binits{R.}},
\bauthor{\bsnm{Jensen}, \binits{K.M.{\O}.}}:
\batitle{Extracting structural motifs from pair distribution function data of nanostructures using explainable machine learning}.
\bjtitle{npj Computational Materials}
\bvolume{8}(\bissue{1}),
\bfpage{1}--\blpage{11}
(\byear{2022})
\doiurl{10.1038/s41524-022-00896-3}
\end{barticle}
\endbibitem

%%% 35
\bibitem[\protect\citeauthoryear{Kl{\o}ve et~al.}{2023}]{klove2023machinelearningbased}
\begin{barticle}
\bauthor{\bsnm{Kl{\o}ve}, \binits{M.}},
\bauthor{\bsnm{Sommer}, \binits{S.}},
\bauthor{\bsnm{Iversen}, \binits{B.B.}},
\bauthor{\bsnm{Hammer}, \binits{B.}},
\bauthor{\bsnm{Dononelli}, \binits{W.}}:
\batitle{A {{Machine-Learning-Based Approach}} for {{Solving Atomic Structures}} of {{Nanomaterials Combining Pair Distribution Functions}} with {{Density Functional Theory}}}.
\bjtitle{Advanced Materials}
\bvolume{35}(\bissue{13}),
\bfpage{2208220}
(\byear{2023})
\doiurl{10.1002/adma.202208220}
\end{barticle}
\endbibitem

%%% 36
\bibitem[\protect\citeauthoryear{Liu et~al.}{2021}]{liu2021validation}
\begin{barticle}
\bauthor{\bsnm{Liu}, \binits{C.-H.}},
\bauthor{\bsnm{Wright}, \binits{C.J.}},
\bauthor{\bsnm{Gu}, \binits{R.}},
\bauthor{\bsnm{Bandi}, \binits{S.}},
\bauthor{\bsnm{Wustrow}, \binits{A.}},
\bauthor{\bsnm{Todd}, \binits{P.K.}},
\bauthor{\bsnm{O'Nolan}, \binits{D.}},
\bauthor{\bsnm{Beauvais}, \binits{M.L.}},
\bauthor{\bsnm{Neilson}, \binits{J.R.}},
\bauthor{\bsnm{Chupas}, \binits{P.J.}},
\bauthor{\bsnm{Chapman}, \binits{K.W.}},
\bauthor{\bsnm{Billinge}, \binits{S.J.L.}}:
\batitle{Validation of non-negative matrix factorization for rapid assessment of large sets of atomic pair distribution function data}.
\bjtitle{Journal of Applied Crystallography}
\bvolume{54}(\bissue{3}),
\bfpage{768}--\blpage{775}
(\byear{2021})
\doiurl{10.1107/S160057672100265X}
\end{barticle}
\endbibitem

%%% 37
\bibitem[\protect\citeauthoryear{Thatcher et~al.}{2022}]{thatcher2022nmfmapping}
\begin{barticle}
\bauthor{\bsnm{Thatcher}, \binits{Z.}},
\bauthor{\bsnm{Liu}, \binits{C.H.}},
\bauthor{\bsnm{Yang}, \binits{L.}},
\bauthor{\bsnm{McBride}, \binits{B.C.}},
\bauthor{\bsnm{Thinh~Tran}, \binits{G.}},
\bauthor{\bsnm{Wustrow}, \binits{A.}},
\bauthor{\bsnm{Karlsen}, \binits{M.A.}},
\bauthor{\bsnm{Neilson}, \binits{J.R.}},
\bauthor{\bsnm{Ravnsb{\ae}k}, \binits{D.B.}},
\bauthor{\bsnm{Billinge}, \binits{S.J.L.}}:
\batitle{{{nmfMapping}}: A cloud-based web application for non-negative matrix factorization of powder diffraction and pair distribution function datasets}.
\bjtitle{Acta Crystallographica. Section A, Foundations and Advances}
\bvolume{78}(\bissue{Pt 3}),
\bfpage{242}--\blpage{248}
(\byear{2022})
\doiurl{10.1107/S2053273322002522}
\end{barticle}
\endbibitem

%%% 38
\bibitem[\protect\citeauthoryear{Guo et~al.}{2023}]{guo2023endtoend}
\begin{botherref}
\oauthor{\bsnm{Guo}, \binits{G.}},
\oauthor{\bsnm{Goldfeder}, \binits{J.}},
\oauthor{\bsnm{Lan}, \binits{L.}},
\oauthor{\bsnm{Ray}, \binits{A.}},
\oauthor{\bsnm{Yang}, \binits{A.H.}},
\oauthor{\bsnm{Chen}, \binits{B.}},
\oauthor{\bsnm{Billinge}, \binits{S.J.}},
\oauthor{\bsnm{Lipson}, \binits{H.}}:
Towards {{End-to-End Structure Solutions}} from {{Information-Compromised Diffraction Data}} via {{Generative Deep Learning}}.
arXiv
(2023).
\doiurl{10.48550/arXiv.2312.15136}
\end{botherref}
\endbibitem

%%% 39
\bibitem[\protect\citeauthoryear{Guo et~al.}{2024}]{guo2024diffusion}
\begin{botherref}
\oauthor{\bsnm{Guo}, \binits{G.}},
\oauthor{\bsnm{Saidi}, \binits{T.}},
\oauthor{\bsnm{Terban}, \binits{M.}},
\oauthor{\bsnm{Billinge}, \binits{S.J.}},
\oauthor{\bsnm{Lipson}, \binits{H.}}:
Diffusion models are promising for ab initio structure solutions from nanocrystalline powder diffraction data
(arXiv:2406.10796)
(2024)
\doiurl{10.48550/arXiv.2406.10796}
{\href{https://arxiv.org/abs/2406.10796}{{arXiv:2406.10796}}}
\end{botherref}
\endbibitem

%%% 40
\bibitem[\protect\citeauthoryear{Baltru{\v{s}}aitis et~al.}{2018}]{baltruvsaitis2018multimodal}
\begin{barticle}
\bauthor{\bsnm{Baltru{\v{s}}aitis}, \binits{T.}},
\bauthor{\bsnm{Ahuja}, \binits{C.}},
\bauthor{\bsnm{Morency}, \binits{L.-P.}}:
\batitle{Multimodal machine learning: A survey and taxonomy}.
\bjtitle{IEEE transactions on pattern analysis and machine intelligence}
\bvolume{41}(\bissue{2}),
\bfpage{423}--\blpage{443}
(\byear{2018})
\end{barticle}
\endbibitem

%%% 41
\bibitem[\protect\citeauthoryear{Hastie et~al.}{2009}]{hastie2009elements}
\begin{bbook}
\bauthor{\bsnm{Hastie}, \binits{T.}},
\bauthor{\bsnm{Tibshirani}, \binits{R.}},
\bauthor{\bsnm{Friedman}, \binits{J.H.}}:
\bbtitle{The Elements of Statistical Learning: Data Mining, Inference, and Prediction}
vol. \bseriesno{2}.
\bpublisher{Springer},
\blocation{New York, NY, USA}
(\byear{2009})
\end{bbook}
\endbibitem

%%% 42
\bibitem[\protect\citeauthoryear{Carbone et~al.}{2024}]{carbone2024accurate}
\begin{barticle}
\bauthor{\bsnm{Carbone}, \binits{M.R.}},
\bauthor{\bsnm{Maffettone}, \binits{P.M.}},
\bauthor{\bsnm{Qu}, \binits{X.}},
\bauthor{\bsnm{Yoo}, \binits{S.}},
\bauthor{\bsnm{Lu}, \binits{D.}}:
\batitle{Accurate, {{Uncertainty-Aware Classification}} of {{Molecular Chemical Motifs}} from {{Multimodal X-ray Absorption Spectroscopy}}}.
\bjtitle{The Journal of Physical Chemistry A}
\bvolume{128}(\bissue{10}),
\bfpage{1948}--\blpage{1957}
(\byear{2024})
\doiurl{10.1021/acs.jpca.3c06910}
\end{barticle}
\endbibitem

%%% 43
\bibitem[\protect\citeauthoryear{Szymanski et~al.}{2024}]{szymanski_integrated_2024-2}
\begin{barticle}
\bauthor{\bsnm{Szymanski}, \binits{N.J.}},
\bauthor{\bsnm{Fu}, \binits{S.}},
\bauthor{\bsnm{Persson}, \binits{E.}},
\bauthor{\bsnm{Ceder}, \binits{G.}}:
\batitle{Integrated analysis of {{X-ray}} diffraction patterns and pair distribution functions for machine-learned phase identification}.
\bjtitle{npj Computational Materials}
\bvolume{10}(\bissue{1}),
\bfpage{1}--\blpage{9}
(\byear{2024})
\doiurl{10.1038/s41524-024-01230-9}
\end{barticle}
\endbibitem

%%% 44
\bibitem[\protect\citeauthoryear{Louppe et~al.}{2013}]{louppe2013understanding}
\begin{botherref}
\oauthor{\bsnm{Louppe}, \binits{G.}},
\oauthor{\bsnm{Wehenkel}, \binits{L.}},
\oauthor{\bsnm{Sutera}, \binits{A.}},
\oauthor{\bsnm{Geurts}, \binits{P.}}:
Understanding variable importances in forests of randomized trees.
Advances in neural information processing systems
\textbf{26}
(2013)
\end{botherref}
\endbibitem

%%% 45
\bibitem[\protect\citeauthoryear{Wilke et~al.}{2001}]{wilke2001oxidation}
\begin{barticle}
\bauthor{\bsnm{Wilke}, \binits{M.}},
\bauthor{\bsnm{Farges}, \binits{F.}},
\bauthor{\bsnm{Petit}, \binits{P.-E.}},
\bauthor{\bsnm{Brown~Jr}, \binits{G.E.}},
\bauthor{\bsnm{Martin}, \binits{F.}}:
\batitle{Oxidation state and coordination of fe in minerals: An fe k-xanes spectroscopic study}.
\bjtitle{American Mineralogist}
\bvolume{86}(\bissue{5-6}),
\bfpage{714}--\blpage{730}
(\byear{2001})
\end{barticle}
\endbibitem

%%% 46
\bibitem[\protect\citeauthoryear{Brown and Shannon}{1973}]{brown1973empirical}
\begin{barticle}
\bauthor{\bsnm{Brown}, \binits{I.}},
\bauthor{\bsnm{Shannon}, \binits{R.}}:
\batitle{Empirical bond-strength--bond-length curves for oxides}.
\bjtitle{Acta Crystallographica Section A: Crystal Physics, Diffraction, Theoretical and General Crystallography}
\bvolume{29}(\bissue{3}),
\bfpage{266}--\blpage{282}
(\byear{1973})
\end{barticle}
\endbibitem

%%% 47
\bibitem[\protect\citeauthoryear{Brown and Altermatt}{1985}]{brown1985bond}
\begin{barticle}
\bauthor{\bsnm{Brown}, \binits{I.D.}},
\bauthor{\bsnm{Altermatt}, \binits{D.}}:
\batitle{Bond-valence parameters obtained from a systematic analysis of the inorganic crystal structure database}.
\bjtitle{Acta Crystallographica Section B: Structural Science}
\bvolume{41}(\bissue{4}),
\bfpage{244}--\blpage{247}
(\byear{1985})
\end{barticle}
\endbibitem

%%% 48
\bibitem[\protect\citeauthoryear{Shannon}{1976}]{shannon1976revised}
\begin{barticle}
\bauthor{\bsnm{Shannon}, \binits{R.D.}}:
\batitle{Revised effective ionic radii and systematic studies of interatomic distances in halides and chalcogenides}.
\bjtitle{Acta crystallographica section A: crystal physics, diffraction, theoretical and general crystallography}
\bvolume{32}(\bissue{5}),
\bfpage{751}--\blpage{767}
(\byear{1976})
\end{barticle}
\endbibitem

%%% 49
\bibitem[\protect\citeauthoryear{Henderson et~al.}{2014}]{henderson2014x}
\begin{barticle}
\bauthor{\bsnm{Henderson}, \binits{G.S.}},
\bauthor{\bsnm{De~Groot}, \binits{F.M.}},
\bauthor{\bsnm{Moulton}, \binits{B.J.}}:
\batitle{X-ray absorption near-edge structure (xanes) spectroscopy}.
\bjtitle{Reviews in Mineralogy and Geochemistry}
\bvolume{78}(\bissue{1}),
\bfpage{75}--\blpage{138}
(\byear{2014})
\end{barticle}
\endbibitem

%%% 50
\bibitem[\protect\citeauthoryear{Waychunas}{1987}]{waychunas_synchrotron_1987-1}
\begin{barticle}
\bauthor{\bsnm{Waychunas}, \binits{G.A.}}:
\batitle{Synchrotron radiation {{XANES}} spectroscopy of {{Ti}} in minerals; effects of {{Ti}} bonding distances, {{Ti}} valence, and site geometry on absorption edge structure}.
\bjtitle{American Mineralogist}
\bvolume{72}(\bissue{1-2}),
\bfpage{89}--\blpage{101}
(\byear{1987})
\end{barticle}
\endbibitem

%%% 51
\bibitem[\protect\citeauthoryear{Virtanen et~al.}{2020}]{virtanen2020scipy}
\begin{barticle}
\bauthor{\bsnm{Virtanen}, \binits{P.}},
\bauthor{\bsnm{Gommers}, \binits{R.}},
\bauthor{\bsnm{Oliphant}, \binits{T.E.}},
\bauthor{\bsnm{Haberland}, \binits{M.}},
\bauthor{\bsnm{Reddy}, \binits{T.}},
\bauthor{\bsnm{Cournapeau}, \binits{D.}},
\bauthor{\bsnm{Burovski}, \binits{E.}},
\bauthor{\bsnm{Peterson}, \binits{P.}},
\bauthor{\bsnm{Weckesser}, \binits{W.}},
\bauthor{\bsnm{Bright}, \binits{J.}}, \betal:
\batitle{Scipy 1.0: fundamental algorithms for scientific computing in python}.
\bjtitle{Nature methods}
\bvolume{17}(\bissue{3}),
\bfpage{261}--\blpage{272}
(\byear{2020})
\end{barticle}
\endbibitem

%%% 52
\bibitem[\protect\citeauthoryear{Farges et~al.}{1997}]{farges_ti_1997}
\begin{barticle}
\bauthor{\bsnm{Farges}, \binits{F.}},
\bauthor{\bsnm{Brown}, \binits{G.E.}},
\bauthor{\bsnm{Rehr}, \binits{J.J.}}:
\batitle{Ti \${{K}}\$-edge {{XANES}} studies of {{Ti}} coordination and disorder in oxide compounds: {{Comparison}} between theory and experiment}.
\bjtitle{Physical Review B}
\bvolume{56}(\bissue{4}),
\bfpage{1809}--\blpage{1819}
(\byear{1997})
\doiurl{10.1103/PhysRevB.56.1809}
\end{barticle}
\endbibitem

%%% 53
\bibitem[\protect\citeauthoryear{Jackson et~al.}{2005}]{jackson2005multi}
\begin{barticle}
\bauthor{\bsnm{Jackson}, \binits{W.E.}},
\bauthor{\bsnm{Farges}, \binits{F.}},
\bauthor{\bsnm{Yeager}, \binits{M.}},
\bauthor{\bsnm{Mabrouk}, \binits{P.A.}},
\bauthor{\bsnm{Rossano}, \binits{S.}},
\bauthor{\bsnm{Waychunas}, \binits{G.A.}},
\bauthor{\bsnm{Solomon}, \binits{E.I.}},
\bauthor{\bsnm{Brown~Jr}, \binits{G.E.}}:
\batitle{Multi-spectroscopic study of fe (ii) in silicate glasses: Implications for the coordination environment of fe (ii) in silicate melts}.
\bjtitle{Geochimica et Cosmochimica Acta}
\bvolume{69}(\bissue{17}),
\bfpage{4315}--\blpage{4332}
(\byear{2005})
\end{barticle}
\endbibitem

%%% 54
\bibitem[\protect\citeauthoryear{Aur et~al.}{1983}]{aur1983local}
\begin{barticle}
\bauthor{\bsnm{Aur}, \binits{S.}},
\bauthor{\bsnm{Kofalt}, \binits{D.}},
\bauthor{\bsnm{Waseda}, \binits{Y.}},
\bauthor{\bsnm{Egami}, \binits{T.}},
\bauthor{\bsnm{Wang}, \binits{R.}},
\bauthor{\bsnm{Chen}, \binits{H.}},
\bauthor{\bsnm{Teo}, \binits{B.-K.}}:
\batitle{Local structure of amorphous mo50ni50 determined by anomalous x-ray scattering using synchroton radiation}.
\bjtitle{Solid state communications}
\bvolume{48}(\bissue{2}),
\bfpage{111}--\blpage{115}
(\byear{1983})
\end{barticle}
\endbibitem

%%% 55
\bibitem[\protect\citeauthoryear{Kofalt et~al.}{1986}]{kofalt1986differential}
\begin{barticle}
\bauthor{\bsnm{Kofalt}, \binits{D.}},
\bauthor{\bsnm{Nanao}, \binits{S.}},
\bauthor{\bsnm{Egami}, \binits{T.}},
\bauthor{\bsnm{Wong}, \binits{K.}},
\bauthor{\bsnm{Poon}, \binits{S.}}:
\batitle{Differential anomalous-x-ray-scattering study of icosahedral and amorphous pd 58.8 u 20.6 si 20.6}.
\bjtitle{Physical review letters}
\bvolume{57}(\bissue{1}),
\bfpage{114}
(\byear{1986})
\end{barticle}
\endbibitem

%%% 56
\bibitem[\protect\citeauthoryear{Waseda}{1984}]{waseda1984novel}
\begin{bbook}
\bauthor{\bsnm{Waseda}, \binits{Y.}}:
\bbtitle{Novel Application of Anomalous (resonance) X-ray Scattering for Structural Characterization of Disordered Materials}.
\bpublisher{Springer},
\blocation{New York, NY, USA}
(\byear{1984})
\end{bbook}
\endbibitem

%%% 57
\bibitem[\protect\citeauthoryear{Waseda}{2002}]{waseda2002anomalous}
\begin{bbook}
\bauthor{\bsnm{Waseda}, \binits{Y.}}:
\bbtitle{Anomalous X-ray Scattering for Materials Characterization: Atomic-scale Structure Determination}
vol. \bseriesno{179}.
\bpublisher{Springer},
\blocation{New York, NY, USA}
(\byear{2002})
\end{bbook}
\endbibitem

%%% 58
\bibitem[\protect\citeauthoryear{Petkov et~al.}{2000}]{petkovLocalStructureIn02000}
\begin{barticle}
\bauthor{\bsnm{Petkov}, \binits{V.}},
\bauthor{\bsnm{Jeong}, \binits{I.-K.}},
\bauthor{\bsnm{{Mohiuddin-Jacobs}}, \binits{F.}},
\bauthor{\bsnm{Proffen}, \binits{T.}},
\bauthor{\bsnm{Billinge}, \binits{S.J.L.}},
\bauthor{\bsnm{Dmowski}, \binits{W.}}:
\batitle{Local structure of {{In0}}.{{5Ga0}}.{{5As}} from joint high-resolution and differential pair distribution function analysis}.
\bjtitle{Journal of Applied Physics}
\bvolume{88}(\bissue{2}),
\bfpage{665}--\blpage{672}
(\byear{2000})
\doiurl{10.1063/1.373718}
\end{barticle}
\endbibitem

%%% 59
\bibitem[\protect\citeauthoryear{Petkov et~al.}{2018}]{petkovApplicationDifferentialResonant2018a}
\begin{barticle}
\bauthor{\bsnm{Petkov}, \binits{V.}},
\bauthor{\bsnm{Shastri}, \binits{S.}},
\bauthor{\bsnm{Kim}, \binits{J.-W.}},
\bauthor{\bsnm{Shan}, \binits{S.}},
\bauthor{\bsnm{Luo}, \binits{J.}},
\bauthor{\bsnm{Wu}, \binits{J.}},
\bauthor{\bsnm{Zhong}, \binits{C.-J.}}:
\batitle{Application of differential resonant high-energy {{X-ray}} diffraction to three-dimensional structure studies of nanosized materials: {{A}} case study of {{Pt}}\textendash{{Pd}} nanoalloy catalysts}.
\bjtitle{Acta Crystallographica Section A: Foundations and Advances}
\bvolume{74}(\bissue{5}),
\bfpage{553}--\blpage{566}
(\byear{2018})
\doiurl{10.1107/S2053273318009282}
\end{barticle}
\endbibitem

%%% 60
\bibitem[\protect\citeauthoryear{Glatzel et~al.}{2009}]{glatzel2009electronic}
\begin{bchapter}
\bauthor{\bsnm{Glatzel}, \binits{P.}},
\bauthor{\bsnm{Smolentsev}, \binits{G.}},
\bauthor{\bsnm{Bunker}, \binits{G.}}:
\bctitle{The electronic structure in 3d transition metal complexes: Can we measure oxidation states?}
In: \bbtitle{Journal of Physics: Conference Series},
vol. \bseriesno{190},
p. \bfpage{012046}
(\byear{2009}).
\bcomment{IOP Publishing}
\end{bchapter}
\endbibitem

%%% 61
\bibitem[\protect\citeauthoryear{Natoli}{1984}]{natoli1984distance}
\begin{bchapter}
\bauthor{\bsnm{Natoli}, \binits{C.}}:
\bctitle{Distance dependence of continuum and bound state of excitonic resonances in x-ray absorption near edge structure (xanes)}.
In: \bbtitle{EXAFS and Near Edge Structure III: Proceedings of an International Conference, Stanford, CA, July 16--20, 1984},
pp. \bfpage{38}--\blpage{42}
(\byear{1984}).
\bcomment{Springer}
\end{bchapter}
\endbibitem

%%% 62
\bibitem[\protect\citeauthoryear{Rehr and Albers}{2000}]{rehr2000theoretical}
\begin{barticle}
\bauthor{\bsnm{Rehr}, \binits{J.J.}},
\bauthor{\bsnm{Albers}, \binits{R.C.}}:
\batitle{Theoretical approaches to x-ray absorption fine structure}.
\bjtitle{Reviews of modern physics}
\bvolume{72}(\bissue{3}),
\bfpage{621}
(\byear{2000})
\end{barticle}
\endbibitem

%%% 63
\bibitem[\protect\citeauthoryear{Timoshenko et~al.}{2020}]{timoshenko2020silver}
\begin{barticle}
\bauthor{\bsnm{Timoshenko}, \binits{J.}},
\bauthor{\bsnm{Roese}, \binits{S.}},
\bauthor{\bsnm{H{\"o}vel}, \binits{H.}},
\bauthor{\bsnm{Frenkel}, \binits{A.I.}}:
\batitle{Silver clusters shape determination from in-situ {{XANES}} data}.
\bjtitle{Radiation Physics and Chemistry}
\bvolume{175},
\bfpage{108049}
(\byear{2020})
\doiurl{10.1016/j.radphyschem.2018.11.003}
\end{barticle}
\endbibitem

%%% 64
\bibitem[\protect\citeauthoryear{Soper et~al.}{1977}]{soper1977neutron}
\begin{barticle}
\bauthor{\bsnm{Soper}, \binits{A.K.}},
\bauthor{\bsnm{Neilson}, \binits{G.W.}},
\bauthor{\bsnm{Enderby}, \binits{J.E.}},
\bauthor{\bsnm{Howe}, \binits{R.A.}}:
\batitle{A neutron diffraction study of hydration effects in aqueous solutions}.
\bjtitle{Journal of Physics C: Solid State Physics}
\bvolume{10}(\bissue{11}),
\bfpage{1793}--\blpage{1801}
(\byear{1977})
\doiurl{10.1088/0022-3719/10/11/014}
\end{barticle}
\endbibitem

%%% 65
\bibitem[\protect\citeauthoryear{Enderby}{1995}]{enderby1995ion}
\begin{barticle}
\bauthor{\bsnm{Enderby}, \binits{J.E.}}:
\batitle{Ion solvation via neutron scattering}.
\bjtitle{Chemical Society Reviews}
\bvolume{24}(\bissue{3}),
\bfpage{159}
(\byear{1995})
\doiurl{10.1039/cs9952400159}
\end{barticle}
\endbibitem

%%% 66
\bibitem[\protect\citeauthoryear{Zhao et~al.}{1998}]{zhao1998neutron}
\begin{barticle}
\bauthor{\bsnm{Zhao}, \binits{J.}},
\bauthor{\bsnm{Gaskell}, \binits{P.H.}},
\bauthor{\bsnm{Cluckie}, \binits{M.M.}},
\bauthor{\bsnm{Soper}, \binits{A.K.}}:
\batitle{A neutron diffraction, isotopic substitution study of the structure of {{Li2O}}{$\cdot$}{{2SiO2}} glass}.
\bjtitle{Journal of Non-Crystalline Solids}
\bvolume{232--234},
\bfpage{721}--\blpage{727}
(\byear{1998})
\doiurl{10.1016/S0022-3093(98)00554-7}
\end{barticle}
\endbibitem

%%% 67
\bibitem[\protect\citeauthoryear{Petkov et~al.}{2000}]{petkovApplicationAtomicPair2000}
\begin{barticle}
\bauthor{\bsnm{Petkov}, \binits{V.}},
\bauthor{\bsnm{Billinge}, \binits{S.J.L.}},
\bauthor{\bsnm{Heising}, \binits{J.}},
\bauthor{\bsnm{Kanatzidis}, \binits{M.G.}}:
\batitle{Application of {{Atomic Pair Distribution Function Analysis}} to {{Materials}} with {{Intrinsic Disorder}}. {{Three-Dimensional Structure}} of {{Exfoliated-Restacked WS2}}:\, {{Not Just}} a {{Random Turbostratic Assembly}} of {{Layers}}}.
\bjtitle{Journal of the American Chemical Society}
\bvolume{122}(\bissue{47}),
\bfpage{11571}--\blpage{11576}
(\byear{2000})
\doiurl{10.1021/ja002048i}
\end{barticle}
\endbibitem

%%% 68
\bibitem[\protect\citeauthoryear{Farrow et~al.}{2011}]{farrow2011nyquist}
\begin{barticle}
\bauthor{\bsnm{Farrow}, \binits{C.L.}},
\bauthor{\bsnm{Shaw}, \binits{M.}},
\bauthor{\bsnm{Kim}, \binits{H.}},
\bauthor{\bsnm{Juh\'as}, \binits{P.}},
\bauthor{\bsnm{Billinge}, \binits{S.J.L.}}:
\batitle{Nyquist-shannon sampling theorem applied to refinements of the atomic pair distribution function}.
\bjtitle{Phys. Rev. B}
\bvolume{84},
\bfpage{134105}
(\byear{2011})
\doiurl{10.1103/PhysRevB.84.134105}
\end{barticle}
\endbibitem

%%% 69
\bibitem[\protect\citeauthoryear{Pan et~al.}{2021}]{pan2021benchmark}
\begin{barticle}
\bauthor{\bsnm{Pan}, \binits{H.}},
\bauthor{\bsnm{Ganose}, \binits{A.M.}},
\bauthor{\bsnm{Horton}, \binits{M.}},
\bauthor{\bsnm{Aykol}, \binits{M.}},
\bauthor{\bsnm{Persson}, \binits{K.A.}},
\bauthor{\bsnm{Zimmermann}, \binits{N.E.R.}},
\bauthor{\bsnm{Jain}, \binits{A.}}:
\batitle{Benchmarking coordination number prediction algorithms on inorganic crystal structures}.
\bjtitle{Inorganic Chemistry}
\bvolume{60},
\bfpage{1590}--\blpage{1603}
(\byear{2021})
\doiurl{10.1021/acs.inorgchem.0c02996}
\end{barticle}
\endbibitem

%%% 70
\bibitem[\protect\citeauthoryear{Pedregosa et~al.}{2011}]{pedregosa2011scikit}
\begin{barticle}
\bauthor{\bsnm{Pedregosa}, \binits{F.}},
\bauthor{\bsnm{Varoquaux}, \binits{G.}},
\bauthor{\bsnm{Gramfort}, \binits{A.}},
\bauthor{\bsnm{Michel}, \binits{V.}},
\bauthor{\bsnm{Thirion}, \binits{B.}},
\bauthor{\bsnm{Grisel}, \binits{O.}},
\bauthor{\bsnm{Blondel}, \binits{M.}},
\bauthor{\bsnm{Prettenhofer}, \binits{P.}},
\bauthor{\bsnm{Weiss}, \binits{R.}},
\bauthor{\bsnm{Dubourg}, \binits{V.}}, \betal:
\batitle{Scikit-learn: Machine learning in python}.
\bjtitle{the Journal of machine Learning research}
\bvolume{12},
\bfpage{2825}--\blpage{2830}
(\byear{2011})
\end{barticle}
\endbibitem

%%% 71
\bibitem[\protect\citeauthoryear{Štrumbelj and Kononenko}{2013}]{strumbelj2014feature}
\begin{barticle}
\bauthor{\bsnm{Štrumbelj}, \binits{E.}},
\bauthor{\bsnm{Kononenko}, \binits{I.}}:
\batitle{Explaining prediction models and individual predictions with feature contributions.}
\bjtitle{Knowledge and Information Systems}
\bvolume{41},
\bfpage{647}--\blpage{665}
(\byear{2013})
\end{barticle}
\endbibitem

%%% 72
\bibitem[\protect\citeauthoryear{Farges et~al.}{1997}]{farges1997ti}
\begin{barticle}
\bauthor{\bsnm{Farges}, \binits{F.}},
\bauthor{\bsnm{Brown}, \binits{G.E.}},
\bauthor{\bsnm{Rehr}, \binits{J.}}, \betal:
\batitle{Ti k-edge xanes studies of ti coordination and disorder in oxide compounds: Comparison between theory and experiment}.
\bjtitle{Physical Review B}
\bvolume{56}(\bissue{4}),
\bfpage{1809}
(\byear{1997})
\end{barticle}
\endbibitem

\end{thebibliography}
%\printbibliography %Prints bibliography

%%%%%%%%%%%%%%%%%%%%%%% merge with SI %%%%%%%%%%%%%%%%%%%%%%%%%%%%%
\pagebreak
% \widetext
\begin{center}
\textbf{\large Supplemental Information for "Interpretable Multimodal Machine Learning Analysis of X-ray Absorption Near-Edge Spectra and Pair Distribution Functions"}
\end{center}
%%%%%%%%%% Merge with supplemental materials %%%%%%%%%%
%%%%%%%%%% Prefix a "S" to all equations, figures, tables and reset the counter %%%%%%%%%%
\setcounter{equation}{0}
\setcounter{figure}{0}
\setcounter{table}{0}
\setcounter{page}{1}
\setcounter{section}{0}
\makeatletter
\renewcommand{\thefigure}{S\arabic{figure}}
\renewcommand{\thetable}{S\arabic{table}}
%%%%%%%%%% Prefix a "S" to all equations, figures, tables and reset the counter %%%%%%%%%%
\section{Datasets}
\begin{figure}[h]
    \centering
    \includegraphics[width=0.8\textwidth]{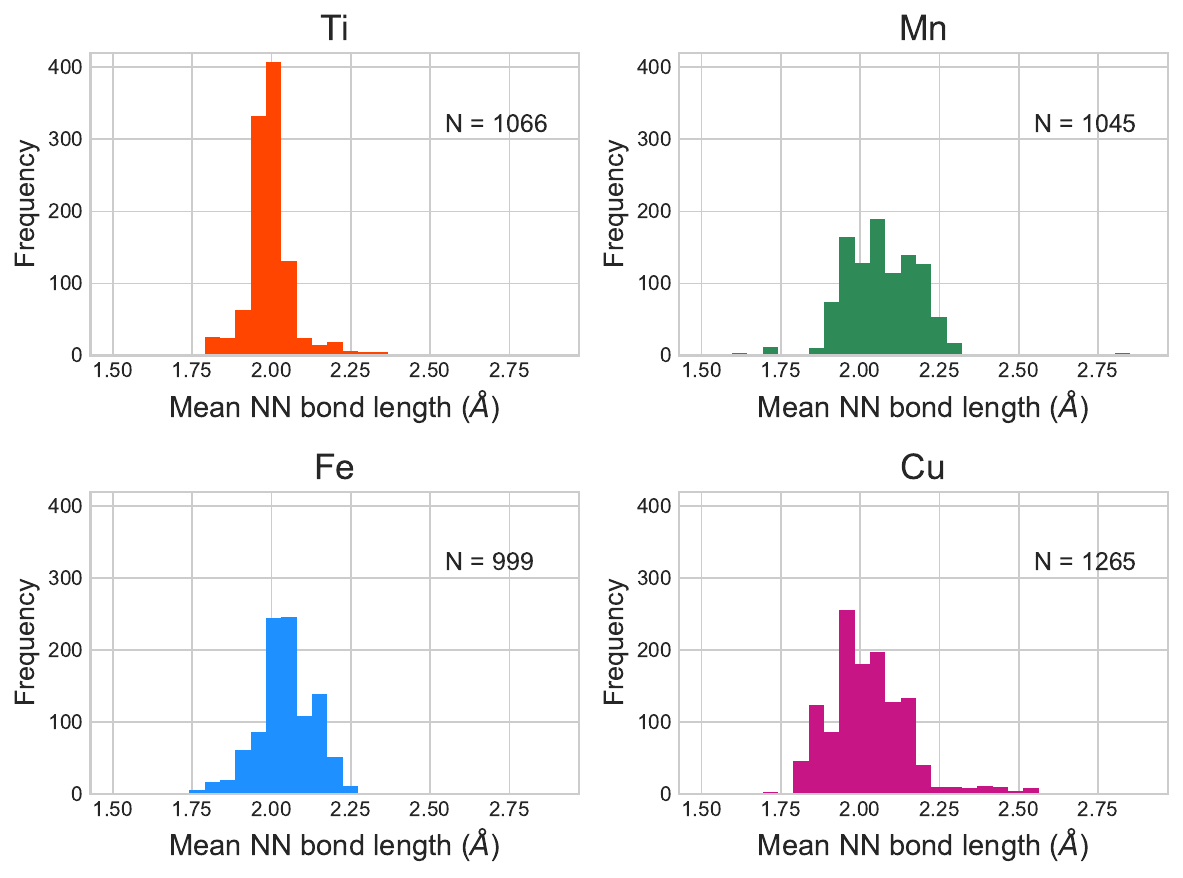}
    \caption{Histograms of mean nearest-neighbor bond lengths for all four datasets (Ti, Mn, Fe, and Cu).}
    \label{fig:SI_BL_dist}
\end{figure}

Tables~\ref{tab:SI_cs_dataset}, ~\ref{tab:SI_cn_dataset}, and \ref{tab:SI_bl_dataset} show dataset sizes and class makeup of each dataset (Ti, Mn, Fe, and Cu) for the three prediction tasks: oxidation state, coordination number, and mean nearest-neighbor bond length, respectively. 
Figure~\ref{fig:SI_BL_dist} shows the distributions of the mean nearest-neighbor bond lengths from all four datasets.

%%% Table: oxidation state dataset %%%
\begin{table}[h]
\caption{Oxidation state: dataset sizes and class makeup.}
\label{tab:SI_cs_dataset}
\begin{tabular}{@{}lrrrr@{}}
\toprule
Element & Total size & CS=2 (1 for Cu) & CS=3 (2 for Cu) & CS=4 (3 for Cu)\\
\midrule
Ti & 892 & 39 & 112 & 741 \\
Mn & 633 & 272 & 225 & 136 \\
Fe & 655 & 258 & 359 & 38 \\
Cu & 1084 & 165 & 796 & 123 \\
\bottomrule
\end{tabular}
\end{table}

%%% Table: coordination number dataset %%%
\begin{table}[h]
\caption{Coordination number: dataset sizes and class makeup.}
\label{tab:SI_cn_dataset}
\begin{tabular}{@{}lrrrr@{}}
\toprule
Element & Total size & CN=4 & CN=5 & CN=6\\
\midrule
Ti & 960 & 64 & 67 & 829 \\
Mn & 930 & 66 & 199 & 665 \\
Fe & 883 & 99 & 171 & 613 \\
Cu & 952 & 542 & 265 & 145 \\
\bottomrule
\end{tabular}
\end{table}
%
%%% Table: bond length dataset %%%
\begin{table}[h]
\caption{Mean nearest-neighbor bond length: dataset sizes.}
\label{tab:SI_bl_dataset}
\begin{tabular*}{0.4\textwidth}{@{\extracolsep{\fill}}ll}
\toprule
Element & Total size \\
\midrule
Ti & 1066 \\
Mn & 1045 \\
Fe & 999 \\
Cu & 1265 \\
\bottomrule
\end{tabular*}
\end{table}

Next, Table~\ref{tab:SI_anions} shows the number of structures in each dataset that contains certain anions.
Our datasets are made up of transition metal oxides, so all structures contain O, but other anions can be present.
%%% Table: counts of structures in the datasets that contain certain anions 
\begin{table}[h]
\caption{Number of structures in each dataset that contain specific anions}
\label{tab:SI_anions}
\begin{tabular*}{0.8\textwidth}{@{\extracolsep{\fill}}lrrrrrrrrrrrr}
\toprule
Metal & H & C & N & F & P & S & Cl & Se & Br & I & \textbf{O} & \textbf{Total} \\
\midrule
Ti & 15 & 9 & 45 & 36 & 231 & 30 & 16 & 5 & 4 & 2 & 1066 & \textbf{1066} \\
\bottomrule
Mn & 28 & 52 & 3 & 119 & 275 & 16 & 5 & 6 & 4 & 2 & 1045 & \textbf{1045} \\
\bottomrule
Fe & 29 & 55 & 8 & 126 & 211 & 29 & 10 & 7 & 0 & 0 & 999 & \textbf{999} \\
\bottomrule
Cu & 82 & 55 & 36 & 29 & 208 & 49 & 61 & 49 & 20 & 9 & 1265 & \textbf{1265} \\
\bottomrule
\end{tabular*}
\end{table}

\section{Energy domains for XANES spectra}
To prepare inputs for training the random forest models, we interpolated all XANES spectra in each metal dataset onto a 100-pt grid such that all spectra y-values (intensity values) correspond to the same x-domains.
These domains depend on the transition metal element (absorbing specties) and were chosen based on where most spectra overlap, mostly consistent with what was used in Torrisi \textit{et al.}~\cite{torrisi_random_2020}.

%
%%% Table: energy domains for XANES %%%
\begin{table}[h]
\caption{Energy domains of interpolated XANES spectra}
\label{tab:SI_XANES_domains}
\begin{tabular*}{0.5\textwidth}{@{\extracolsep{\fill}}ll}
\toprule
    Absorbing element & Energy (eV) \\
    \midrule
    Ti & 4969.000 - 5021.000 \\
    Mn & 6542.896 - 6594.431 \\
    Fe & 7115.617 - 7167.759 \\
    Cu & 8988.142 - 9039.728 \\
\bottomrule
\end{tabular*}
\end{table}

\section{Impurity-based vs. permutation feature importance}
\begin{figure}[b]
    \centering
    \includegraphics[width=\textwidth]{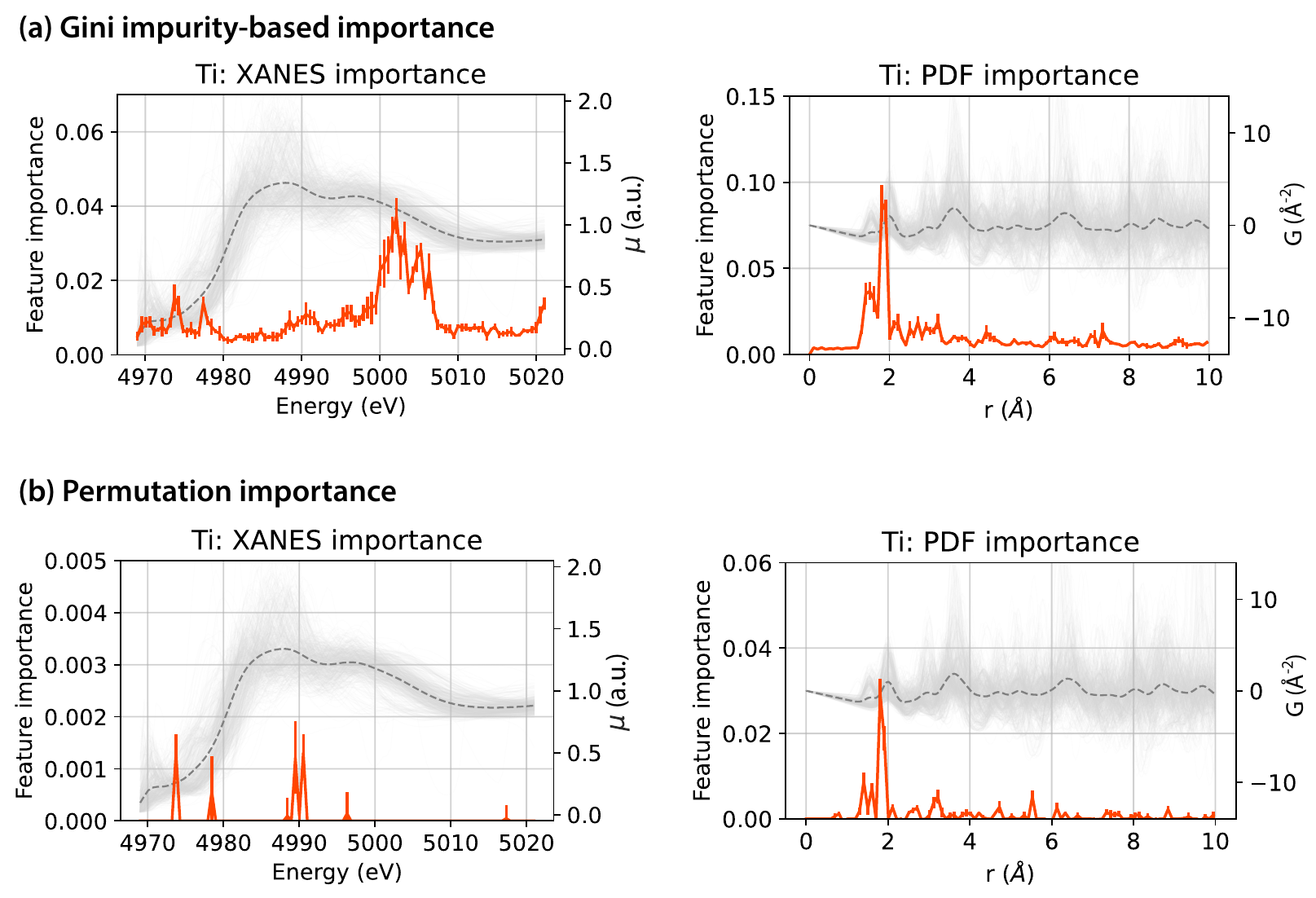}
    \caption{Ti XANES and PDF feature importance computed based on (a) decrease in Gini impurity (chosen method) and (b) permutation method.}\label{fig:SI_Ti_permute}
\end{figure}

\fig{SI_Ti_permute} shows XANES and PDF feature importance for Ti computed by two different methods. 
The top row (a) shows the results from the Gini impurity-based calculation. 
The bottom row (b) shows the feature importance computed by the permutation method. 
Both methods can be easily implemented with scikit-learn~\cite{pedregosa2011scikit}. 

The results from the two methods look very different for XANES. 
While impurity-based importance shows that the region between 5000-5010 eV (post-edge) is important for oxidation state prediction of Ti, the permutation-based importance in this region is completely zero. 
This highlights a limitation of the permutation method, which tends to suppress the importance of strongly correlated features. 
The PDF permutation importance is less affected by this limitation because the PDF features are less strongly correlated, hence both methods produced similar plots.

Due to multicollinearity in our XANES and PDF inputs, we chose to use the Gini impurity-based importance over the permutation method.
We also computed SHAP (SHapley Additive exPlanation) values~\cite{strumbelj2014feature} of the spectra points and found generally good agreement with the impurity-based feature importance. 

\section{Model Performances and feature importance plots}
\subsection{Oxidation state}
Table~\ref{tab:SI_cs} shows test weighted mean F1 scores for oxidation state classification models trained on XANES, PDF, differential PDF (dPDF) and the combined models XANES+PDF and XANES+dPDF for all four metals studied: Ti, Mn, Fe, Cu.
The baseline scores were computed based on a trivial classifier that labels all samples as the modal class.
%The best test score achieved for each element is in bold. For Cu, the XANES-only model and the combined XANES+PDF model performed almost equally well, so both scores are highlighted.
\\
%%% Table 1: oxidation state F1 scores %%%
\begin{table}[h]
\caption{Oxidation state results: Test weighted mean F1 scores for models trained on XANES, PDF, dPDF, XANES+PDF, and XANES+dPDF with standard deviations given in parantheses.}
\label{tab:SI_cs}
\begin{tabular*}{\textwidth}{@{\extracolsep{\fill}}lrrrrrr}
\toprule
Element & Baseline & XANES & PDF & dPDF & XANES+PDF & XANES+dPDF \\
\midrule
Ti & 0.73 & 0.962(5) & 0.86(2) & 0.934(7) & 0.945(5) & 0.967(6) \\
Mn & 0.28 & 0.86(1) & 0.82(1) & 0.879(4) & 0.86(1) & 0.903(4) \\
Fe & 0.40 & 0.83(1) & 0.79(1) & 0.81(1) & 0.87(2) & 0.859(4) \\
Cu & 0.63 & 0.865(5) & 0.764(7) & 0.839(8) & 0.86(1) & 0.875(9) \\
\bottomrule
\end{tabular*}
\end{table}

\begin{figure}[ht]
    \centering
    \includegraphics[width=0.9\textwidth]{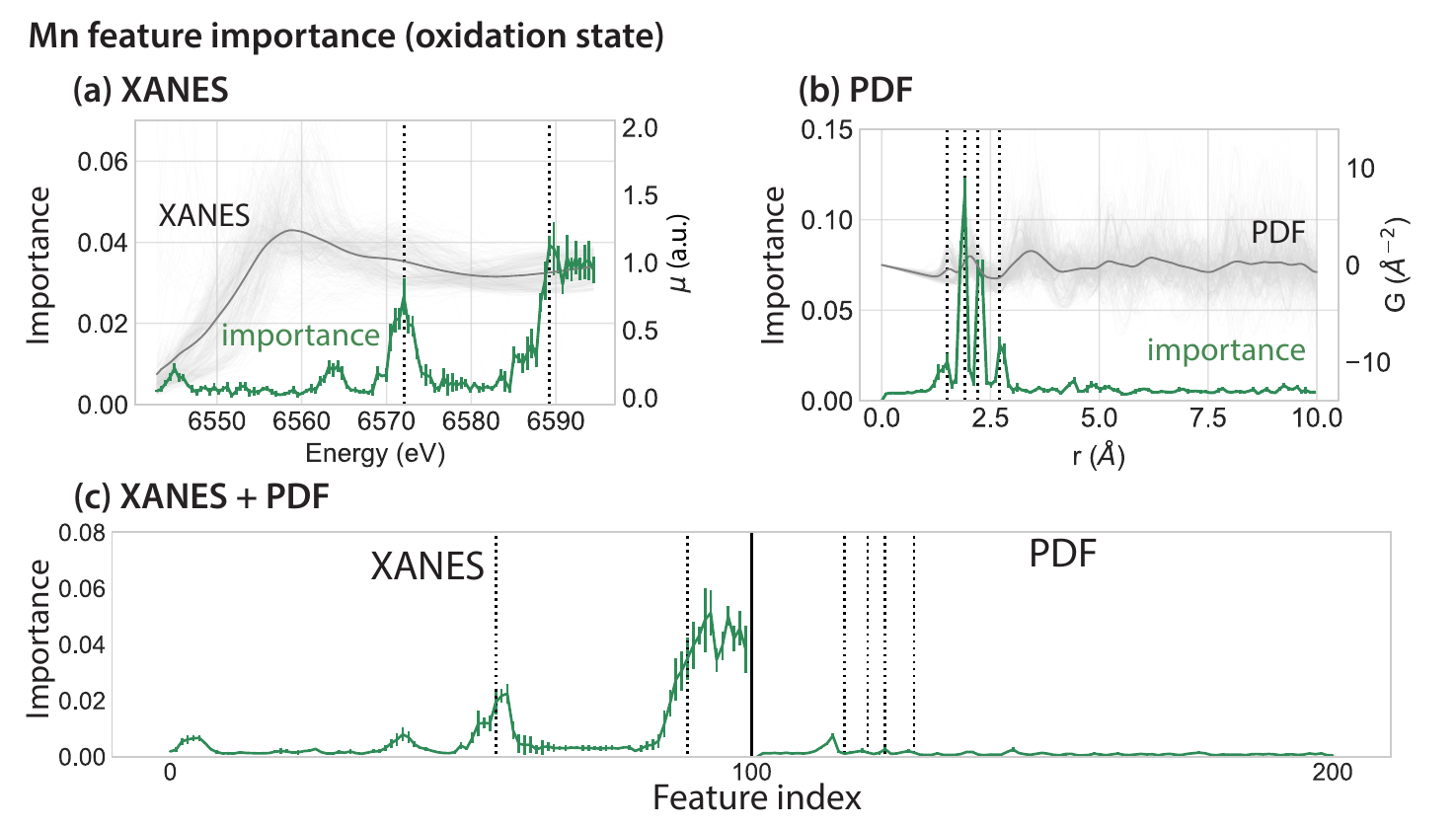}
    \caption{Mn feature importance when using (a) XANES, (b) PDF, (c) XANES + PDF as inputs. 
    (a) and (b) also show the spectra from the training sets and its mean given by the dark grey line. 
    Dotted lines mark the locations of prominent features of XANES and PDF separately. 
   The same lines are drawn on the combined XANES+PDF importance plot to highlight the changes.}
   \label{fig:SI_cs_Mn}
\end{figure}

\begin{figure}[h]
    \centering
    \includegraphics[width=0.9\textwidth]{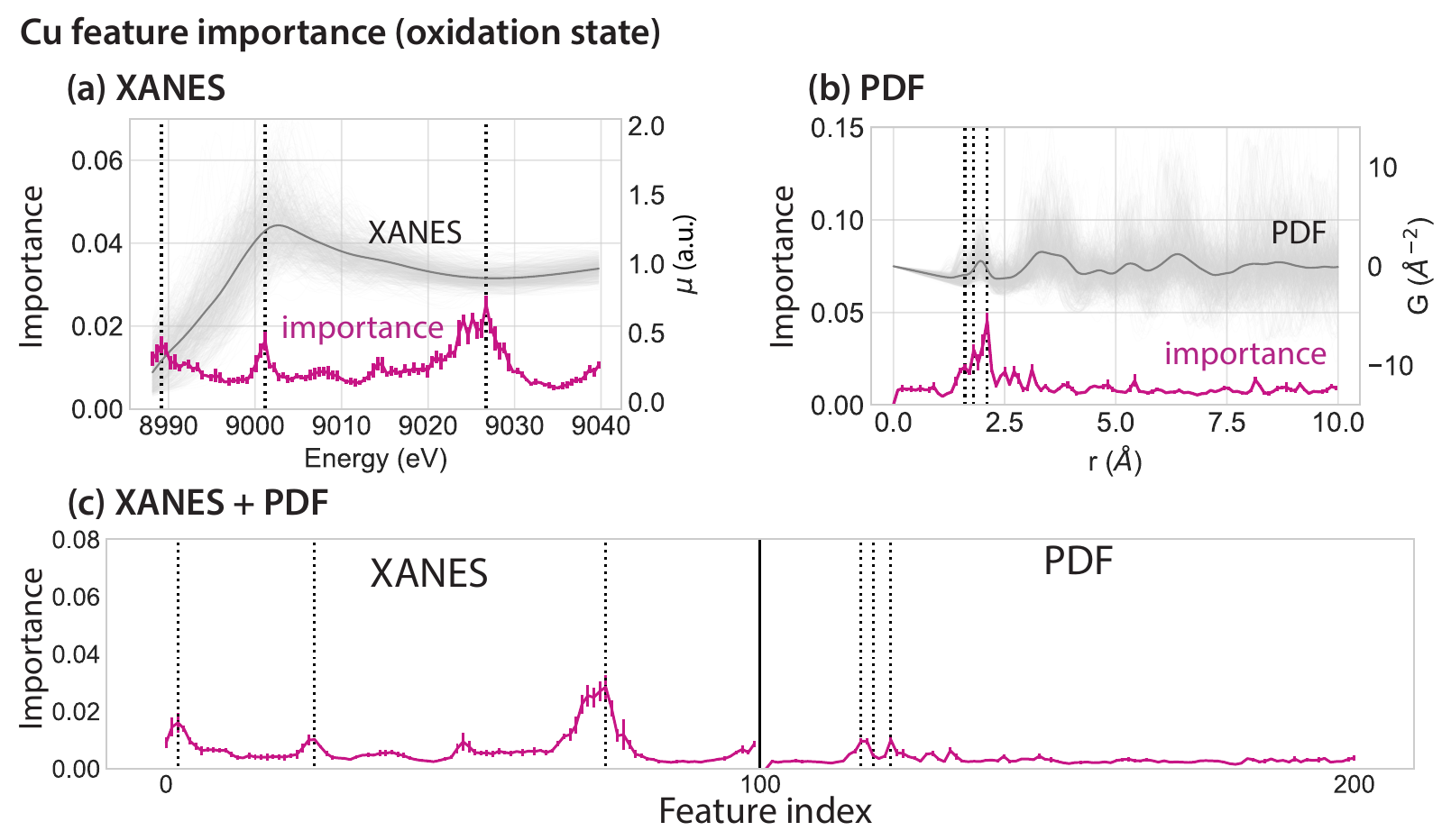}
    \caption{Cu feature importance when using (a) XANES, (b) PDF, (c) XANES + PDF as inputs. 
    (a) and (b) also show the spectra from the training sets and its mean given by the dark grey line. 
    Dotted lines mark the locations of prominent features of XANES and PDF separately. 
   The same lines are drawn on the combined XANES+PDF importance plot to highlight the changes.}\label{fig:SI_cs_Cu}
\end{figure}

Next, in addition to the Ti and Fe feature importance plots provided in the main text, Figs.~\ref{fig:SI_cs_Mn} and ~\ref{fig:SI_cs_Cu} show the feature importance plots for Mn and Cu, respectively.
For each metal, the subplots show feature importance for (a) XANES, (b) PDF, and (c) XANES + PDF combined. 

\subsection{Coordination number}
Table~\ref{tab:SI_cn} shows test weighted mean F1 scores for coordination number classification models trained on XANES, PDF, differential PDF (dPDF) and the combined models XANES+PDF and XANES+dPDF for all four metals studied: Ti, Mn, Fe, Cu. 
The baseline scores were computed based on a trivial classifier that labels all samples as the modal class.
%%% Table: CN scores %%%
\begin{table}[h]
\caption{Coordination number results: Test weighted mean F1 scores for models trained on XANES, PDF, dPDF, XANES+PDF, and XANES+dPDF with standard deviations given in parantheses.}
\label{tab:SI_cn}
\begin{tabular*}{\textwidth}{@{\extracolsep{\fill}}lrrrrrr}
\toprule
Element & Baseline & XANES & PDF & dPDF & XANES+PDF & XANES+dPDF \\
\midrule
Ti & 0.79 & 0.95(1) & 0.87(1) & 0.917(5) & 0.955(2) & 0.953(8) \\
Mn & 0.63 & 0.902(7) & 0.868(3) & 0.891(4) & 0.908(4) & 0.907(8) \\
Fe & 0.56 & 0.934(5) & 0.93(1) & 0.938(7) & 0.970(4) & 0.960(6) \\
Cu & 0.43 & 0.881(9) & 0.80(2) & 0.855(9) & 0.880(8) & 0.905(7) \\
\bottomrule
\end{tabular*}
\end{table}

\begin{figure}[ht]
    \centering
    \includegraphics[width=0.9\textwidth]{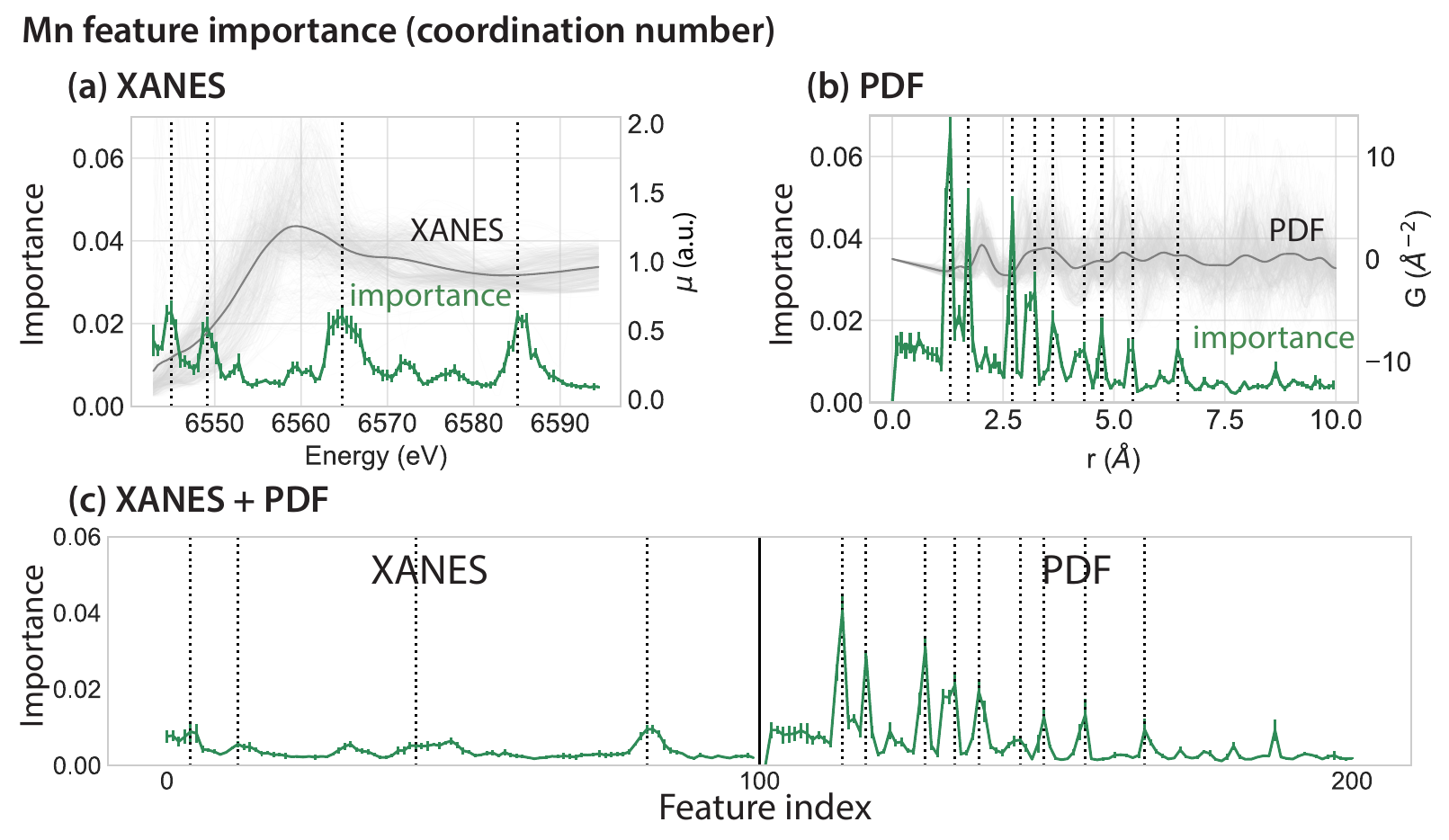}
    \caption{Mn feature importance when using (a) XANES, (b) PDF, (c) XANES + PDF as inputs. 
    (a) and (b) also show the spectra from the training sets and its mean given by the dark grey line. 
    Dotted lines mark the locations of prominent features of XANES and PDF separately. 
   The same lines are drawn on the combined XANES+PDF importance plot to highlight the changes.}
   \label{fig:SI_cn_Mn}
\end{figure}

\begin{figure}[h]
    \centering
    \includegraphics[width=0.9\textwidth]{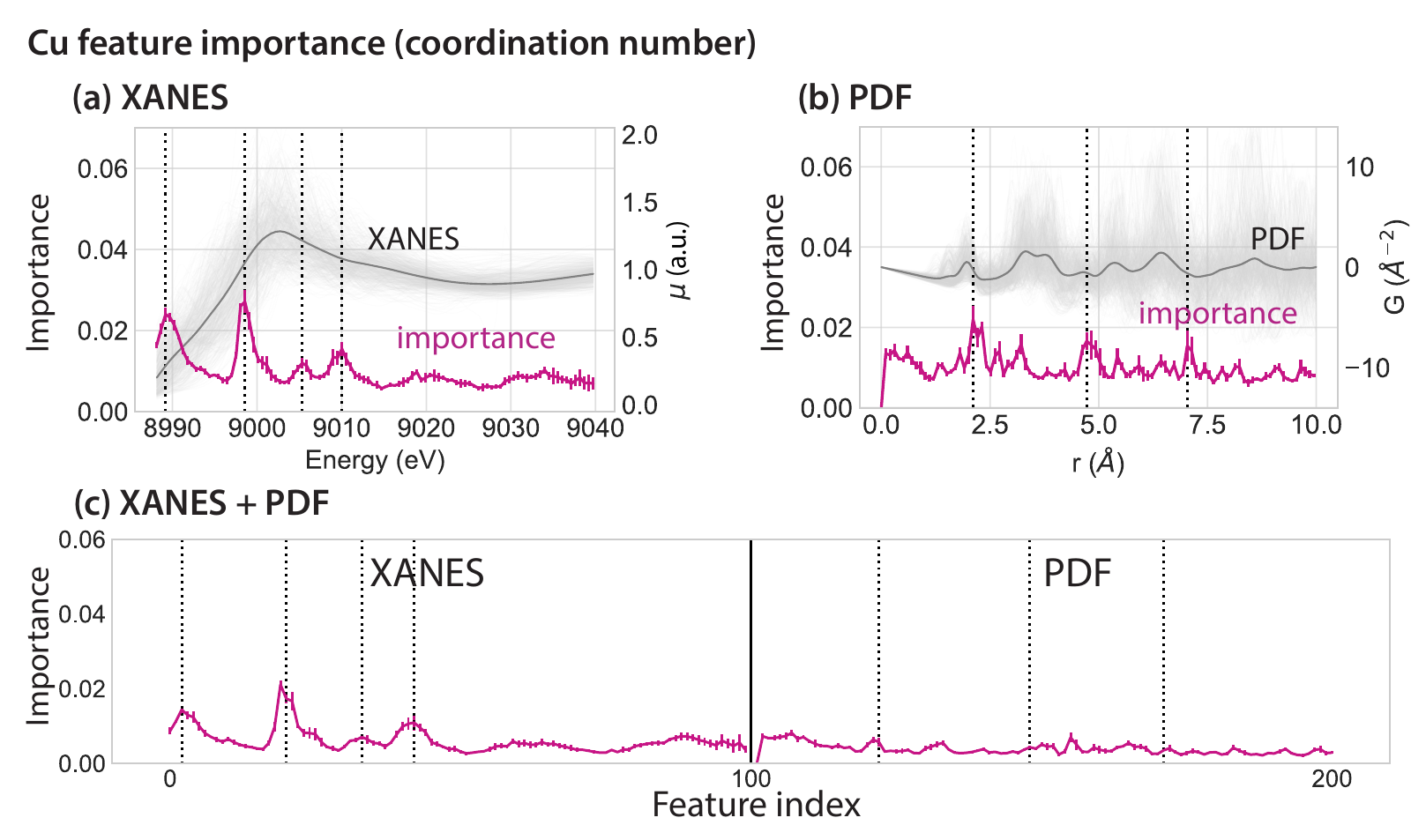}
    \caption{Cu feature importance when using (a) XANES, (b) PDF, (c) XANES + PDF as inputs. 
    (a) and (b) also show the spectra from the training sets and its mean given by the dark grey line. 
    Dotted lines mark the locations of prominent features of XANES and PDF separately. 
   The same lines are drawn on the combined XANES+PDF importance plot to highlight the changes.}\label{fig:SI_cn_Cu}
\end{figure}

Next, in addition to the Ti and Fe feature importance plots provided in the main text, Figs.~\ref{fig:SI_cn_Mn} and ~\ref{fig:SI_cn_Cu} show the feature importance plots for Mn and Cu, respectively.
For each metal, the subplots show feature importance for (a) XANES, (b) PDF, and (c) XANES + PDF combined. 

\clearpage
\subsection{Mean nearest-neighbor bond length}
Table~\ref{tab:SI_bl} shows test RMSEs for mean nearest-neighbor bond length regression models trained on XANES, PDF, differential PDF (dPDF) and the combined models XANES+PDF and XANES+dPDF for all four metals studied: Ti, Mn, Fe, Cu.
The baseline test errors were computed for a trivial model that naively uses the location of the tallest PDF peak in the range 0-4 \AA\ as the mean bond length prediction. 
The naive RMSE scores are roughly 5-8 times the XANES RMSE scores, showing that our random forest models performed much better.
%%% Table 3: bond length RMSE scores %%%
\begin{table}[h]
\centering
\caption{Bond length: Test RMSEs (in \AA) for models trained on XANES, PDF, dPDF, XANES+PDF, and XANES+dPDF with standard deviations given in parantheses (lower numbers indicate better performances).}\label{tab:SI_bl}
\begin{tabular*}{\textwidth}{@{\extracolsep{\fill}}lrrrrrr}
\toprule
Element & Baseline & XANES & PDF & dPDF & XANES+PDF & XANES+dPDF \\
\midrule
Ti & 0.28 & 0.0615(5) & 0.0779(5) & 0.0661(5) & 0.0648(6) & 0.0628(5) \\
Mn & 0.43 & 0.0598(3) & 0.0785(6) & 0.0622(4) & 0.0594(2) & 0.0556(5) \\
Fe & 0.40 & 0.0527(6) & 0.0625(7) & 0.0428(8) & 0.0526(6) & 0.0451(5) \\
Cu & 0.40 & 0.0792(4) & 0.0801(7) & 0.0478(5) & 0.0726(5) & 0.0617(6) \\
\bottomrule
\end{tabular*}
\end{table}

\begin{figure}[b]
    \centering
    \includegraphics[width=\textwidth]{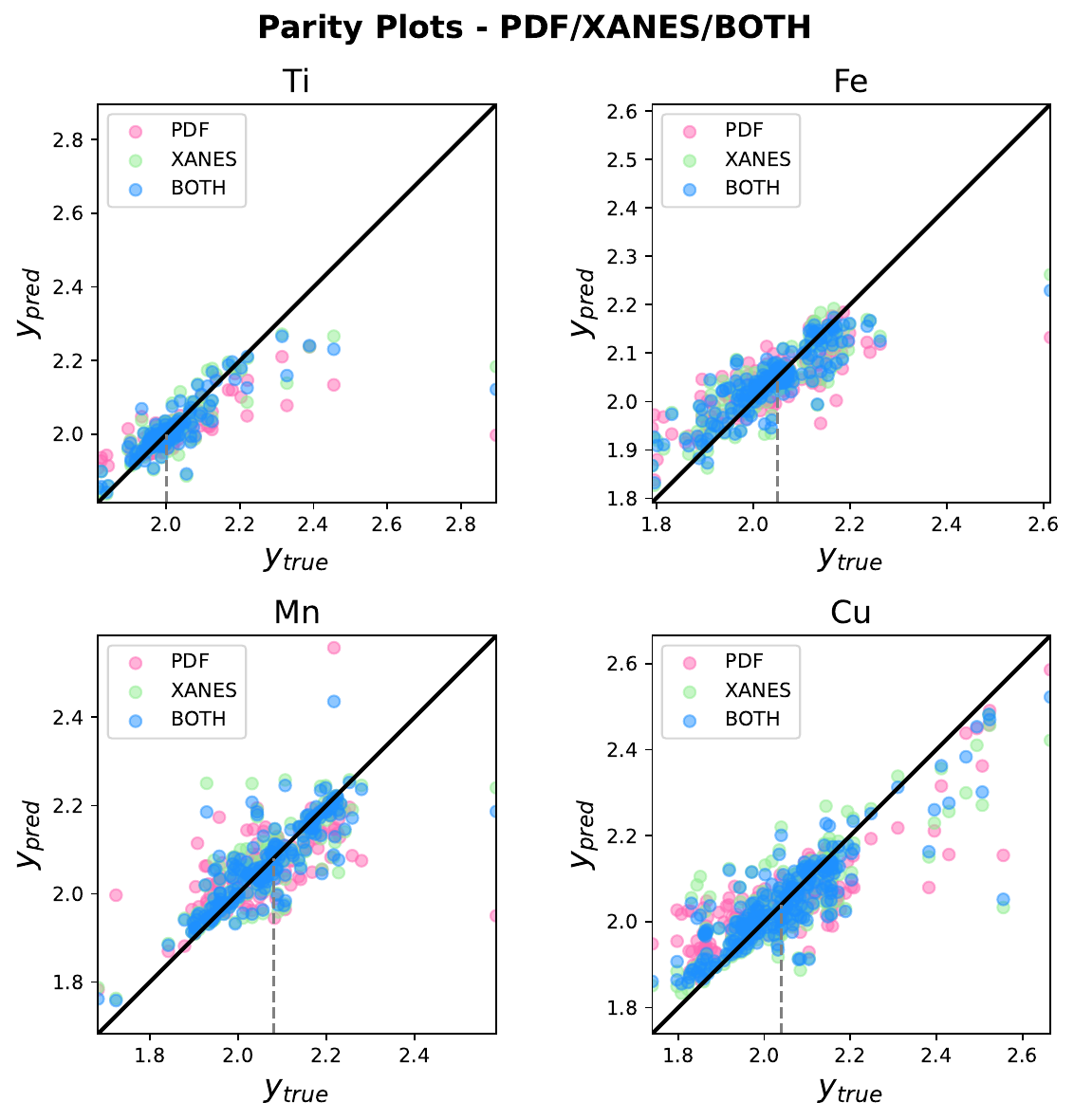}
    \caption{Parity plots (from test datasets) for models using XANES, PDF, and XANES + PDF as inputs.
    $y_{true}$ refers to the true average bond length for a structure and $y_{pred}$ are the predictions made by the models for that sample. 
    Dashed lines mark the location of the mean value of the total average bond length distributions for each metal.}\label{fig:SI_parity}
\end{figure}

For the bond length regression task, we also made parity plots from the test datasets, which provide diagnostics of how well the model predictions agree with the true target values.
\fig{SI_parity} shows that the predictions are most accurate when the true value is close to the mean of all bond lengths (dashed lines).
Different colors here represent three models trained on different types of input: total-PDF (pink), XANES (green), and both XANES+PDF (blue). 

\begin{figure}[ht]
    \centering
    \includegraphics[width=0.9\textwidth]{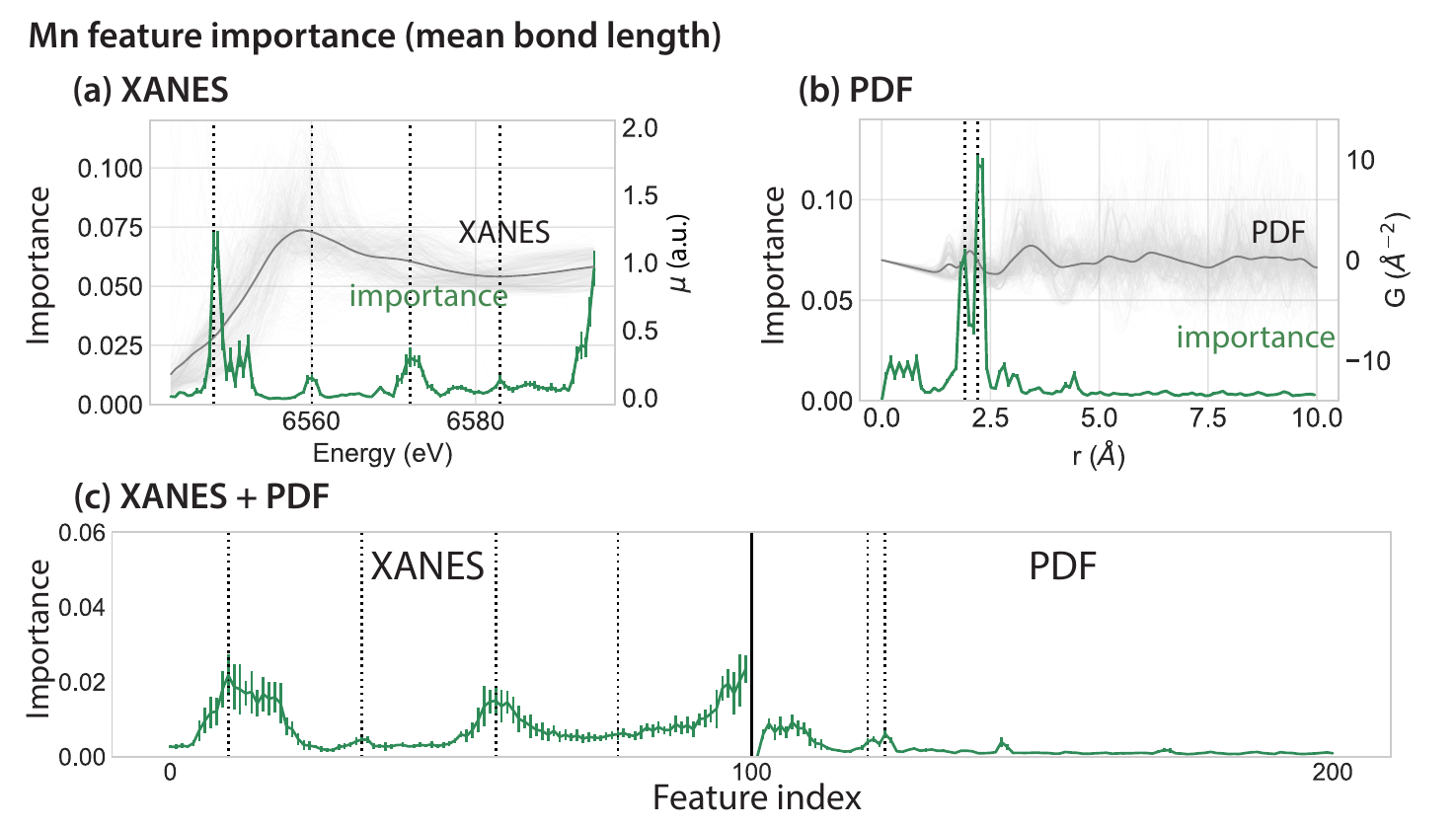}
    \caption{Mn feature importance when using (a) XANES, (b) PDF, (c) XANES + PDF as inputs. 
    (a) and (b) also show the spectra from the training sets and its mean given by the dark grey line. 
    Dotted lines mark the locations of prominent features of XANES and PDF separately. 
   The same lines are drawn on the combined XANES+PDF importance plot to highlight the changes.}\label{fig:SI_bl_Mn}
\end{figure}

\begin{figure}[h]
    \centering
    \includegraphics[width=0.9\textwidth]{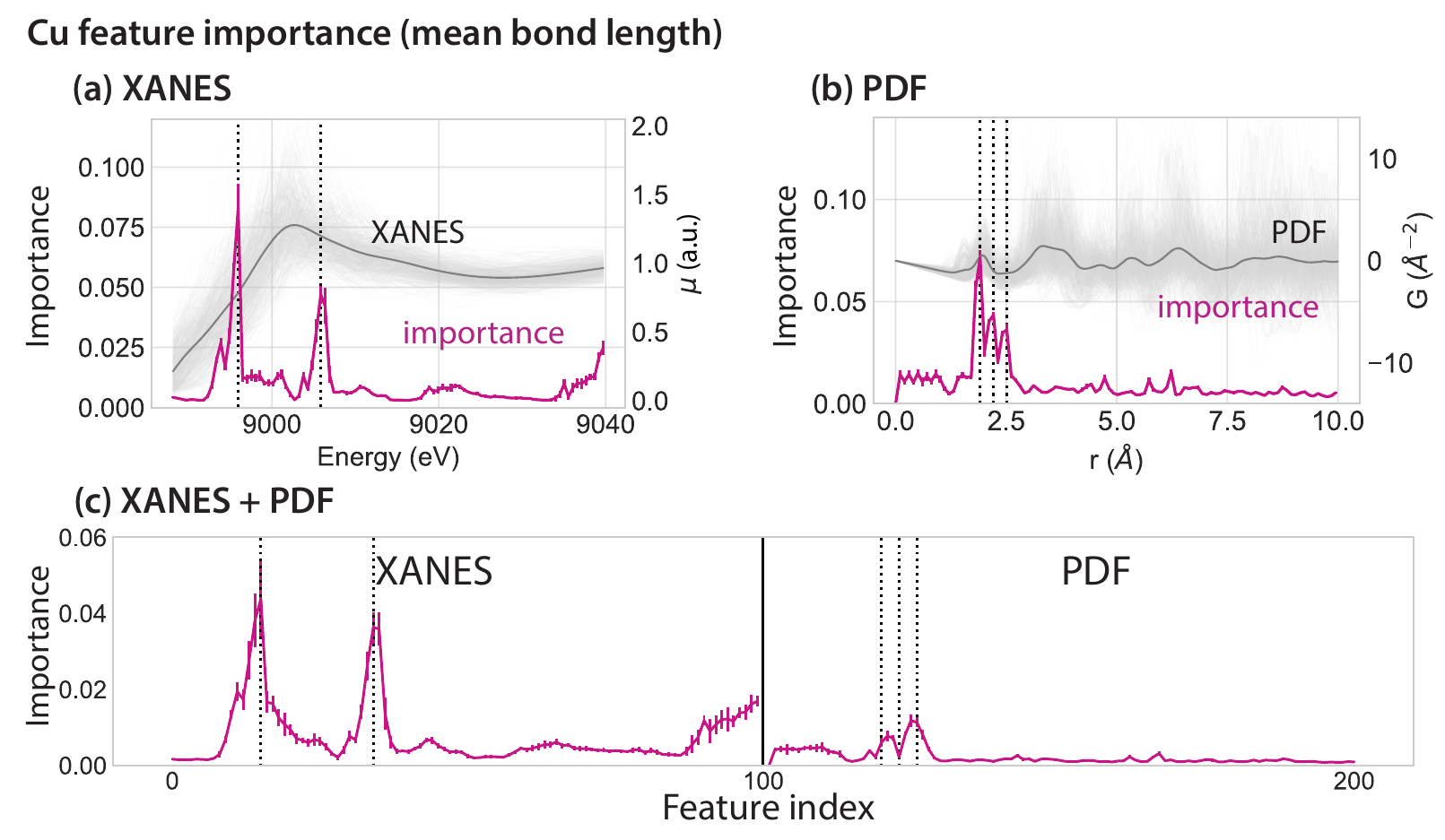}
    \caption{Cu feature importance when using (a) XANES, (b) PDF, (c) XANES + PDF as inputs. 
    (a) and (b) also show the spectra from the training sets and its mean given by the dark grey line. 
    Dotted lines mark the locations of prominent features of XANES and PDF separately. 
   The same lines are drawn on the combined XANES+PDF importance plot to highlight the changes.}\label{fig:SI_bl_Cu}
\end{figure}

Next, in addition to the Ti and Fe feature importance plots provided in the main text, \fig{SI_bl_Mn} and \fig{SI_bl_Cu} show the feature importance plots for Mn and Cu, respectively. 
For each metal, the subplots show feature importance for (a) XANES, (b) PDF, and (c) XANES + PDF combined. 

\clearpage
\section{Differential-PDF results}
\subsection{Oxidation state}
\begin{figure}
    \centering
    \includegraphics[width=0.8\textwidth]{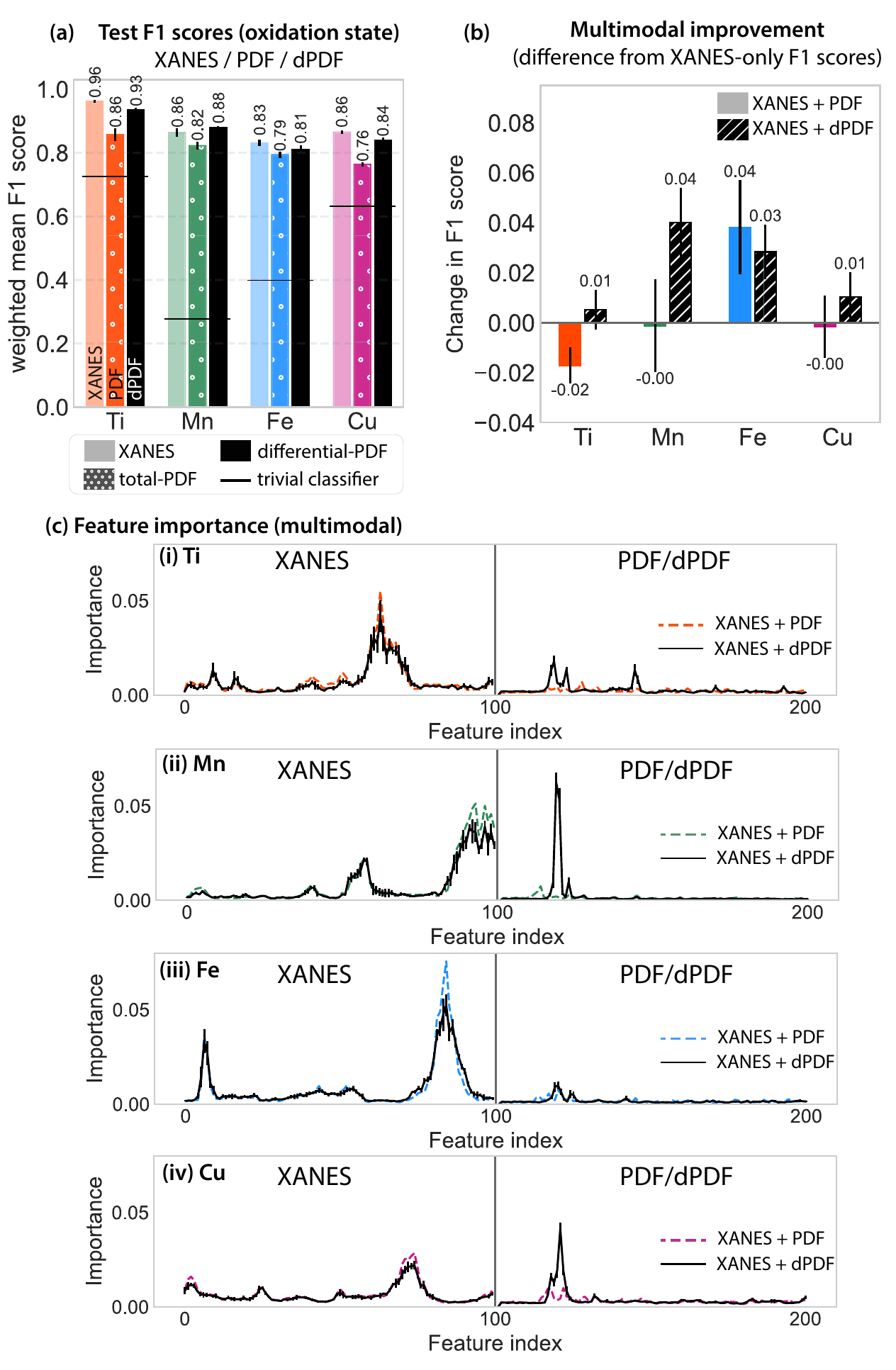}
    \caption{differential-PDF (dPDF) results on oxidation state classification: (a) test F1 scores for XANES, PDF, and dPDF models from left to right on the four datasets: Ti, Mn, Fe, and Cu.
   (b) Change in RMSEs when training the models on multimodal inputs, combining XANES with PDF (solid) or dPDF (black, striped).
   (c) Feature importance plots for the multimodal inputs, XANES+PDF (dashed line) and XANES+dPDF (solid line) for the four metals.}\label{fig:SI_cs_dpdf}
\end{figure}

\fig{SI_cs_dpdf} shows the model performances when trained with differential-PDFs instead of total-PDFs as input for oxidation state classification.
The important takeaways here include: (a) dPDF models performed better than total-PDF models and roughly as well as XANES models, and (b) multimodal XANES+dPDF models (black, hatched bars) performed slightly better than the XANES-only models, by the largest margins for Mn and Fe. 
As the feature importance plots in (c) show, this multimodal improvement in (b) might come from the increased contribution of the PDF input when using differential-PDFs as opposed to the total-PDFs.
At least in Ti, Mn, and Cu, dPDF features contributed more to the multimodal prediction when combined with XANES compared to when using total-PDFs as input.
In Fe, it is unclear from the feature importance that switching to dPDF made a significant difference.

\subsection{Coordination number}
\begin{figure}
    \centering
    \includegraphics[width=0.8\textwidth]{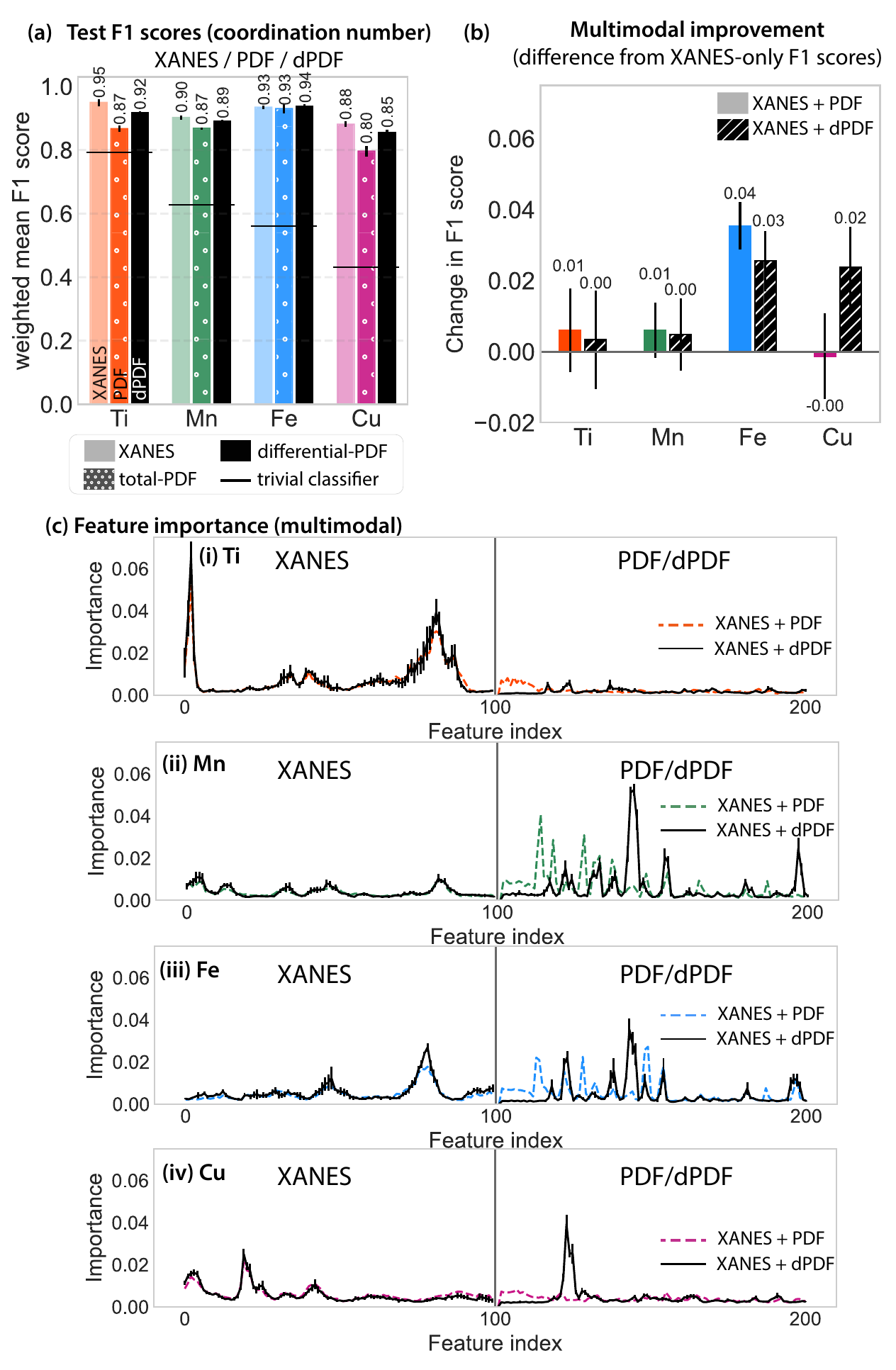}
    \caption{differential-PDF (dPDF) results on coordination number classification: (a) test F1 scores for XANES, PDF, and dPDF models from left to right on the four datasets: Ti, Mn, Fe, and Cu.
   (b) Change in RMSEs when training the models on multimodal inputs, combining XANES with PDF (solid) or dPDF (black, striped).
   (c) Feature importance plots for the multimodal inputs, XANES+PDF (dashed line) and XANES+dPDF (solid line) for the four metals.}\label{fig:SI_cn_dpdf}
\end{figure}

\fig{SI_cn_dpdf} shows the model performances when trained with differential-PDFs instead of total-PDFs as input for coordination number classification.
The important takeaways here include: (a) dPDF models performed better than total-PDF models and roughly as well as XANES models, and (b) multimodal XANES+dPDF models performed slightly better than the XANES-only models for Fe and Cu. 
For most metals except Ti, as the feature importance plots in (c) show, dPDF features contributed more to the multimodal prediction when combined with XANES compared to when using total-PDFs as input.
In contrast, the multimodal model for Ti still extracted most information from XANES (especially from the pre-edge region where a strong correlation is known~\cite{farges1997ti}), so switching from total- to differential-PDF barely made a difference.
On the other hand, although it appears that most information for Mn came from the PDF side of the input, the models trained on total- and differential-PDF performed almost equally well.

\subsection{Mean nearest-neighbor bond length}
\begin{figure}
    \centering
    \includegraphics[width=0.8\textwidth]{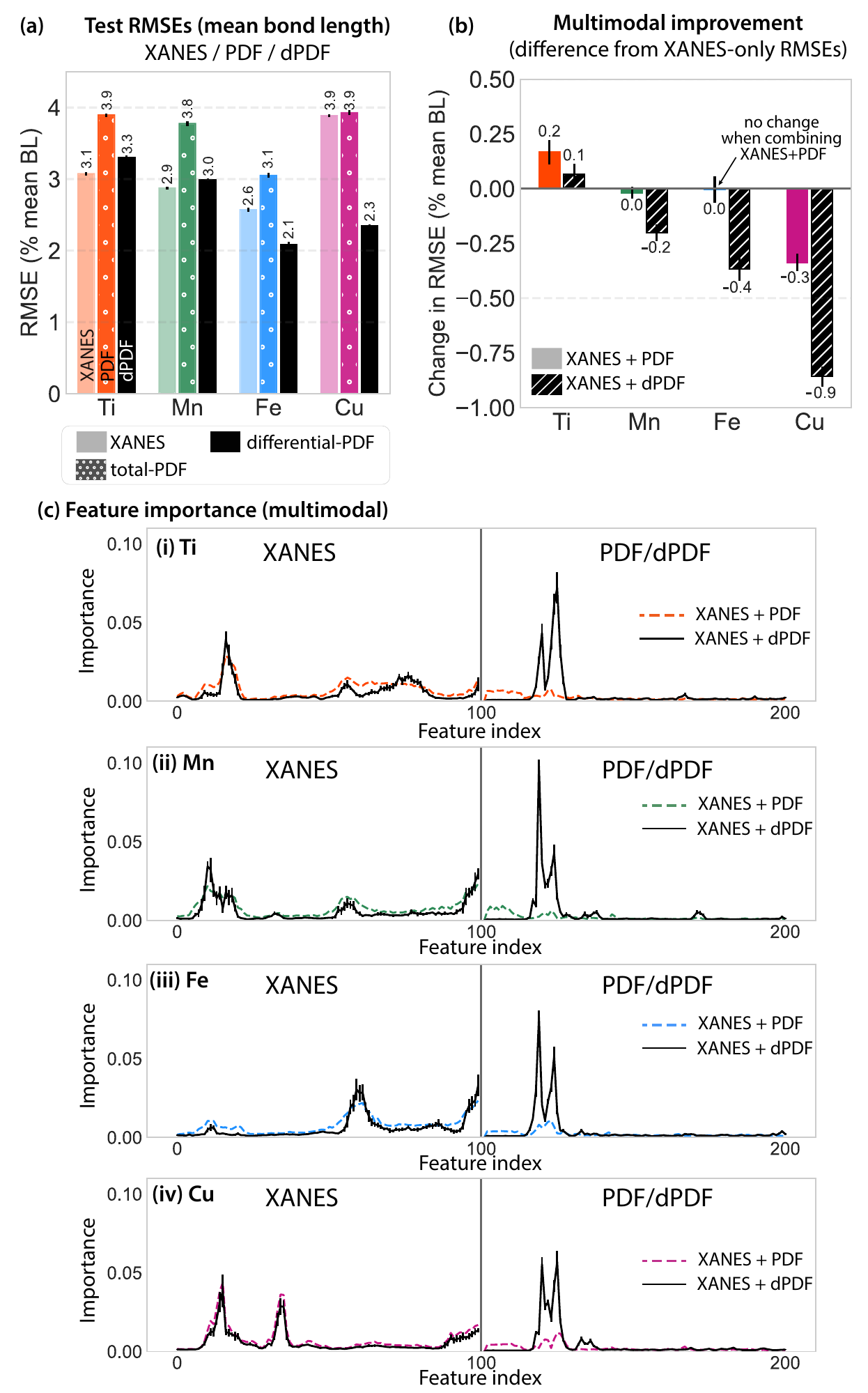}
    \caption{differential-PDF (dPDF) results on bond length regression: (a) test RMSE scores as percentages of mean bond length for XANES, PDF, and dPDF models from left to right on the four datasets: Ti, Mn, Fe, and Cu.
    Note that these are test errors, so shorter bars indicate better performance.
   (b) Change in RMSEs when training the models on multimodal inputs, combining XANES with PDF (solid) or dPDF (black, striped).
   (c) Feature importance plots for the multimodal inputs, XANES+PDF (dashed line) and XANES+dPDF (solid line) for the four metals.}\label{fig:SI_bl_dpdf}
\end{figure}

Using differential-PDFs instead of total-PDFs to train the models made the biggest improvement on bond length regression.
\fig{SI_bl_dpdf} shows the model performances when trained with differential-PDFs instead of total-PDFs as input for the mean nearest-neighbor bond length regression.
The barplots in (a) and (b) and the feature importance plot for Fe provided in the main text (Fig.~3) and reproduced here for easy comparison.
The important takeaways here include: (a) dPDF models performed better than total-PDF models and roughly as well as XANES models, and (b) multimodal XANES+dPDF models yielded lower errors than the XANES models for most metals (Mn, Fe, and Cu). 
For all metals here, as the feature importance plots in (c) show, dPDF features around the nearest-neighbor peak retained their importance (black line) even when combined with XANES, a stark contrast to when using total-PDF (dashed line) where the importance was heavily suppressed.

\clearpage
\section{Training on pre-edge vs post-edge XANES separately}
%cutoff energies = {'Ti': 4980, 'Mn': 6553, 'Fe': 7126, 'Cu': 8996}
Additionally, to investigate and compare the information contained in the pre-edge vs post-edge region of XANES more closely, we split the XANES input into two sections (before and after the edge) and trained the models on each section separately to predict the coordination number.
The splitting points (cutoff energies) for the four metals are 4980 eV for Ti, 6553 eV for Mn, 7126 eV for Fe, and 8996 eV for Cu. 
These values are slightly ($\approx$10 eV) above the K-edge energies of the metals.

\begin{figure}[b]
    \centering
    \includegraphics[width=0.55\textwidth]{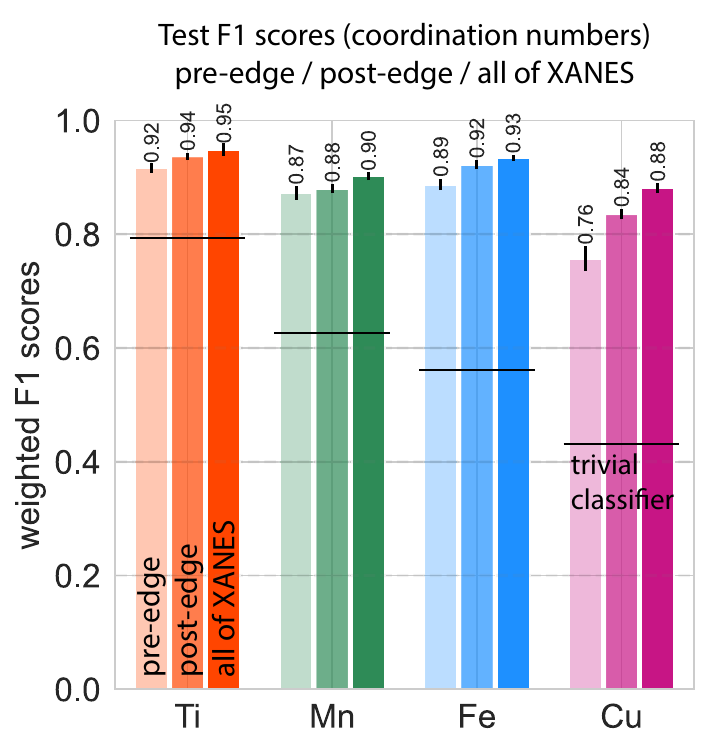}
    \caption{Test weighted mean F1 scores for the coordination number models trained on pre-edge XANES, post-edge XANES, and all of XANES from left to right on the four datasets: Ti, Mn, Fe, and Cu. The horizontal black lines indicate the baseline F1 scores achieved by a trivial classifier that labels all samples as the modal class.}\label{fig:SI_cn_split_xanes}
\end{figure}
\fig{SI_cn_split_xanes} shows that, consistently for all metals, the models performed best when trained with all regions of XANES (energy ranges given in Table~\ref{tab:SI_XANES_domains}).
The models performed slightly worse when only the post-edge region was used as input, but still better than when the input was the pre-edge region alone.
This indicates that in our dataset, which consists of FEFF-generated XANES spectra from the Materials Project database, the post-edge region might contain slightly more information than the pre-edge region.
The feature importance trends for the separate regions are largely the same as when using all regions of XANES, so we did not reproduce the plots here.
Carbone et al. performed a very similar test on their CNN models \cite{carbone_classification_2019} on the same prediction target: coordination number of the transition metal cations in oxides.
They also found that their models performed worse when using only the pre-edge region as the input rather than the whole range of XANES.

\subsection{Combining pre/post-edge XANES with PDF}
\begin{figure}[h]
    \centering
    \includegraphics[width=0.55\textwidth]{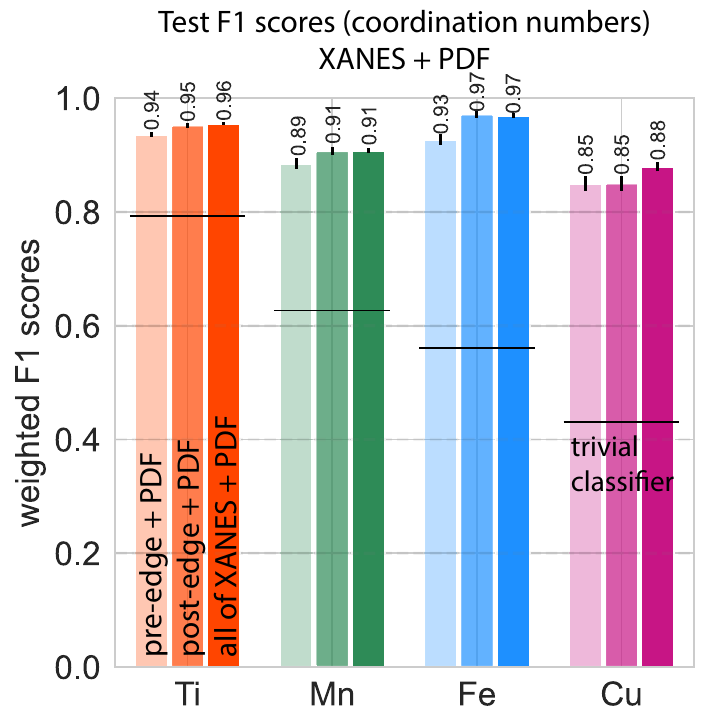}
    \caption{Test weighted mean F1 scores for the combined models (XANES+PDF) trained on pre-edge XANES + PDF, post-edge XANES + PDF, and all of XANES + PDF from left to right on the four datasets: Ti, Mn, Fe, and Cu. The horizontal black lines indicate the baseline F1 scores achieved by a trivial classifier that labels all samples as the modal class.}\label{fig:SI_cn_split_both}
\end{figure}

Next, we combined each region (pre vs post-edge) of XANES with PDF as the model input for coordination number prediction. 
\fig{SI_cn_split_both} shows that the test scores are very similar whether we use only the pre-edge region, the post-edge region, or all of XANES as input, but the post-edge models still performed slightly better than the pre-edge models.

\begin{figure}
    \centering
    \includegraphics[width=\textwidth]{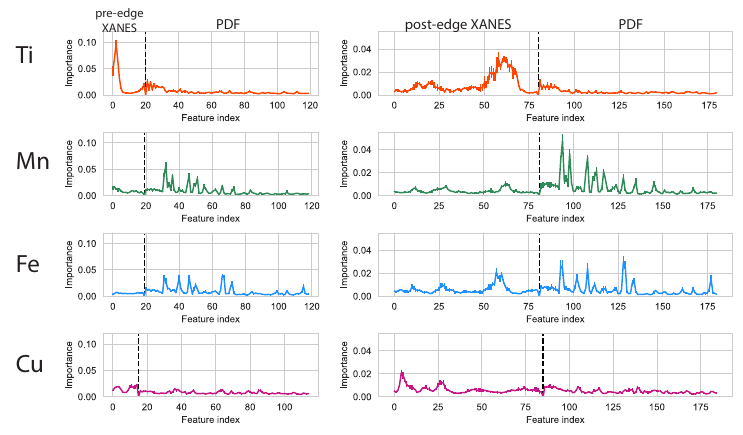}
    \caption{Feature importance plots for the combined models: pre-edge XANES + PDF (left column) and post-edge XANES + PDF (right column) for the four metals: Ti, Mn, Fe, and Cu.}\label{fig:SI_cn_split_both_importance}
\end{figure}

We plotted the respective feature importance plots in \fig{SI_cn_split_both_importance}. 
The left column shows the combined feature importance when training the models with pre-edge XANES and PDF together, and the right column shows results from combining post-edge XANES and PDF. 
Interestingly, Ti is the only metal here where pre-edge XANES heavily dominates the coordination number prediction.
For Ti, the post-edge region of XANES is also relatively more informative than PDF.
In contrast, as seen previously, the Mn and Fe models were able to extract substantial information from many PDF features including those in the higher $r$ range.
The trends are less clear for Cu, but post-edge XANES might contain the most coordination number information relative to the pre-edge region of XANES and the total PDF.


\subsection{Combining pre/post-edge XANES with dPDF}
\begin{figure}
    \centering
    \includegraphics[width=0.55\textwidth]{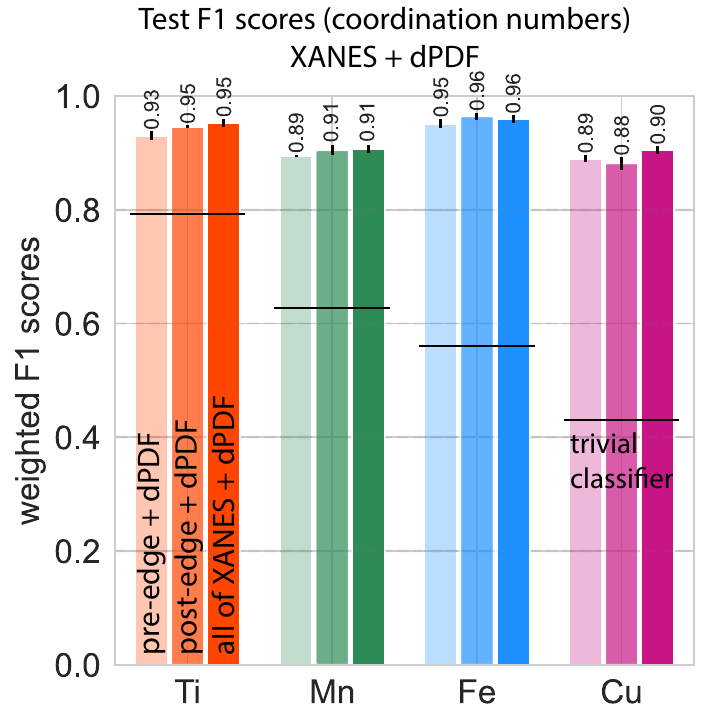}
    \caption{Test weighted mean F1 scores for the combined models (XANES+dPDF) trained on pre-edge XANES + dPDF, post-edge XANES + dPDF, and all of XANES + dPDF from left to right on the four datasets: Ti, Mn, Fe, and Cu. The horizontal black lines indicate the baseline F1 scores achieved by a trivial classifier that labels all samples as the modal class.}\label{fig:SI_cn_split_dboth}
\end{figure}

Finally, we repeat the same process with differential PDFs, separately combining the pre- and post-edge region of XANES with dPDFs as the model input for coordination number prediction. 
As with the total PDFs, \fig{SI_cn_split_dboth} shows that the test scores are very similar whether we use only the pre-edge region, the post-edge region, or all of XANES as input.

\begin{figure}
    \centering
    \includegraphics[width=\textwidth]{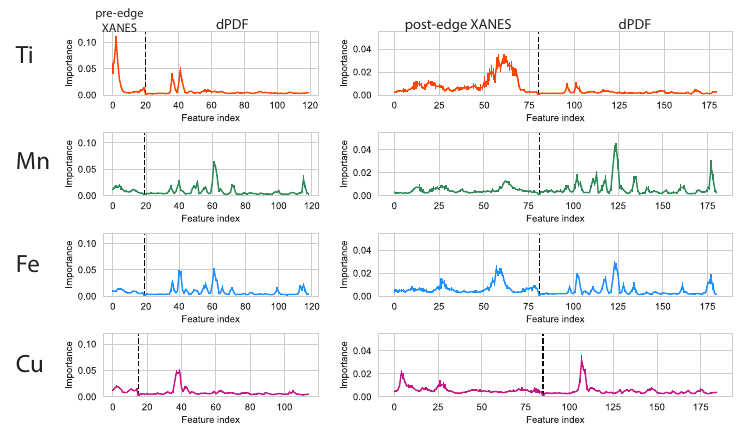}
    \caption{Feature importance plots for the combined models: pre-edge XANES + dPDF (left column) and post-edge XANES + dPDF (right column) for the four metals: Ti, Mn, Fe, and Cu.}\label{fig:SI_cn_split_dboth_importance}
\end{figure}

The respective feature importance plots are shown in \fig{SI_cn_split_dboth_importance}. 
Like before, the left column shows the combined feature importance when training the models with pre-edge XANES and dPDF together, and the right column shows results from combining post-edge XANES and dPDF. 
Like before, Ti is the only metal here where pre-edge XANES heavily dominates the coordination number prediction.
With the species-specific (metal's) dPDFs, Ti and Cu models were able to extract more coordination number information from around the nearest-neighbor PDF peak, although for Ti, information from XANES still heavily dominates.
For Cu, it became clearer that the model gained more information from dPDF than it did from pre-edge XANES.
As with total PDFs, for Mn and Fe, many dPDF features across the whole $r$ range (0-10 \AA) were found to be informative for the coordination number prediction.


\clearpage
%\bibliography{billinge-group-bib,bg-pdf-standards,tn_ml_x-ray}
%\printbibliography %Prints bibliography

\end{document}